\documentstyle[12pt,epsf]{article}
\renewcommand{\theequation}{\thesection.\arabic{equation}}
\renewcommand{\thefootnote}{\fnsymbol{footnote}}
\newlength{\extraspace}
\newlength{\extraspaces}
\newcommand{\be}{\begin{equation}
\addtolength{\abovedisplayskip}{\extraspaces}
\addtolength{\belowdisplayskip}{\extraspaces}
\addtolength{\abovedisplayshortskip}{\extraspace}
\addtolength{\belowdisplayshortskip}{\extraspace}}
\newcommand{\ee}{\end{equation}}
 
\newcommand{\ba}{\begin{eqnarray}
\addtolength{\abovedisplayskip}{\extraspaces}
\addtolength{\belowdisplayskip}{\extraspaces}
\addtolength{\abovedisplayshortskip}{\extraspace}
\addtolength{\belowdisplayshortskip}{\extraspace}}
\newcommand{\ea}{\end{eqnarray}}

\newcommand{\bas}{\begin{eqnarray*}
\addtolength{\abovedisplayskip}{\extraspaces}
\addtolength{\belowdisplayskip}{\extraspaces}
\addtolength{\abovedisplayshortskip}{\extraspace}
\addtolength{\belowdisplayshortskip}{\extraspace}}
\newcommand{\eas}{\end{eqnarray*}}
 
\newcounter{subequation}[equation]
\makeatletter

\expandafter\let\expandafter
\reset@font\csname reset@font\endcsname

\def\subeqnarray{\arraycolsep1pt
    \def\@eqnnum\stepcounter##1{\stepcounter{subequation}%
        {\reset@font\rm(\theequation\alph{subequation})}}
\jot5mm     \eqnarray}

\makeatother

\newcommand{\newappendix}[1]{
\vspace{15mm}
\pagebreak[3]
\addtocounter{section}{1}
\setcounter{equation}{0}
\setcounter{subsection}{0}
\setcounter{footnote}{0}
\renewcommand{\theequation}{\Alph{section}.\arabic{equation}}
\begin{flushleft}
{\large\bf \Alph{section}. #1}
\end{flushleft}
\nopagebreak
\medskip
\nopagebreak}

\newcommand{\newsection}[1]{
\vspace{15mm}
\pagebreak[3]
\addtocounter{section}{1}
\setcounter{equation}{0}
\setcounter{subsection}{0}
\setcounter{footnote}{0}
 
\begin{flushleft}
{\large\bf \thesection. #1}
\end{flushleft}
\nopagebreak
\medskip
\nopagebreak}
 
\newcommand{\newsubsection}[1]{
\vspace{1cm}
\pagebreak[3]
 
\addtocounter{subsection}{1}
\noindent{ \bf \thesection.\arabic{subsection} #1}
\nopagebreak
\vspace{2mm}
\nopagebreak}

\newcommand{\NP}[1]{Nucl.\ Phys.\ {\bf #1}}
\newcommand{\PL}[1]{Phys.\ Lett.\ {\bf #1}}

\newcommand{\C}{\mbox{$\,${\sf I}\hspace{-1.2ex}{\bf C}}}

\newcommand{\R}{\mbox{\rm I\hspace{-.4ex}R}}
\newcommand{\1}{\mbox{1\hspace{-.8ex}1}}
\newcommand{\bra}{\langle}
\newcommand{\ket}{\rangle}
\newcommand{\ra}{\rightarrow}

\newcommand{\rra}{\ \longrightarrow \ }

\newcommand{\is}{ &\!=\!& }
\newcommand{\nonum}{\nonumber \\[1.5mm]}
\newcommand{\sspace}{\makebox[1cm]{ }}
\newcommand{\bspace}{\makebox[2cm]{ }}
\newcommand{\nspace}{\!\!\!\!\!\!\!\!\!\!}

\newcommand{\th}{{\theta}}
\newcommand{\eps}{{\epsilon}}

\newcommand{\sh}{{\rm sh}}
\newcommand{\ch}{{\rm ch}}
\newcommand{\rmd}{{\rm d}}

\newcommand{\cF}{{\cal F}}

\newcommand{\cN}{{\cal N}}
\newcommand{\cO}{{\cal O}}

\newcommand{\kb}{\underline{k}}

\newcommand{\mb}{\underline{m}}
\newcommand{\lb}{\underline{l}}

\newcommand{\gr}{g_{\sc r}}
\newcommand{\Mr}{M_{\sc r}}
\newcommand{\Zr}{Z_{\sc r}}

\begin{document}
%
\begin{titlepage}
%
\renewcommand{\thefootnote}{\fnsymbol{footnote}}
\begin{flushright}
BUTP-2000-01 \\
MPI-PhT/2000-01\\
\end{flushright}
\vspace{1cm}

\begin{center}
{\LARGE The Intrinsic Coupling in}\\[3mm]
{\LARGE Integrable Quantum Field Theories}\vspace{1.6cm}

{\large J. Balog$^1$, M. Niedermaier$^2$, F. Niedermayer$^3$}\\[2mm]
{\large  A. Patrascioiu$^4$,  E. Seiler$^5$, P. Weisz$^5$}\\[7mm]

{\small\sl $^1$Research Institute for Particle and Nuclear Physics}\\
{\small\sl 1525 Budapest 114, Hungary}
\\[2mm]
{\small\sl $^2$Department of Physics, University of Pittsburgh}\\
{\small\sl Pittsburgh, PA 15260, U.S.A.}
\\[2mm]
{\small\sl $^3$Institute for Theoretical Physics, University of Bern}\\
{\small\sl CH-3012 Bern, Switzerland}
\\[2mm]
{\small\sl $^4$Physics Department, University of Arizona}\\
{\small\sl Tucson, AZ 85721, U.S.A.}
\\[2mm]
{\small\sl $^5$Max-Planck-Institut f\"{u}r Physik}\\
{\small\sl 80805 Munich, Germany}
\vspace{1cm}

{\bf Abstract}
\end{center}
\vspace{-5mm}

\begin{quote}
The intrinsic 4-point coupling, defined in terms of a truncated 4-point 
function at zero momentum, provides a well-established measure
for the interaction strength of a QFT. We show that this coupling can be
computed non-perturbatively and to high accuracy from the form factors 
of an (integrable) QFT. The technique is illustrated and tested with 
the Ising model, the XY-model and the O(3) nonlinear sigma-model. 
The results are compared to those from high precision lattice simulations.
\end{quote}
\vfill
\setcounter{footnote}{0}
\end{titlepage}

\newsection{Introduction}
The intrinsic coupling $\gr$, also sometimes called `physical' or
`renormalized' coupling, is a quantity of great interest in a
Quantum Field Theory (QFT), especially for scalar fields. In some cases,
such as the $\Phi^4$ theories, its vanishing implies actually
that the theory is trivial in the sense that the higher correlation
functions of the scalar field can be written as sums of products of
two point functions, as in a free theory \cite{Newman}. On the other
hand, a non-vanishing $\gr$ is not sufficient to assure the non-triviality
of a theory; it only assures that a certain four point vertex
function does not vanish identically, but does not exclude that
it vanishes on shell. 

Aside from that, $\gr$ is certainly a renormalization group invariant
and a characteristic physical quantity of a field theory. In
particular it can be used to check the equivalence or nonequivalence
of different definitions of theories; this will be our main theme in
this paper. $\gr$ is proportional to the connected -- sometimes called
truncated -- four point function at zero momentum, divided by the square 
of the zero momentum two point function and appropriate powers of the 
mass gap to make it dimensionless; details will be given in the body of 
the paper.

We are dealing in this article with two main approaches to the 
construction of a QFT. The first one starts from a suitably 
regularized functional integral and then removes the regularization in
a controlled way. This is a rather general procedure usable for a wide 
variety of models; it has been successfully employed to construct QFTs 
in 2 and 3 dimensions obeying all the required axioms (see for instance
\cite{GliJaff}). Here we will make use of a Euclidean space-time
lattice as a regulator. Removal of the regularization, i.e. taking the
continuum limit in a lattice theory requires the existence of a 
second order phase transition point at which the characteristic length 
(correlation length) of the model diverges. This approach raises the
problem of `universality', i.e. the question whether different
regularizations yield the same QFT after the regulator has been removed.

The other approach studied here is applicable to a large class of
so-called integrable models. It is not based on a Lagrangian,
rather the dynamics is specified in terms of a postulated 
exact `bootstrap' S-matrix, supposed to enjoy a factorization property 
that allows to express all S-matrix elements in terms of the two-particle 
S-matrix \cite{ZZ}. In physical terms this property is linked to the 
existence of an infinite number of conservation laws and the absence 
of particle production. The postulated S-matrices are then
used to set up a system of recursive functional equations for the form
factors; solving this system one can  in principle compute exactly all 
the form factors, in other words continue the S-matrix off the mass 
shell \cite{KaWe,Kar,Smir}. Once the form factors are known, one can 
express the correlation functions of the basic fields as well as other 
(composite) operators by inserting complete sets of scattering states 
between them. This gives the correlation functions as -- hopefully rapidly
converging -- infinite series of convolution products of form factors. 
In particular in this way one can express the intrinsic coupling in terms 
of the form factors. 

In both approaches, in principle one has to verify in the end that
the axioms of a QFT hold. In the lattice approach with a reflection
positive action, such as the standard nearest neighbor action, 
essentially the only nontrivial question besides the existence of a
critical point concerns the restoration of Euclidean (Poincar\'e)
invariance in the continuum limit. In the form factor approach it is 
less obvious whether the axioms hold, in particular for the form factor
expansion of multi-point correlation functions. There exists, however, 
a formal proof (disregarding convergence aspects) of locality \cite{Smir} 
and it is hoped, of course, that the other field theoretic axioms will hold as 
well, because the construction is to a large extent inspired by them. 
It is also not clear from first principles -- though in practice there are
very natural guesses --, which field in one construction should be
identified with which form factor sequence in the other. In any case, 
even assuming that all the axioms hold in both constructions and that one 
has correctly identified the fields, it is a nontrivial question whether 
the two approaches define the same theory, and in particular whether they 
give the same value for $\gr$.

In the lattice approach the intrinsic coupling of the two-dimensional
O$(n)$ models we are discussing here has been widely studied, both by
Monte Carlo simulations \cite{FMPPT,Kim,usgR} and by various expansions 
in a small parameter \cite{CPRVgR,CPRVhT,BC,peliss}. For a more precise 
comparison with the form factor approach, we also carried out our own 
high precision Monte Carlo simulations which are reported in this 
paper.

In the form factor approach the series for $\gr$, being a low energy 
quantity, is expected to converge very rapidly. Our results give every 
indication that these hopes are fully justified, though  the actual
computations turn out to be surprisingly intricate. In the present study 
we want to develop this computational framework, outline the 
computations and compare their results, where possible, to those 
obtained numerically from the lattice approach by the 
different methods mentioned above.

Remarkably in all the examples considered the first
non-trivial term in the series, which contains only one and two particle
intermediate states, appears to give about $98\%$ of the full answer (!).
Moreover for this dominant contribution a general model-independent 
expression in terms of the 1-and 3-particle form factors and the 
derivative of the S-matrix can be obtained. 

In this paper we discuss three models which can be viewed as the O($n$)
nonlinear sigma-models for $n=1,2,3$. Though formally members of the
O($n$) series of nonlinear sigma-models, the physics of these systems,
their form factor description, and not the least our motivation to 
study them is very different: The $n=1$ case is just the massive 
continuum limit of the Ising model. Here the spin form factors are very 
simple and we were able to push the computation of the series up to all 
terms with a total particle number (summed over the three intermediate 
states) of less or equal 8. The extremely rapid decay of the terms is 
manifest and we use the observed pattern as a guideline for the other 
systems. The final result amounts to a determination of $\gr$ with an 
estimated precision of better than $0.001 \%$. 

The $n=2$ case is better known as the XY-model. Here we rely on a
bootstrap description of the model, to which we hope to return in more 
detail elsewhere \cite{usO2}. Not all the form factors are known
explicitly, but the specific version of the 3-particle spin form factor 
needed for the dominant contribution can be found by elementary 
techniques. We compare this leading order result with that 
obtained by lattice techniques
and find reasonable agreement, which can be taken as 
support for the proposed bootstrap description.

Finally the $n=3$ model is the first with a nonabelian symmetry 
group. The evaluation of $\gr$ here is in part motivated by 
the controversy about the absence or presence of a Kosterlitz-Thouless
type phase transition; see \cite{usgR} for a more thorough discussion.

Let us remark that the form factor bootstrap has also been applied to 
the computation of $\gr$ in the sinh-Gordon model; in this model
the intrinsic coupling is especially interesting because of its 
relevance to the issue of ``triviality'' versus ``weak-strong-duality''.
For details see the accompanying paper \cite{MNSinh}.

The article is organized as follows. In the next section we describe
the form factor construction of Green's functions in terms of form
factors generally and derive the formula for the dominant 
contribution to the coupling. Further we prepare the ground for the 
computation of the sub-leading terms in the specific models. We then give 
a few generalities about the Monte Carlo simulations, and move on to 
discuss the three models as outlined above one by one in more detail,
comparing the results of the form factor construction to those obtained 
by the the lattice definition of the models; for the latter the values of
$\gr$ are estimated by high temperature expansions as well as Monte 
Carlo simulations.
\newpage
\newsection{Construction of Green functions in terms of form factors}

In this section we will consider a general massive QFT described in 
terms of its generalized form factor sequences 
by which we mean matrix elements of local operators between physical
states. We will restrict our
attention to the case of $d=2$ dimensions (although the extension 
to arbitrary $d$ is often straightforward).
The application of the representation to the
integrable models where the form
factors are explicitly known will be the subject of the next chapter.

\newsubsection{Generalities}

Our first goal is to construct the Euclidean correlation functions 
(Schwinger functions) from the generalized form factors. The Schwinger 
functions are convenient because they have simpler properties
than the Wightman functions and also because it facilitates the
comparison with lattice results later. For points $x_k \in \R^2$,
$k =1,\ldots ,L$, we denote by $(x_{k1},x_{k2})$ their components 
and by $\iota x_k = (-i x_{k2},x_{k1})$ a Wick rotated version.
For definiteness we will consider here correlation functions of $n$ 
scalar fields $\Phi^a(x),\,\,\,a=1,\dots,n$ (the generalization to 
other types of fields is straightforward). Then 
\ba
&& S^{a_1\ldots a_L}(x_1,\ldots,x_L) = \bra \Phi^{a_1}(x_1) \ldots
\Phi^{a_L}(x_L)\ket\;,\nonum
&& S^{a_1\ldots a_L}(x_1,\ldots,x_L) = W^{a_1\ldots a_L}
(\iota x_1,\ldots,\iota x_L)
\;\;\;\mbox{for}\;\;\;x_{12} > \ldots > x_{L2}\;.
\label{g1}
\ea
The first equation is the usual operator interpretation of the 
Schwinger functions. The second equation (\ref{g1}) then indicates the 
relation of the Schwinger function to the corresponding Wightman function 
for points $(z_1,\ldots,z_L) = (\iota x_1,\ldots, \iota x_L)$
in the ``primitive tube'' of analyticity.%
\footnote{We use the signature $(+,-)$ for (complexified) Minkowski
space in which case $(z_1,\ldots,z_L)$ is in the primitive tube iff 
$- {\rm Im}(z_k - z_{k+1})\in V^+,\;k =1, \ldots, L-1$, where $V^+$
is the forward light cone.} 
Outside the primitive tube the Schwinger functions can in 
principle likewise be obtained from the Wightman functions by analytic
continuation and are then found to be completely symmetric 
in all variables. In a form factor expansion however the 
primitive domain is preferred in that only there the convergence
of the momentum space integrals is manifest through exponential
damping factors (c.f. below). We thus mimic the effect of 
the analytic continuation by performing the symmetrization by 
hand
\ba
&&S^{a_1\ldots a_L}(x_1,\ldots,x_L)= \sum_{s \in {\cal S}_L}  
S_{\Theta}^{a_{s1}\ldots a_{sL}}(x_{s1},\ldots,x_{sL})\;,
\nonumber\\
&&S_{\Theta}^{a_1\ldots a_L}(x_1,\ldots,x_L):=    
\Theta(x_1,\ldots,x_L)W^{a_1\ldots a_L}(\iota x_1,\ldots,\iota x_L)\;,
\label{g2}
\ea
where $\Theta(x_1,\ldots, x_L)$ is a generalized step function 
that vanishes unless $x_{12} \geq \ldots \geq x_{L2}$ holds and the 
sum is over all elements of the permutation group ${\cal S}_L$. 
The functions $S_{\Theta}^{a_1\ldots a_L}(x_1,\ldots,x_L)$ 
are expected to have a 
convergent expansion in terms of form factors in the interior 
of their support region (as well as for certain points on the boundary). 
The cases of interest here are $L =2$ and $L =4$. Formally inserting a 
resolution of the identity in terms of asymptotic multi-particle
states $\1 = \sum_{\mb} |\mb \ket \bra \mb|$ 
one obtains 
\be
S_{\Theta}^{a_1a_2}(x_1,x_2)=\Theta(x_1,x_2) 
\sum_{\mb} e^{-(x_1 - x_2)_2 E_{\mb}}
e^{i(x_1 - x_2)_1 P_{\mb}}\bra 0|\Phi^{a_1}(0)|\mb\ket \bra
\mb|\Phi^{a_2}(0)|0\ket\;,     
\label{g3}
\ee
and 
\ba
S_{\Theta}^{a_1a_2a_3a_4}(x_1,x_2,x_3,x_4)
\is\,\Theta(x_1,x_2,x_3,x_4)\sum_{\kb,\lb,\mb} 
e^{-(x_1 - x_2)_2 E_{\kb} } e^{i(x_1 - x_2)_1 P_{\kb} }\,
\nonum &&
e^{-(x_2 - x_3)_2 E_{\lb} } e^{i(x_2 - x_3)_1 P_{\lb} }\,
e^{-(x_3 - x_4)_2 E_{\mb} } e^{i(x_3 - x_4)_1 P_{\mb} }\,
\nonum &&
\bra 0|\Phi^{a_1}(0)|\kb\ket \bra \kb|\Phi^{a_2}(0)|\lb\ket\,
\bra \lb|\Phi^{a_3}(0)|\mb \ket \bra \mb|\Phi^{a_4}(0)|0\ket\,.
\label{g4}
\ea
The states $|\mb\ket$ are assumed to be improper eigenstates
of the momentum operator $P_{\mu}$, and $E_{\mb}, P_{\mb}$ 
denote the eigenvalues of $P_0,P_1$ on $|\mb\ket$, respectively.
To write down an explicit parameterization
of the complete set of states $|\mb\ket$
requires of course the full knowledge of the spectrum of 
stable particles. This is a basic input assumption for the
integrable models dealt with in the next section.
Here for simplicity of notation we will consider the case
where there is only one multiplet of stable particle states of
mass $M$. An explicit parameterization will then be given 
in subsection 2.3.

We introduce their (dimensionless) Fourier transforms $V$ by
\ba
&& (2\pi)^2 \delta^{(2)}(k_1 + \ldots + k_L) \, M^{-2(L-1)}\,
V^{a_1\ldots a_L}(k_1,\ldots,k_L) 
\nonum
&& \sspace = \int \rmd^2x_1 \ldots \rmd^2x_L \,
S_{\Theta}^{a_1\ldots a_L}(x_1,\ldots,x_L)\,
e^{i(k_1x_1 + \ldots + k_L x_L)}\,,
\label{g5}
\ea
taking into account the translation invariance of $S_{\Theta}$. 
The Fourier transform of the full Schwinger function is then obtained 
by symmetrization
\be
\tilde{S}^{a_1\ldots a_L}(k_1,\ldots,k_L) = 
(2\pi)^2 \delta^{(2)}(k_1 + \ldots + k_L)
M^{-2(L-1)}
\sum_{s \in {\cal S}_L} V^{a_{s1}\ldots a_{sL}}(k_{s1},\ldots,k_{sL})\;,
\label{g5b} 
\ee
whereon rotational invariance gets restored. 
The desired representation of the two and four point functions
in terms of form factors is given by  
\ba
V^{a_1a_2}(k_1,k_2) \is   \sum_{\mb}V_{\mb}^{a_1a_2}(k_1,k_2)\,,
\nonum
V^{a_1a_2a_3a_4}(k_1,k_2,k_3,k_4) \is  \sum_{\kb,\lb,\mb} 
V_{\kb\lb\mb}^{a_1a_2a_3a_4}(k_1,k_2,k_3,k_4)\,,
\label{g6}
\ea
where
\ba
V_{\mb}^{a_1a_2}(k_1,k_2) \is  2\pi M^2\, 
\frac{ \delta(P_{\mb} + k_{11})}{E_{\mb} - i k_{12}}
\bra 0|\Phi^{a_1}(0)|\mb\ket \bra \mb|\Phi^{a_2}(0)|0\ket\;,  
\nonum
V_{\kb\lb\mb}^{a_1a_2a_3a_4}(k_1,k_2,k_3,k_4) \is  (2\pi M^2)^3\, 
\frac{ \delta(P_{\kb} + k_{11})}{E_{\kb} - i k_{12}}
 \frac{ \delta(P_{\lb} + k_{11}+ k_{21})}{E_{\lb} - i k_{12} - i k_{22}}
 \frac{ \delta(P_{\mb} - k_{41})}{E_{\mb} + i k_{42}}
\nonum && \bra 0|\Phi^{a_1}(0)|\kb\ket \bra \kb|\Phi^{a_2}(0)|\lb\ket\,
\bra \lb|\Phi^{a_3}(0)|\mb \ket \bra \mb|\Phi^{a_4}(0)|0\ket\,,
\label{g7}
\ea
with the understanding that the sum of the momenta $k_j$ vanishes.
Further we denote by $V^{a_1a_2}_m(k_1,k_2)$ and 
$V^{a_1a_2a_3a_4}_{klm}(k_1,k_2,k_3,k_4)$ the quantities
(\ref{g7}) with the integrations over the rapidities performed,
the measure being inherited from Eq.~(\ref{statesum}) below.

The key assumption of the form factor approach in this context is 
that the matrix elements in (\ref{g7}) can be computed exactly 
via solutions of a recursive system of functional equations, the 
so-called form factor equations or Smirnov axioms. Symbolically
\be
\bra \lb| \Phi^a(0) |\mb \ket \;\;\;\longleftrightarrow
\;\;\; {\cal F}^a_{b_1\ldots b_l a_1 \ldots a_m}(\omega_1,\ldots
,\omega_l|\th_1,\ldots,\th_m) =: {\cal F}^a_{BA}(\omega|\th)\;.
\label{g8}
\ee
The rhs, for which we shall often use the indicated shorthand 
notation, is called a generalized form factor, the special 
case with either $l=0$ or $m=0$ are the form factors proper. 
The form factors are meromorphic functions in the rapidities,
while the generalized form factors are distributions. 
The form factors can be computed, at least in principle, as 
solutions of the before mentioned system of functional equations.
The generalized form factors can then be obtained from them
by means of an explicit, though cumbersome, combinatorial 
formula. We shall later just state the special cases of this
formula required. A discussion of the general formula can e.g.~be 
found in the appendix of \cite{MNmod}. 

Implicit in the products of matrix elements in (\ref{g7}) of course
are appropriate index contractions. For definiteness let us note them 
explicitly
\ba
\bra 0 |\Phi^a(0)|\mb\ket \bra \mb|\Phi^b(0) |0\ket
& \; \longleftrightarrow \;& I^{ab}_m(\th)\;,
\nonum
\bra 0|\Phi^a(0)|\kb\ket \bra \kb|\Phi^b(0)|\lb\ket\,
\bra \lb|\Phi^c(0)|\mb \ket \bra \mb|\Phi^d(0)|0\ket
&\; \longleftrightarrow \;& I_{klm}^{abcd}(\omega|\xi|\th)\;,
\label{g9}
\ea
where
\ba
&& I_m^{ab}(\th) = \sum_A \cF^a_A(\th)\,\cF^b_{A^T}(\th^T)\;,
\nonum
&& I_{klm}^{abcd}(\omega|\xi|\th)= \sum_{A,B,C}
\cF^a_A(\omega)\,\cF^b_{A^TB}(\omega^T|\xi)\,
\cF^c_{B^TC}(\xi^T|\th)\,\cF^d_{C^T}(\th^T)\,.
\label{g10}
\ea
Here $A^T =(a_k,\ldots,a_1)$, $\omega^T = (\omega_k,\ldots,\omega_1)$,
etc. The construction is such that $I^{ab}_m(\th)$ is a completely 
symmetric function in $\th= (\th_1,\ldots,\th_m)$. Similarly
$I_{klm}^{abcd}(\omega|\xi|\th)$ is symmetric
in each of the sets of variables $\omega = (\omega_1,
\ldots,\omega_k)$, $\xi = (\xi_1,\ldots,\xi_l)$ and
$\th = (\th_1,\ldots,\th_m)$, individually.

\newsubsection{The intrinsic coupling}

As surveyed in the introduction the intrinsic coupling is defined 
in terms of the zero momentum limit of a connected 4-point
function. We may assume that the Schwinger functions of the scalar 
fields with an odd number of arguments vanish;
then the connected $L=2,4$ Schwinger functions of interest here are 
\ba
\tilde{S}^{a_1a_2}_c(k_1,k_2) \!\is \!\tilde{S}^{a_1a_2}(k_1,k_2)\;,\nonum
\tilde{S}^{a_1a_2a_3a_4}_c(k_1,k_2,k_3,k_4) \!\is\!  
\tilde{S}^{a_1a_2a_3a_4}(k_1,k_2,k_3,k_4) -
\tilde{S}^{a_1a_2}(k_1,k_2)\tilde{S}^{a_3a_4}(k_3,k_4) 
\nonum \!&-&\! \tilde{S}^{a_1a_3}(k_1,k_3)\tilde{S}^{a_2a_4}(k_2,k_4) -
\tilde{S}^{a_1a_4}(k_1,k_4)\tilde{S}^{a_2a_3}(k_2,k_3)\,. 
\label{b5}
\ea
Making explicit the overall delta-functions arising from translational
invariance we introduce the Green functions by  
\be
\tilde{S}_c^{a_1\ldots a_L}(k_1,\ldots,k_L)=(2\pi)^2 
\delta^{(2)}(k_1+\ldots+k_L)\,G^{a_1\ldots a_L}(k_1,\ldots,k_L)\,,
\label{gcdef}
\ee
where the constraint $k_1 + \ldots + k_L =0$ in the arguments
of $G^{a_1\ldots a_L}$ will always be understood. 

In the following we will now assume that the theory is O$(n)$
invariant and thus for the 2-point function we can write
\be
G^{a_1a_2}(k,-k)=\delta^{a_1a_2}G(k)\,.
\ee 
The intrinsic coupling is then defined by
\be
\gr=-\cN \frac{M^2}{G(0)^2}\frac{1}{n^2}\sum_{a,b}G^{aabb}(0,0,0,0)\,,
\label{grdef1}
\ee
where we leave the choice of positive constant $\cN$ for later.
 
Performing the symmetrization (\ref{g2}) and the Fourier transform 
one recovers the familiar expression for $G(k)$ in terms of the
spectral density
\be
G(k) = \int_0^{\infty} \rmd\mu\, 
\rho(\mu) \frac{1}{\mu^2 + k^2}\,,
\label{b6}
\ee
where 
\be
\rho(\mu) =
\sum_{\mb}\delta(\mu-\sqrt{E_{\mb}^2-P_{\mb}^2})\,4\pi E_{\mb}
\delta(P_{\mb})\frac{1}{n}\sum_a
\bra 0|\Phi^a(0)|\mb\ket \bra \mb|\Phi^a(0)|0\ket\,. 
\label{b7}
\ee

In order to compute $\tilde{S}^{a_1a_2a_3a_4}(k_1,k_2,k_3,k_4)$ 
the symmetrized sum (\ref{g5b}), (\ref{g6}) 
has to be performed. For reasons that will become clear 
immediately we first single out 
the partial sum with $\lb=0$. Taking into account the ${\cal S}_4$
permutations  one finds
\ba
&& (2\pi)^2 \delta^{(2)}(k_1+k_2+k_3+k_4) M^{-6}\,
\sum_{\kb,\mb}\sum_{s \in {\cal S}_4} V^{a_1a_2a_3a_4}_{\kb 0\mb}
(k_{s 1},k_{s 2},k_{s 3},k_{s 4})
\nonum
&& = \tilde{S}^{a_1a_2}(k_1,k_2)\tilde{S}^{a_3a_4}(k_3,k_4) 
+ \tilde{S}^{a_1a_3}(k_1,k_3)\tilde{S}^{a_2a_4}(k_2,k_4)
+ \tilde{S}^{a_1a_4}(k_1,k_4)\tilde{S}^{a_2a_3}(k_2,k_3) 
\nonum
&& + (2\pi)^2 \delta^{(2)}(k_1+k_2+k_3+k_4)M^{-6}\,\sum_{s \in {\cal S}_4}
\Omega^{a_{s1}a_{s2}a_{s3}a_{s4}}(k_{s 1},k_{s 2},k_{s 3},k_{s 4})\,.
\label{b8}
\ea
Here 
\be
\Omega^{a_1a_2a_3a_4}(k_1,k_2,k_3,k_4) =
-\frac{\pi}{2}\delta(k_{11}+k_{21}) M^6
\,\delta^{a_1a_2}\delta^{a_3a_4}H(k_1,k_2)G(k_4)\;,
\label{b9}
\ee
with 
\be
H(k_1,k_2) = \int_0^\infty\,\rmd\mu\rho(\mu)\,
{\mu^2+k_{11}^2+k_{12}k_{22}
\over(\mu^2+k_1^2)(\mu^2+k_2^2)}{1\over\sqrt{\mu^2+k_{11}^2}}\,.
\label{b10}
\ee
In obtaining (\ref{b8}) we defined the second denominator in 
(\ref{g7}) for $\lb=0$ with the $i\epsilon$ prescription as:
$-i(k_{12}+k_{22}+i\epsilon)$. Here and later the distributional identity
$$
{1\over x+i\epsilon}={\cal P}{1 \over x}-i\pi\delta(x)\,,
$$
will be heavily used, where ${\cal P}$ is the Principal Value prescription.
 
One observes that the first three terms in (\ref{b8}) are precisely the
ones removed by the definition of the connected 4-point function. 
Remarkably there is a remainder, the $\Omega$ term, which is present 
even in the free theory. Typically the spectral densities are decreasing or
bounded by a constant as $\mu \ra \infty$. The functions $G(k_4)$ and 
$H(k_1,k_2)$ are then regular at $k_i=0$. Inserting finally (\ref{b8}) 
into (\ref{g5b}), (\ref{g6}) one obtains for 
the Green function (\ref{gcdef})
\ba
&\nspace & M^6 G^{a_1a_2a_3a_4}(k_1,k_2,k_3,k_4) \nonum
&\nspace &  = \sum_{\kb,\lb\neq 0,\mb} \sum_{s\in {\cal S}_4}
V^{a_{s1}a_{s2}a_{s3}a_{s4}}_{\kb\lb\mb}(k_{s 1},k_{s 2},k_{s 3},k_{s 4})
+ \sum_{s \in {\cal S}_4}   
\Omega^{a_{s1}a_{s2}a_{s3}a_{s4}}(k_{s 1},k_{s 2},k_{s 3},k_{s 4})\,.
\label{b13}
\ea

On general grounds one expects the vertex function to be real analytic.
In particular there must also be terms involving the 
delta function in the $V_{\kb\lb\mb}$ above
which cancel those of the $\Omega$-term (we will demonstrate
this explicitly in the computation of $V_{121}$ in appendix A). 
Thus, provided the coupling is well-defined (finite)
at all, the result will be independent of the way the zero momentum limit
is taken. It is therefore desirable to find a convenient limiting procedure
that simplifies the computation. To this end we first
observe that in (\ref{g7}) the $k_{j1}$ and $k_{j2}$ components enter 
asymmetrically. In particular as long as the intermediate state is
not the vacuum (i.e. $\lb\neq 0$ in the second formula, and
recalling that we are assuming that
none of the operators involved has a non-zero vacuum expectation value) 
one can put $k_{j2} =0$, $j =1,2,3,4$. We now compute
the 4-point vertex function at zero momentum through the limiting
procedure:
\be
G^{a_1a_2a_3a_4}(0,0,0,0)=\lim_{\kappa_j\to0}
G^{a_1a_2a_3a_4}(k_1,k_2,k_3,k_4)\bigg|_{k_j = (M {\rm sh} \kappa_j,0)}
\label{b1}
\ee
where $\sum_{j=1}^4 \sh \kappa_j =0$, and the limit is taken such 
that
\be
|\kappa_i| \neq |\kappa_j| 
\,\,\,\,\mbox{for}
\,\, i \neq j\,\,,\mbox{and}\,\, 
|\kappa_i-\kappa_j|
\neq |\kappa_k-\kappa_l|\,\,\,\mbox{for distinct pairs}\,.
\label{kappas}
\ee
In view of (\ref{b9}) it is clear that the limit prescription in 
(\ref{b1}), which we will use in the following, has just been designed
such that the $\Omega$ term does not have to be considered in 
computation of the coupling. 

Before embarking on further computations let us comment on a few 
structural issues. 
On physical grounds one expects the intrinsic coupling to be both finite 
(in a theory with a mass gap) and positive 
(for $\cN > 0$) when the interaction is repulsive.  
Mathematically however it is a quite challenging problem to actually
prove this, whatever non-perturbative definition of the theory one adopts.
In the context of constructive (lattice) QFT such results seem to
be available only for a single phase $\Phi^4_2$ theory 
(see e.g~\cite{GliJaff} for a survey).
In the present context we wish to define the theory strictly terms of its 
form factors. Mathematically speaking one should then try to prove in 
particular that the right hand side of (\ref{b13}) defines a real 
analytic function. For the dominant low particle contributions we 
demonstrate in appendix A explicitly that all non-analytic 
(e.g. distributional) 
terms indeed cancel out. We have not attempted to prove this in general,
nor can we estimate the rate of convergence of the sums in (\ref{b13}) on
general grounds. 
In all the examples considered later however the series appears 
to be rapidly convergent; the terms are alternating in sign and decrease
in magnitude very quickly with increasing particle numbers.

\newsubsection{State parameterization}

Here we assume that the single particle spectrum consists 
only of an O$(n)$ vector multiplet of mass $M$. 
The one particle states
$|a,\alpha\ket$ are thus specified by an internal ``isospin" label $a$
and the rapidity $\alpha$ (i.e. the spatial momentum of
the state is $p=M\sh\alpha$). The states are normalized according to 
\be
\bra a,\alpha |b,\beta\ket =4\pi\delta_{ab}\delta(\alpha-\beta)\,.
\ee
The condensed notation for the sum over states now becomes 
\ba
&&\sum_{\mb}|\mb\ket\bra\mb |  \;\;\longleftrightarrow \;\;
|0\ket\bra 0|+
\sum_{m=1}^{\infty}\sum_{a_1,\ldots a_m}
\nonumber\\
&&\int_{-\infty}^{\infty}\frac{\rmd\th_1}{4\pi}
\int_{-\infty}^{\th_1}\frac{\rmd\th_2}{4\pi}\ldots
\int_{-\infty}^{\th_{m-1}}\frac{\rmd\th_m}{4\pi}
|a_1,\theta_1;\ldots;a_m,\theta_m\ket^{\rm in} 
\,\phantom{}^{\rm in}\bra a_1,\theta_1;\ldots;a_m,\theta_m |\,.
\label{statesum}
\ea
It is often convenient (for a fixed $m$) to perform the change of
variables
\be
u_j = \th_j - \th_{j+1}\,,\;\;\;j =1,\ldots, m-1\;,\sspace
\Lambda = \frac{1}{2}\ln
\left(\frac{\sum_j e^{\th_j}}{\sum_j e^{-\th_j}}\right)\,,
\label{trans2}
\ee
since in terms of these variables the total energy and momentum 
of the states take on a simpler form:
\be
(E_{\mb},\, P_{\mb})\;\; \longleftrightarrow \;\; \bigg(M \sum_{j =1}^m
\ch \th_j,\,
M \sum_{j =1}^m \sh \th_j\bigg) = 
\left(M^{(m)}(u)\,\ch\Lambda,\,M^{(m)}(u)\,\sh\Lambda\right) \;,
\ee
where the
eigenvalues $M_{\mb} = \sqrt{E_{\mb}^2 - P_{\mb}^2}$ of the mass 
operator are given by
\be
M^{(m)}(u) = M\bigg[ m + 2 \sum_{i < j} 
\ch (u_i + \ldots + u_{j-1})\bigg]^{1/2}\;.
\label{trans4}
\ee
Correspondingly the integration measures in (\ref{statesum}) above are
replaced by
\be
\int_{-\infty}^{\infty}\frac{\rmd\th_1}{4\pi}
\int_{-\infty}^{\th_1}\frac{\rmd\th_2}{4\pi}\ldots
\int_{-\infty}^{\th_{m-1}}\frac{\rmd\th_m}{4\pi}  
\;\; \rra\;\; \int_0^{\infty} \frac{\rmd^{m-1}u}{(4\pi)^{m-1}}
\int_{-\infty}^{\infty} \frac{\rmd \Lambda}{4\pi}\;.
\label{trans5}
\ee
For later reference we also display the inverse transformation
\ba
&& \th_j = u_j + \ldots + u_{m-1} + u_m + \Lambda\,,\;\;\;
j =1,\ldots, m\;,\nonum
&& \mbox{where} \;\;\;u_m := \frac{1}{2}\ln
\left(\frac{1+ \sum_{j=1}^{m-1} e^{-u_j - \ldots - u_{m-1}}}%
{1+ \sum_{j=1}^{m-1} e^{u_j + \ldots + u_{m-1}}}\right)\,.
\label{trans3}
\ea

\newsubsection{The two point function}

The spectral function (\ref{b7}) appearing in the representation of the
two point function can be written as a sum of contributions
of fixed particle number $m$
\be
\rho(\mu)=\sum_{0<m\,\,{\rm odd}} \rho^{(m)}(\mu)\,,
\label{spectdec}
\ee
where only odd numbers of particles contribute due to our assumption
that the fields $\Phi^a$ are parity odd.
We normalize the fields $\Phi^a$ by 
\be
\bra 0|\Phi^a(0)|b,\alpha\ket=\delta^a_b\,,
\ee
rendering the 1-particle contribution to the spectral density simply 
\be
\rho^{(1)}(\mu)=\delta(\mu-M)\,.
\ee
The $m\ge3$-particle contribution to the spin spectral function
(\ref{b7}) is given by
\be   
\rho^{(m)}(\mu)= \int_0^{\infty}\frac{\rmd^{m-1} u}{(4\pi)^{m-1}}
\,\delta(\mu-M^{(m)}(u))I_m(u)\,,
\ee  
with
\be
I_m(u):= {1\over n}\sum_a\sum_{a_1,\dots,a_m}
|{\cal F}^a_{a_1\dots a_m}(\theta_1,\dots,\theta_m)|^2\,,
\ee
which equals $I^{11}_m(\th)$ under the integral. 
The function ${\cal F}^a$ featuring here corresponds to the matrix element 
of $\Phi^a$ between vacuum and an $m$-particle in-state as in 
(\ref{g8})
\be
{\cal F}^a_{a_1\dots a_m}(\theta_1,\dots,\theta_m)=
\bra 0|\Phi^a(0)|a_1,\theta_1;\ldots;a_m,\theta_m\ket^{\rm in}\,,
\sspace \theta_m<\ldots<\theta_1\,.
\label{calF}
\ee

The inverse 2-point function has a low momentum expansion
of the form 
\be
G(k)^{-1} = 
\Zr^{-1}[\Mr^2 + k^2 + O(k^4)]\,, 
\ee
with
\be
\Mr^2 = M^2 \frac{\gamma_2}{\delta_2}\;,\sspace
\Zr = \frac{\gamma_2^2}{\delta_2}\;,
\label{norm2} 
\ee
where $\gamma_2, \delta_2$ are spectral moments:
\be
\gamma_2 = M^2 \int_0^{\infty}
\frac{\rmd\mu}{\mu^2} \rho(\mu)\;,\sspace 
\delta_2 = M^4 \int_0^{\infty}
\frac{\rmd\mu}{\mu^4} \rho(\mu)\;.
\label{norm1}
\ee

\newsubsection{The intrinsic coupling revisited}

In (\ref{grdef1}) we left open the choice of the normalization constant
$\cN$ because for different models different choices are convenient.
In analytical and numerical lattice computations 
(at fixed cutoff) it is often easier to compute the
second moment mass $\Mr$ instead of the (exponential) spectral mass $M$
(in lattice units). For ease of comparison with these techniques we
thus choose $\cN=\Mr^2/M^2$, i.e. we define the intrinsic coupling by
\be
\gr=-\frac{\Mr^2}{G(0)^2}\frac{1}{n^2}\sum_{a,b}G^{aabb}(0,0,0,0)\,.
\label{grdef} 
\ee
Using O$(n)$ symmetry it follows  
\be
G^{a_1a_2a_3a_4}(0,0,0,0) = M^{-6}\gamma_4(\delta^{a_1a_2}
\delta^{a_3a_4} + \delta^{a_1a_3}\delta^{a_2a_4} +
\delta^{a_1a_4} \delta^{a_2a_3})\;,
\ee
and hence we can write (\ref{grdef}) as
\be
\gr=-\frac{n+2}{n}\frac{\gamma_4}{\gamma_2\delta_2}\,.
\label{grx}
\ee

These spectral moments have, corresponding to the 
decomposition (\ref{spectdec}),
an expansion in contributions arising from states 
with a fixed (odd) number of particles
\be
\gamma_2=1+\sum_{3\geq m\,\,{\rm odd}} \gamma_{2;m}\,,\sspace
\delta_2=1+\sum_{3\geq m\,\,{\rm odd}} \delta_{2;m}\,.
\label{gamdel}
\ee
Similarly, corresponding to the sum in (\ref{b13}) we have
\be
\gamma_4=\sum_{k,l>0,m}\gamma_{4;klm}\,,\sspace 
\gamma_{4;klm}=\gamma_{4;mlk}\,,
\label{gamma4exp}
\ee
where the sum goes over odd integers $k,m$
and positive even integers $l$. 
To avoid writing many O$(n)$ indices we will use
\be
\gamma_{4;klm}=\frac{1}{3}{\rm Lim}_{\kappa_j\to0}\sum_{s\in {\cal S}_4}
v_{klm}(\kappa_{s1},\kappa_{s2},\kappa_{s3},\kappa_{s4})\,,\,\,\,\,\,
\ee
where
\be
v_{klm}(\kappa_1,\kappa_2,\kappa_3,\kappa_4)=
V_{klm}^{1111}(k_1,k_2,k_3,k_4)\bigg|_{k_j=(M\sh\kappa_j,0)}\,,
\ee
and the symbol ${\rm Lim}$ above means taking the limit $\kappa_j\to0$
with the $\kappa_j$ satisfying $\sum_j\sh\kappa_j=0$ and the 
constraints in Eq.~(\ref{kappas}).

\newpage
\newsection{The nonlinear O{\boldmath $(n)$} sigma-models}

As outlined in the introduction the form factor bootstrap (FFB)
construction of an integrable quantum field theory starts from postulates
of the on shell properties of the theory. By integrable here it is meant
that the theory has an infinite set of conserved charges which
entail that there is no particle production. This property usually is 
characteristic of non-relativistic quantum mechanics, remarkably here 
it holds for relativistic quantum field theories (QFTs) 
(assuming that the FFB approach does indeed define a QFT). 
In 4-dimensions absence of particle production 
implies that the theory is free but in 
two dimensions this is not so. In addition to the absence of particle 
production, one postulates the spectrum of stable particle states
and their 2-particle S-matrix which has to satisfy the so-called
Yang-Baxter (or factorization) equation (\ref{s1}). 

In principle one could proceed without reference to a Lagrangian, 
but often contact to a Lagrangian description is desirable. Thus 
typically postulates of specific S-matrices are motivated by studies of 
associated Lagrangian QFTs. Unfortunately in most cases one cannot
solve the QFTs to the extent necessary to really derive the candidate
S-matrix, rather one has patches of partial information. This is in 
particular the case for the O$(n)$
nonlinear sigma models formally described by a set of spin fields
$\sigma^a,\,\,\,\,a=1\dots n \ge2$, with the constraint $\sigma^2=1$
and Lagrangian density $\propto (\partial_{\mu}\sigma)^2$.
There is a wealth of information on these models which will be
recalled when we study the various cases in the following sections,
and for an overview we refer the reader to our previous paper \cite{usgR}. 
In particular the spectrum of stable particles is thought to consist
of an O$(n)$ vector multiplet of mass $M$ without further bound states
(i.e of the form of the spectrum considered in subsection 2.3). 
The S-matrix element (for $n\geq2$) has the decomposition 
\be
S_{ab;cd}(\theta)=\sigma_1(\theta)\,\delta_{ab}\delta_{cd}
+\sigma_2(\theta)\,\delta_{ac}\delta_{bd}
+\sigma_3(\theta)\,\delta_{ad}\delta_{bc}\,,
\label{OnS}
\ee
where the center of mass energy is given by $\sqrt s=2M\ch\th/2$.

Classically the theories have an infinite
set of local and non-local conserved charges. 
One can argue that there are no anomalies which 
obstruct the existence of such charges in the quantum theory.
In the case of the non-local charges for $n\ge 3$ the
construction of L\"{u}scher \cite{luschernonloc} 
is closely connected to the usual
perturbative renormalizability and the (perturbative) asymptotic
freedom of the model. Knowledge of the action of the non-local charges
on the asymptotic states then restricts the S-matrix to the form
postulated by Zamolodchikov and Zamolodchikov \cite{ZZ} for $n\geq3$
\ba
\sigma_1(\theta)&=& \frac{-2\pi i\theta}{(i\pi-\theta)}
\cdot\,\frac{s_2(\theta)}{(n-2)\theta-2\pi i}\,,\nonum
\sigma_2(\theta)&=&(n-2)\theta
\cdot\,\frac{s_2(\theta)}{(n-2)\theta-2\pi i}\,,\label{S123}\\
\sigma_3(\theta)&=&\,\,\,-2\pi i\,\,\,\,\,
\cdot\,\frac{s_2(\theta)}{(n-2)\theta-2\pi i}\,,\nonumber
\ea
i.e. the invariant amplitudes are all given in terms of one amplitude
which we have chosen here to be the invariant amplitude $s_2(\theta)$ 
in the symmetric traceless (``isospin 2") channel. The amplitude 
$s_2(\theta)$ is off-hand determined only up to so called CDD factors, 
which were initially \cite{ZZ} fixed by selecting the solution with the 
minimal number of poles and zeros in the physical strip. This 
solution for $s_2(\th)$ is given by 
\be
s_2(\theta)=-\exp\Big\{2i\int_0^\infty\frac{d\omega}{\omega}
\sin(\theta\omega)\,\tilde K_n(\omega)\Big\}
\label{S0}
\ee
with
\be
\tilde K_n(\omega)=\frac{e^{-\pi\omega}+e^{-2\pi \frac{\omega}{n-2}}}
{1+e^{-\pi\omega}}\,.
\label{tildeK}
\ee
The proposed identification of (\ref{S123}) -- (\ref{tildeK})
with the S-matrix of the O($n$) sigma-model passes several non trivial
tests. First, the leading terms of its large $n$-expansion coincide 
with those obtained in leading orders of a field theoretical large $n$ 
computation. Second, in the determination of the exact $M/\Lambda$ ratio 
a consistency condition arises when matching the results of a
perturbative computation against that obtained via the thermodynamic Bethe 
ansatz \cite{HMN}. This consistency condition is also sensitive to 
the CDD factor; the minimal bootstrap solution (\ref{S123}) --
(\ref{tildeK}) passes the test. 

We note that the above formulae have a smooth $n\to2$ limit. 
A study of the possible relation
of the so defined FFB O(2) model to the continuum limit of
the lattice XY model (from the massive phase) will be the
topic of a future publication \cite{usO2}.

Further we remark that the S-matrix for the case 
$n=1$ (Ising model) can also be written in the form (\ref{OnS})
by setting
\be
\sigma_1(\th)=\sigma_2(\th)=0\,,\,\,\,\sigma_3(\th)=-1\,,\sspace n=1\,.
\ee
The representation (\ref{OnS}) is of course redundant in this case, 
but it does allow us in the following to discuss all $n\ge1$
simultaneously. For example
in all cases we have an expansion at low energies of the form
\be
S_{ab;cd}(\th)=-\delta_{ad}\delta_{bc}+i\theta D_{ab;cd}
+{\cal O}(\th^2)\,,
\ee
in sharp contrast to a weak perturbation of a free field theory.

Having all the on-shell information covers all the 
physical information on the theory one observes from scattering
of the stable particles, but off shell information 
is being explored if the system is probed by external sources 
weakly coupled to local operators with given quantum numbers.

\newsubsection{Derivation of the leading term in the FF expansion 
of {\boldmath $\gr$}}

For the leading $1$-$2$-$1$ particle contribution to $\gr$ a general
model-independent expression can be given in terms of the derivative 
the S-matrix and the 3-particle form factor. For notational reasons
we restrict attention here to the O$(n)$ models considered 
later in more detail. The extension to a general integrable QFT without
bound states is described in appendix A. For the O$(n)$ models the 
formula reads
\ba
\gamma_{4;121} = 4 i \sum_{j=1}^3 
\frac{\rmd\sigma_j(\th)}{\rmd\th}\bigg|_{\th=0} +
\frac{1}{8\pi}
\int_0^\infty \rmd u \Bigg\{
\frac{1}{\ch^2 u} f_c(u) f_c(-u)
-\frac{64}{u^2}\Bigg\}\,.
\label{grlead}
\ea
Here $f_c(\th)$ is a particular version of the 3-particle form 
factor ${\cal F}^a_{bcd}(\th_1,\th_2,\th_3)$ of the local field $\Phi^a$, 
supposed to correspond to the renormalized spin field $\sigma^a_{\sc r}$ 
in a Lagrangian construction. Explicitly
\be
f_c(\th) := {\cal F}^1_{1cc}(i\pi,\th,-\th)\;.
\label{f3}
\ee
In order to derive (\ref{grlead}) consider first more generally 
the $(1,l,1)$ contribution in (\ref{b13}) with $l\ge2$. 
Using (\ref{g7}) and switching to the explicit notation introduced 
in subsection 2.3 one can perform the integrals over the rapidities of the  
`${\underline 1}$' particles. Then one decomposes the rapidity measure for 
the intermediate `$\lb$' particle contribution according to 
(\ref{trans2}). Using (\ref{trans5}) the $\Lambda$ integration can be 
performed and by means of (\ref{trans3}) one arrives at 
\ba
v_{1l1}(\kappa_1,\kappa_2,\kappa_3,\kappa_4) 
=&& \frac{\pi}{2}\frac{1}{\ch^2 \kappa_1\,\ch^2\kappa_4}\,
\frac{1}{l!}\int \frac{\rmd^l\th}{(4\pi)^l} 
\frac{\delta(\th_1,\ldots,\th_l,\kappa_1,\kappa_2)}%
{\sum_{j =1}^l \ch\th_j}\, I_{1l1}(-\kappa_1|\th|\kappa_4)
\nonum 
=&& \frac{1}{8\ch^2\kappa_1\ch^2\kappa_4}
\int\frac{\rmd^{l-1}u}{(4\pi)^{l-1}}
\frac{1}{\ch^2 \Lambda_*}\frac{M^2}{M^{(l)}(u)^2}
I_{1l1}(-\kappa_1|\th|\kappa_4) \,,
\label{dom1}
\ea
where $I_{1l1}(-\kappa_1|\th|\kappa_4) :=
I^{1111}_{1l1}(-\kappa_1|\th|\kappa_4)$ is a product of generalized
form factors as in (\ref{g9}), (\ref{g10}). Explicitly the 
correspondence to the matrix elements is 
\ba
I_{1l1}(-\kappa_1|\th|\kappa_4) \is
\sum_{b_1,\ldots,b_l} \langle 1,-\kappa_1|\Phi^1(0)|b_1,\th_1;
\ldots;b_l,\th_l\rangle^{\rm in}
\nonumber\\
&& \sspace \times   
\phantom{}^{\rm in}\langle b_1,\th_1;\ldots;b_l,\th_l
|\Phi^1(0)|1,\kappa_4\rangle\,.
\label{dom2}
\ea
In the first expression we introduced the notation 
\be
\delta(\th_1,\ldots,\th_l) = \delta\bigg(\sum_{j=1}^l \sh\th_j\bigg)\;,
\label{dom3}
\ee
in the second one $\Lambda_*$ is defined by
\be
\sh\Lambda_* = - \frac{M}{M^{(l)}(u)}(\sh \kappa_1 + \sh \kappa_2)\,.
\label{dom3a}
\ee
As remarked before the generalized form factors can be expressed 
in terms of form factors of the same operator and delta distributions
by an explicit combinatorial formula. We shall usually just display the 
specific version needed. A discussion of the general formula can be
found in the appendix of \cite{MNmod}. For the generalized form factor 
entering (\ref{dom2}) the formula reads
\ba
&&\langle a_1,-\kappa_1|\Phi^{a_2}(0) | b_1,\th_1;b_2,\th_2 \rangle^{\rm
in}\bigg|_{\th_1>\th_2}
={\cal F}_{a_1b_1b_2}^{a_2}(-\kappa_1+i\pi-i\epsilon,\th_1,\th_2)
\nonum
&&+4\pi\delta_{a_1b_1}\delta_{a_2b_2}\delta(\kappa_1+\th_1)
+4\pi S_{b_1b_2;a_2a_1}(\th_1-\th_2)\delta(\kappa_1+\th_2)\,.
\ea
Substituting this in Eq.~(\ref{dom1}) we obtain
\ba
&& v_{121}=
\frac{1}{64\pi\ch^2\kappa_1\ch^2\kappa_4}\cdot\Big\{
\int_{-\infty}^\infty \rmd\alpha_1\frac{1}{\ch\bar\alpha_2(
\ch\alpha_1+\ch\bar\alpha_2)}\nonum
&&\times 
{\cal F}^1_{1xy}(i\pi-\kappa_1-i\epsilon,\alpha_1,\bar\alpha_2)
{\cal F}^1_{1xy}(i\pi-\kappa_4-i\epsilon,-\alpha_1,-\bar\alpha_2)
\nonum
&&+\frac{8\pi}{\ch\kappa_3(\ch\kappa_3+\ch\kappa_4)}
{\cal F}^1_{111}(i\pi-\kappa_1-i\epsilon,\kappa_4,\kappa_3)
\label{start}\\
&&+\frac{8\pi}{\ch\kappa_2(\ch\kappa_1+\ch\kappa_2)}
{\cal F}^1_{111}(i\pi-\kappa_4-i\epsilon,\kappa_1,\kappa_2)\Big\}\,.
\nonumber
\ea
Here we used the simplifications discussed above and 
the real analyticity property (\ref{ffeqs}e) below.  
Moreover, we changed the
integration variable from the difference of the two rapidities to
one of the rapidities ($\alpha_1$). The other rapidity ($\bar\alpha_2$) 
is then the solution of the transcendental equation
\be
\sh\alpha_1+\sh\bar\alpha_2+\sh\kappa_1+\sh\kappa_2=0
\label{baralpha}
\ee
and is an analytic function of $\alpha_1$.
There are no contributions from terms involving
delta-functions such as $\delta(\kappa_1+\kappa_4)$ appearing in
$I_{121}$ since we are taking the limit (\ref{b1}) where
these delta-functions vanish. (These terms are however crucial
to cancel corresponding singularities in the $\Omega$ term; 
c.f.~appendix A).

The form factors appearing in (\ref{start}) obey a system of 
functional equations which allow one to further simplify the 
expression. Let us recall these equations in the form relevant to the 
three-particle form factor ${\cal F}^d_{abc}(\alpha,\beta,\gamma)$ 
in the O$(n)$ model.
\begin{subeqnarray}
{\cal F}^d_{abc}(\alpha,\beta,\gamma)&=& S_{bc;yx}(\beta-\gamma) 
{\cal F}^d_{axy}(\alpha,\gamma,\beta)\,,\\
{\cal F}^d_{abc}(\alpha,\beta,\gamma)&=& 
{\cal F}^d_{cab}(\gamma+2\pi i,\alpha,\beta)\,,\\
{\cal F}^d_{abc}(\alpha,\beta,\gamma)&=& 
{\cal F}^d_{abc}(\alpha+\lambda,\beta+\lambda,\gamma+\lambda)\,,\\
{\cal F}^d_{abc}(\alpha,\beta,\gamma)&=& 
{\cal F}^d_{cba}(-\gamma,-\beta,-\alpha)\,,\\{}
[{\cal F}^d_{abc}(\alpha,\beta,\gamma)]^* &=& 
{\cal F}^d_{abc}(-\alpha^*,-\beta^*,-\gamma^*)\,.
\label{ffeqs}
\end{subeqnarray}
Here the S-matrix appearing in the exchange axiom (\ref{ffeqs}a)
is the O$(n)$ S-matrix (\ref{OnS}). (\ref{ffeqs}d) and 
(\ref{ffeqs}e) express the parity invariance and real
analyticity property of the form factors, respectively.
The homogeneous axioms (\ref{ffeqs}) are supplemented by the 
inhomogeneous residue equation
\be
\lim_{\alpha\to\beta+i\pi}
(\alpha-\beta-i\pi){\cal F}^d_{abc}(\alpha,\beta,\gamma)=
2i\big\{\delta_{ab}\delta_{cd}-
S_{bc;ad}(\beta-\gamma)\big\}\,.
\label{res}
\ee 

We now take advantage of the analytic properties of the form factors and 
change the integration contour in (\ref{start}) from the real axis 
to a curve ${\cal C}$ which is arbitrary except that it has to 
stay within the \lq physical strip' $0<{\rm Im}\,\alpha_1<\pi/2$. Along this
contour we can put $\epsilon=0$ and also the limit $\kappa_i\to0$ can
safely be taken. The integrated part of (\ref{start}) then simplifies
to
\be
v_{121}^{(II)}=
\frac{1}{128\pi}
\int_{{\cal C}}\frac{\rmd\alpha}{\ch^2\alpha}
f_b(\alpha)f_b(-\alpha)\,,
\label{VC}
\ee
where we introduced the shorthands
\ba
f^d_{abc}(\alpha)&:=&{\cal F}^d_{abc}(i\pi,\alpha,-\alpha)\,,
\nonum
f^1_{1bc}(\alpha)&=:&\delta_{bc} f_b(\alpha) \sspace \mbox{(no sum)}\,.
\label{ytheta}
\ea
Of course, one has to take into account the contribution of those
singular points of the integrand that get crossed when deforming
the contour of integration. There are two such singular points:
\be
\alpha_1=\kappa_4+i\epsilon
\qquad\qquad{\rm and}\qquad\qquad 
\bar\alpha_2=-\kappa_1-i\epsilon\,,
\label{sing}
\ee
which never coincide if (\ref{kappas}) holds.

Applying Cauchy's theorem one can evaluate the contribution from the
first singular point using the residue axiom (\ref{res}). This gives 
$$
-\frac{1}{16\ch^2\kappa_1\ch^2\kappa_4
\ch\bar\alpha_2(\ch\alpha_1+\ch\bar\alpha_2)}
\Big\{
{\cal F}^1_{111}(i\pi-\kappa_1-i\epsilon,\alpha_1,\bar\alpha_2)
-{\cal F}^1_{111}(i\pi-\kappa_1-i\epsilon,\bar\alpha_2,
\alpha_1)\Big\}\,,
$$
where in the second term we also used the exchange axiom 
(\ref{ffeqs}a). After taking the limit $\epsilon\to0$,
which is possible if (\ref{kappas}) holds, the contribution
of the first singular point becomes
$$
\frac{1}{16\ch^2\kappa_1\ch^2\kappa_4
\ch\kappa_3(\ch\kappa_3+\ch\kappa_4)}
\Big\{   
{\cal F}^1_{111}(i\pi-\kappa_1,\kappa_3,\kappa_4)
-{\cal F}^1_{111}(i\pi-\kappa_1,\kappa_4,\kappa_3)
\Big\}\,.
$$
The contribution of the second singularity is similar:
$$
\frac{1}{16\ch^2\kappa_1\ch^2\kappa_4
\ch\kappa_2(\ch\kappa_1+\ch\kappa_2)}
\Big\{
{\cal F}^1_{111}(i\pi-\kappa_4,\kappa_2,\kappa_1)
-{\cal F}^1_{111}(i\pi-\kappa_4,\kappa_1,\kappa_2)
\Big\}\,.
$$
Putting together the contribution of the singular points and
the last two terms of (\ref{start}) the non-integrated contribution
can be written as
$$
v_{121}^{(I)}
\stackrel{\cdot}{=}
\frac{1}{8\ch^2\kappa_1\ch^2\kappa_4
\ch\kappa_3(\ch\kappa_3+\ch\kappa_4)}
\Big\{
{\cal F}^1_{111}(i\pi-\kappa_1,\kappa_3,\kappa_4)
+{\cal F}^1_{111}(i\pi-\kappa_1,\kappa_4,\kappa_3)
\Big\}\,,
$$
where $\stackrel{\cdot}{=}$ indicates equality after
the symmetrization over the elements of the permutation group
${\cal S}_4$ has been carried out.

We now use the Smirnov axioms (\ref{ffeqs}) and (\ref{res}) to simplify 
the non-integrated part in the (symmetrized) $\kappa_i\to0$ limit. 
It is convenient to first introduce the reduced form factor 
${\cal G}^d_{abc}(\alpha,\beta,\gamma)$ by 
\be
{\cal F}^d_{abc}(\alpha,\beta,\gamma)=
T_3(\alpha,\beta,\gamma)
{\cal G}^d_{abc}(\alpha,\beta,\gamma)\,.
\label{red}
\ee
Here and in the following we set 
\be
T_N(\th_1,\dots,\th_N) :=\prod_{1\le i<j\le N}T(\th_i-\th_j)\,,
\ee
where $T$ is basically the tanh-function $T(\th) :=\tanh\th/2$.
Note $T(\th)$ has a simple pole at $\th=i\pi$, 
$T(i\pi-\theta)=-T(i\pi+\theta)=-2/\theta +{\cal O}(\theta)$, 
and a simple zero at $\theta=0$.
The advantage of the representation (\ref{red}) is that
the singularities are carried by the $\tanh$ factors and
the reduced form factor ${\cal G}^d_{abc}$ is analytic everywhere 
in the physical strip. In particular, for small $\alpha,\beta$
and $\gamma$ it can be expanded as
\be
{\cal G}^d_{abc}(i\pi+\alpha,\beta,\gamma)=
J^d_{abc}+(\alpha-\gamma)K^d_{abc}+
(\beta-\gamma)L^d_{abc}+\dots\,,
\label{tildeG}
\ee
where the dots stand for terms higher order in $\alpha,\beta$ and $\gamma$.
We can compute the constant tensors appearing in the
expansion (\ref{tildeG}) using the form factors equations. From the residue
axiom (\ref{res}) we can immediately fix 
\be
J^d_{abc} = 
i\big(\delta_{ab}\delta_{cd}+\delta_{ac}\delta_{bd}\big)\,,
\sspace K^d_{abc}+L^d_{abc} = D_{bc;ad}\,.
\label{res1}
\ee
To determine the expansion coefficients individually we employ the 
exchange relation (\ref{ffeqs}a) and find
\be
K^d_{abc}=D_{bc;ad}-D_{bc;da}\;,\sspace 
L^d_{abc}=D_{bc;da}\,.
\ee
Using the expansion (\ref{tildeG}),
for small $\kappa$ the non-integrated contribution becomes
\be
v^{(I)}_{121}
\stackrel{\cdot}{=}
\frac{1}{4}\frac{\kappa_3-\kappa_4}{(\kappa_1+\kappa_3)(\kappa_1+\kappa_4)}
\Bigl\{2i
+(\kappa_3-\kappa_4)D
+O(\kappa^2)\Bigr\}\,,
\ee
where
\be
D=D_{11;11}=-i\sum_{j=1}^3 
\frac{\rmd\sigma_j(\th)}{\rmd\th}\bigg|_{\th=0}\,.
\label{D}
\ee
This can be simplified by noting that upon averaging over the
permutations
\be
\frac{1}{\kappa_1+\kappa_4}
\stackrel{\cdot}{=}
\frac{1}{\kappa_1+\kappa_3}
\stackrel{\cdot}{=}0\,,
\ee
and similarly
\be
\frac{\kappa_4}{\kappa_1+\kappa_4}
\stackrel{\cdot}{=}
\frac{\kappa_3}{\kappa_1+\kappa_3}
\stackrel{\cdot}{=}
-\frac{\kappa_3}{\kappa_1+\kappa_4}
\stackrel{\cdot}{=}
-\frac{\kappa_4}{\kappa_1+\kappa_3}
\stackrel{\cdot}{=}
\frac{1}{2}\,.
\ee
After this simplification we have for the non-integrated contribution
\be
v^{(I)}_{121}
\stackrel{\cdot}{=}-\frac{1}{2}D\,,
\label{nonint}
\ee
and hence the non-integrated part of the leading contribution to the 
four-point coupling eventually becomes
\be
\gamma_{4;121}^{(I)}=-4D\,,
\label{gammaA}
\ee
For the Ising model the S-matrix is constant and therefore
$\gamma_{4;121}^{(I)}=0$ for $n=1$. For $n\geq2$ we use (\ref{S0}) 
and find
\be
\gamma_{4;121}^{(I)}=-\frac{4}{\pi}+8
\int_0^\infty \rmd\omega \tilde K_n(\omega)\,.
\label{gammaAOn}
\ee

The integrated part (\ref{VC}) (which is 
in this form rather useful for numerical evaluation)
can be written in an alternative form 
using the the residue axiom which implies 
\be
f^d_{abc}(\th)
=-\frac{4}{\th}\,J^d_{abc}+
\cO(1)
=-\frac{4i}{\th}\big(\delta_{ab}\delta_{cd}
+\delta_{ac}\delta_{bd}\big)+O(1)\,.
\ee
Using this 
we can explicitly subtract the singular part in (\ref{VC}) and
shift the contour back to the real axis. Noting also that the 
integrand is an even function of $\alpha$ we arrive at
\be
\gamma^{(II)}_{4;121}=
\frac{1}{8\pi}
\int_0^\infty \rmd u \Bigg\{
\frac{1}{\ch^2 u}
f_b(u)f_b(-u)-\frac{64}{u^2}\Bigg\}\,.
\label{VCreal}
\ee
The extension of the formula (\ref{grlead}) to general integrable models 
without bound states is described in appendix A. 

\newsubsection{The three particle form factor}

Only the special three-particle form factor $f^d_{abc}(\th)$ in 
(\ref{ytheta}) is necessary to compute the leading contribution
(\ref{grlead}) to $\gr$. It turns out to obey an autonomous system 
of functional equations (in a single variable) that derives from 
the form factor equations satisfied by 
${\cal F}^d_{abc}(\alpha,\beta,\gamma)$. Solving it allows one to compute
$f_b(\th)$ -- and hence to evaluate (\ref{grlead}) -- in situations 
where the general form factors are not known. 

We begin by noting that the functions $f^d_{abc}(\theta)$ are real analytic, 
i.e.~$[f^d_{abc}(\theta)]^*=f^d_{abc}(-\theta^*)$, in the physical
strip $0\leq{\rm Im}\,\theta\leq\pi$, with simple poles at $\theta=0$
and $\theta=\frac{i\pi}{2}$. Moreover, using (\ref{ffeqs}b,d) one can 
easily deduce that it is symmetric in its last two indices,
\be
f^d_{abc}(\theta)=f^d_{acb}(\theta)\,.
\label{Smy1}
\ee
Using (\ref{ffeqs}a) one obtains
\be
f^d_{abc}(\theta)=S_{bc;yx}(2\theta)\,f^d_{axy}(-\theta)\,,
\label{Smy2}
\ee
and finally  combining (\ref{ffeqs}a--d) results in 
\be
f^d_{abc}(i\pi-\theta)=S_{ca;yx}(\theta)\,
S_{yb;zl}(2\theta)\,S_{lx;vw}(\theta)\,
f^d_{wvz}(i\pi+\theta)\,.
\label{Smy3}
\ee
These are the consequences of the homogeneous form factor axioms;
they are supplemented by the residue equations
\ba
{\rm Res}\,f^d_{abc}(0)&=&-4i\big(\delta_{ab}\delta_{cd}+
\delta_{ac}\delta_{bd}\big)\,,\label{resy1}\\
{\rm Res}\,f^d_{abc}\left(\frac{i\pi}{2}\right)&=&i\left\{
\delta_{bc}\delta_{ad}-S_{ca;bd}\left(\frac{i\pi}{2}\right)
\right\}\,.\label{resy2}
\ea
In view of Eq.~(\ref{Smy1}) we can parameterize $f$ as 
\be
f^d_{abc}(\theta)=k(\theta)\delta_{ad}\delta_{bc}+
l(\theta)\Big[\delta_{ac}\delta_{bd}+\delta_{ab}\delta_{cd}
\Big]\,.
\label{kl}
\ee
Then the contribution $\gamma_{4;121}^{(II)}$ is given by
\be
\gamma_{4;121}^{(II)}=\frac{1}{8\pi}\int_0^\infty\,
\rmd u\,\Bigg\{\frac {n k(u)k(-u)+
2k(u)l(-u)+2k(-u)l(u)+4l(u)l(-u)}
{\ch^2 u}-\frac{64}{u^2}\Bigg\}\,.
\label{gammaC1}
\ee
In terms of the two functions $k(\th)$ and $l(\th)$ Eq.~(\ref{Smy2}) can 
be written as  
\ba
k(\theta)&=&\,\left[s_2(2\theta)+n\sigma_1(2\theta)\right]k(-\theta)
+2\sigma_1(2\theta)\,l(-\theta)\,,\nonum  
l(\theta)&=&\,s_2(2\theta)\,l(-\theta)\,,
\label{Smykl2}
\ea
while (\ref{Smy3}) becomes 
\ba
k(i\pi-\th)&=&\left[
A_{11}(\th)k(i\pi+\th)+A_{12}(\th)l(i\pi+\th)\right]
a(\th)s_2(\th)^2s_2(2\th)\,, \nonum
l(i\pi-\th)&=&\left[
A_{21}(\th)k(i\pi+\th)+A_{22}(\th)l(i\pi+\th)\right]
a(\th)s_2(\th)^2s_2(2\th)\,.\label{Smykl4}
\ea
Here 
\be
a(\th)=\frac{(n-2)\th+2i\pi}{(i\pi-\th)(i\pi-2\th)[(n-2)\th-i\pi]
[(n-2)\th-2i\pi]^2}\,,
\ee
and
\ba
A_{11}(\th)&=&(\th-i\pi)\left[ 2(n-2)^2\th^3+(n-2)(n-4)\th^2 i\pi
+(n+2)\th \pi^2 -2 i\pi^3\right]\,,\nonum
A_{12}(\th)&=&-4(n-2)i\pi\th (\th-i\pi)(\th+i\pi)\,,\nonum
A_{21}(\th)&=&-2(n-4)i\pi^3\th\,,\nonum
A_{22}(\th)&=&A_{11}(-\th)\,.
\ea  
The matrix $A(\th)$ satisfies 
\be
A(\th)^{-1}=A(-\th)a(\th)a(-\th)\,,\sspace 
{\rm det}A(\th)=\frac{1}{a(\th)a(-\th)}\,.
\label{detA}
\ee
The functional equations (\ref{Smykl2}), (\ref{Smykl4}) still contain the 
transcendental function $s_2(\th)$. It can be eliminated 
by the following standard procedure. We introduce the function 
$u(\theta)$ as the unique solution of
\ba
u(\theta)&=&s_2(\theta)\,u(-\theta)\,,\label{u1}\\
u(i\pi-\theta)&=&-u(i\pi+\theta)\,,
\label{u2}
\ea
subject to the normalization condition
\be
u(i\pi-\theta)=\frac{1}{\theta}+{\cal O}(\theta)\,.
\ee
Using the results of Appendix D one can immediately write
down the solution
\be
u(\theta)=-\frac{1}{2}T(\theta)\,e^{\Delta(\theta)}\,,
\label{solu}
\ee
where in (\ref{Fourier2}) of course the kernel $\tilde K_n(\omega)$ 
defined in (\ref{tildeK}) has to be used. Introducing 
\be
Y(\theta)=\frac{2i}{u_0}\,u(i\pi-\theta)\,u(i\pi+\theta)\,
u(2\theta)\,,\sspace 
u_0=u^\prime(0)=-\frac{1}{4}\,e^{\Delta(0)}\,.
\label{Y}
\ee
we parameterize $k,l$ as 
\be
k(\theta)=Y(\theta)\,K(\theta)
\qquad\qquad
{\rm and}
\qquad\qquad
l(\theta)=Y(\theta)\,L(\theta)\;.
\label{KL}
\ee
Rewriting then the functional equations (\ref{Smykl2}), (\ref{Smykl4}) 
in terms of $K$ and $L$, the new system involves only rational coefficient 
functions. Explicitly they read 
\ba
K(\theta)&=&\frac{-1}{[(n-2)\th-i\pi](i\pi-2\th)}
\Bigr\{[(n-2)\th+i\pi](i\pi+2\th)
K(-\theta)+4i\pi\th L(-\theta)\Bigr\}\,,\nonum
L(\theta)&=&L(-\theta)\,,\label{SmyKL2}
\ea
and
\ba
K(i\pi-\th)&=&\left[
A_{11}(\th)K(i\pi+\th)+A_{12}(\th)L(i\pi+\th)\right]a(\th)\,,\nonum
L(i\pi-\th)&=&\left[
A_{21}(\th)K(i\pi+\th)+A_{22}(\th)L(i\pi+\th)\right]a(\th)\,,
\label{SmyKL3}
\ea
respectively. The first equation of (\ref{SmyKL2}) can be used to
eliminate $L(\theta)$ in favor of $K(\theta)$ via 
\be
L(\theta)=\frac{1}{4i\pi\th}\Big\{[i\pi-(n-2)\th]
(i\pi-2\theta)K(\th)
-[i\pi+(n-2)\th](i\pi+2\theta)K(-\th)\Big\}\,,
\label{solL}
\ee
and (\ref{solL}) also solves the second equation of (\ref{SmyKL2}).
Inserting (\ref{solL}) into (\ref{SmyKL3}) results in a single 
linear functional equation for $K(\th)$. The normalization of the 
solution is fixed by the residue equations (\ref{resy2}). 

We expect that this procedure can be used to compute $f_b(\th)$ 
and hence the leading contribution to the coupling for all O$(n)$
models. For the O(2) model we demonstrate this in section 6.



\newsubsection{Sub-leading contributions} 

In order to achieve higher accuracy and to obtain some clue on 
the rate of convergence of the series (\ref{gamma4exp}) 
we will compute some of the sub-leading terms as well. 
It turns out that the 1-2-1 term indeed gives the numerically most
important contribution to the coupling. 
But based on the computation of the sub-leading terms the numerical result 
can also be endowed with an {\em intrinsic} error estimate. 
Our results indicate that the next important contributions to 
the coupling are $(1,2,3)+ (3,2,1)$ and $(1,4,1)$. Its explicit 
evaluation is deferred to appendices C and B. The difficulty in the 
evaluation lies in the rapidly varying nature of the integrands,
which have in the multidimensional phase space many zeros and
(integrable) singularities. To deal with these we have either
decomposed the integrand into appropriate parts or avoided the
singularities by shifting some contours of integration into the
complex plane. 

The $(1,4,1)$ contribution is the $l=4$ case of Eq.~(\ref{dom1})
and will be evaluated in appendix B. Here we prepare the ground for 
the evaluation of the $(1,2,3)+ (3,2,1)$ terms. More generally let 
us examine the $(1,2,m) + (m,2,1)$ contribution 
and to this end return to (\ref{g7}). 
Performing the internal rapidity integrations one obtains
\ba
&& v_{12m}(\kappa_1,\kappa_2,\kappa_3,\kappa_4) =
\frac{\pi^2}{\ch^2 \kappa_1} \frac{1}{m!} \int \frac{\rmd\xi_1
\rmd\xi_2}{(4\pi)^2}
\frac{\delta(\xi_1,\xi_2,\kappa_1,\kappa_2)}{\ch\xi_1 + \ch\xi_2}
\nonum
&& \sspace \times 
\int\frac{\rmd^m\th}{(4\pi)^m}\frac{\delta(\th_1,\ldots,\th_m,-\kappa_4)}%
{\sum_{j=1}^m \ch\th_j}\;I_{12m}(-\kappa_1|\xi_2,\xi_1|\th)\;.
\label{sdom1}
\ea   
Next one spells out $I_{12m} := I_{12m}^{1111}$ by inserting the 
formula expressing the generalized form factors in terms of ordinary 
form factors. Taking advantage of the S-matrix exchange relations 
many of terms contribute equally upon integration and one 
ends up with four terms
\be
v_{12m}=v_{12m}^{(I)}+v_{12m}^{(II)}+v_{12m}^{(III)}+v_{12m}^{(IV)},
\label{v12m}
\ee
with
\ba
& \nspace &v_{12m}^{(I)}(\kappa_1,\kappa_2,\kappa_3,\kappa_4)
\sim{1\over 16(4\pi)^m m!}\sum_{b_1,b_2}\sum_{a_1,...,a_m}
\int\rmd^2 \beta
{\delta(\beta_1,\beta_2,\kappa_1,\kappa_2)
\over \ch\beta_1+\ch\beta_2}
\nonumber\\
&\nspace &\quad \int\rmd^m \alpha
{\delta(\alpha_1,...,\alpha_m,-\kappa_4)
\over \sum_{i=1}^k\ch\alpha_i}
{\cal F}^1_{1b_1b_2}(-\kappa_1+i\pi_-,\beta_1,\beta_2)
\nonumber\\
&\nspace &\quad  
{\cal F}^1_{b_2b_1a_1a_2...a_m}(\beta_2+i\pi_-,\beta_1+i\pi_-,  
\alpha_1,...,\alpha_m)
{\cal F}^1_{a_1a_2...a_m}(\alpha_1,...,\alpha_m)^*
\,,\\
\nonumber\\
& \nspace &v_{12m}^{(II)}(\kappa_1,\kappa_2,\kappa_3,\kappa_4)
\sim{1\over 8(4\pi)^{m-1} (m-1)!}\sum_{b_1,b_2}\sum_{a_2,...,a_m}
\int\rmd^2 \beta
{\delta(\beta_1,\beta_2,\kappa_1,\kappa_2)
\over \ch\beta_1+\ch\beta_2}
\nonumber\\
&\nspace &\quad \int\rmd^{m-1} \alpha
{\delta(\beta_1,\alpha_2...,\alpha_m,-\kappa_4)
\over  \ch\beta_1+\sum_{i=2}^m\ch\alpha_i}
{\cal F}^1_{1b_1b_2}(-\kappa_1+i\pi_-,\beta_1,\beta_2)
\nonumber\\
&\nspace &\quad 
{\cal F}^1_{b_2a_2...a_m}(\beta_2+i\pi_-,
\alpha_2,...,\alpha_m)
{\cal F}^1_{b_1a_2...a_m}(\beta_1,\alpha_2...,\alpha_m)^*
\,,\\
\nonumber\\
& \nspace &v_{12m}^{(III)}(\kappa_1,\kappa_2,\kappa_3,\kappa_4)
\sim{1\over 16(4\pi)^{m-2}(m-2)!}\sum_{b_1,b_2}\sum_{a_3,...,a_m}
\int\rmd^2 \beta
{\delta(\beta_1,\beta_2,\kappa_1,\kappa_2)
\over \ch\beta_1+\ch\beta_2}
\nonumber\\
&\nspace &\quad \int\rmd^{m-2} \alpha
{\delta(\alpha_3,...,\alpha_m,\kappa_3)
\over  \ch\beta_1+\ch\beta_2+\sum_{i=3}^m\ch\alpha_i}
{\cal F}^1_{1b_1b_2}(-\kappa_1+i\pi_-,\beta_1,\beta_2)
\nonumber\\
&\nspace &\quad 
{\cal F}^1_{a_3...a_m}(\alpha_3,...,\alpha_m)
{\cal F}^1_{b_1b_2a_3...a_m}(\beta_1,\beta_2,\alpha_3,...,\alpha_m)^*
\,,\\
\nonumber\\
&\nspace &v_{12m}^{(IV)}(\kappa_1,\kappa_2,\kappa_3,\kappa_4) 
\sim{1\over 16(4\pi)^{m-1} m!}\sum_{a_1,...,a_m}
\int\rmd^m \alpha
{\delta(\alpha_1,...,\alpha_m,-\kappa_4)
\over \sum_{i=1}^m\ch\alpha_i}
\nonumber\\
&\nspace &\quad 
{\cal F}^1_{11a_1a_2...a_m}(-\kappa_2+i\pi_-,-\kappa_1+i\pi_-,
\alpha_1,...,\alpha_m)
{\cal F}^1_{a_1a_2...a_m}(\alpha_1,...,\alpha_m)^*\,,
\label{v12mx}
\ea
where
$\pi_-$ stands for $\pi-\epsilon$. All integrals range from
$-\infty$ to $+\infty$.

Further details of the computation of the 1-2-3 contribution are
given in Appendix C. Note that for the numerical evaluation
of the $k+l+m=6$ contributions we need the analytic expressions for
the 5-particle form factor of the spin operator. Unfortunately
these are at present only known for $n=1$ and $n=3$. For the Ising 
model all the form factors are explicitly 
known and in this case we have also computed the $k+l+m=8$ contributions.
After a preparatory next section where we discuss the 
definition and measurement of the intrinsic coupling in the lattice
regularization, we will discuss the cases $n=1,2,3$ in turn.
\newpage
\newsection{Lattice computations of {\boldmath $\gr$}} 

In the subsequent sections we will compare the results of the form factor
bootstrap coupling $\gr$ with those obtained from the lattice theory.
As noted earlier, in the framework of the lattice regularization there 
are two methods to compute $\gr$ in the O($n$) models: high temperature 
(= strong coupling) expansions and Monte Carlo simulations. 
Both approaches usually 
take the standard lattice action on a square lattice 
\be
S=-\beta\sum_{x,\mu}\sigma(x)\cdot\sigma(x+\hat{\mu})\,,
\label{SA}
\ee
as the starting point, where $\sigma(x)\cdot\sigma(x)=
\sum_a\sigma^a(x)\sigma^a(x)=1$. 

The lattice definition of $\gr(\beta)$ is as in Eq.~(\ref{grdef})
\be
\gr(\beta)=-\frac{1}{\xi_2^2G_2(0)^2}\frac{1}{n^2}\sum_{a,b}G_4^{aabb}\,,
\ee
where all quantities are defined analogously to the continuum theory
\ba
G_2(k)&=&\frac{1}{n}\sum_a\sum_x
e^{ikx}\langle\sigma^a(x)\sigma^a(0)\rangle\,,
\\
G_4^{a_1a_2a_3a_4}&=&\sum_{x_1,x_2,x_3}\Bigl\{
\langle\sigma^{a_1}(x_1)\sigma^{a_2}(x_2)
\sigma^{a_3}(x_3)\sigma^{a_4}(0)\rangle
\nonumber\\
&&-\Bigl[\langle\sigma^{a_1}(x_1)\sigma^{a_2}(x_2)\rangle
\langle\sigma^{a_3}(x_3)\sigma^{a_4}(0)\rangle+2\,\,{\rm perms}\Bigr]
\Bigr\}\,,
\ea
and $\xi_2$ is the second moment correlation length
\be
\xi_2^2 =\frac{\mu_2}{4G_2(0)}\,,\sspace 
\mu_2 =\frac{1}{n} \sum_a\sum_x x^2\langle\sigma^a(x)\sigma^a(0)\rangle\,.
\ee
The coupling from the lattice regularization is defined as the
continuum limit
\be
\gr=\lim_{\beta\to\beta_{\rm c}}\gr(\beta)\,,
\label{contlim}
\ee
where $\beta_{\rm c}$ is a critical point where the correlation length
diverges (in lattice units).

Butera and Comi \cite{BC} have produced long high temperature series for 
$G_2(0),\mu_2,$ and $G_4$ in the O$(n)$ model with standard action,
and Pelissetto and Vicari \cite{peliss} have reanalyzed these series 
to compute estimates for the intrinsic coupling $\gr$ for $n\leq 4$.
Similar computations have been performed previously by Campostrini et
al \cite{CPRVhT}.

Our Monte Carlo simulations were of course done on a finite lattice,
more precisely a square lattice of size $L$ (points) in each direction and 
periodic boundary conditions, both with the standard action (\ref{SA}) 
and the fixed point action of ref.~\cite{FP}. The infinite volume 
lattice coupling $\gr(\beta)$ is then obtained as the limit 
\be
\gr(\beta)=\lim_{L\to\infty}\gr(\beta,L)\,,
\ee
of a finite volume coupling $\gr(\beta,L)$ which 
is proportional to Binder's cumulant $u_L$:
\ba
\gr(\beta,L)&=&\left({L\over\xi^{\rm eff}(\beta,L)}\right)^2 u_L\,,
\nonum
u_L&=&1+{2\over n}-
{\langle (\Sigma^2)^2\rangle \over \langle \Sigma^2\rangle^2}\,,
\ea
where $\Sigma^a=\sum_x \sigma^a(x)$. 
In this definition $\xi^{\rm eff}(\beta,L)$ is an effective correlation
length which converges to the second moment correlation length 
$\xi_2$ in the limit $L\to\infty$. 
In our computations we used the particular definition 
(as e.g. in ref.~\cite{FMPPT}):
\be
\xi^{\rm eff}(\beta,L)={1\over 2 \sin(\pi/L)} 
\sqrt{{G_2(0)\over G_2(k_0)}-1}\,,
\label{xisecmom}
\ee
where $k_0=(2\pi/L,0)$. 

In our analysis of the Monte Carlo data we shall make the working 
assumption that one is allowed to replace the limiting 
procedure $\lim_{\beta\to\beta_{\rm c}}\lim_{L\to\infty}$ by 
\ba
\gr&=&\lim_{z\to\infty}\hat{g}_{\rm R}(z)\,,\sspace 
z:=L/\xi^{\rm eff}(\beta,L)\,,
\nonum
\hat{g}_{\rm R}(z)&:=&\lim_{\beta\to\beta_{\rm c},\,z\,\,{\rm fixed}}
\gr(\beta,L)\,.
\ea
That is we attempt to first take the continuum limit at fixed
physical volume and afterwards take the physical volume to infinity.
The $z\to\infty$ limit of $\hat{g}_{\rm R}(z)$ is expected 
to be reached exponentially; 
for example in the leading order $1/n$ expansion
\cite{CPRVgR}
\be
\hat{g}_{\rm R}(z)=\hat{g}_{\rm R}(\infty)
\Bigl(1-c\sqrt{z}\exp(-z)+....\Bigl)\,.
\label{grz}
\ee
The situation may however be slightly more complicated due
to our particular definition of $\xi^{\rm eff}$. Indeed in the 
continuum limit at fixed physical volume we expect
$G_2(0)/G_2(k_0)\to G(0)/G(k)$ where $k\sim K_0=(2\pi \Mr/z,0)$ and
the continuum expressions are in finite physical volume.
On the other hand for the continuum two point function
defined in infinite volume
\be
\frac{1}{K_0^2}\Bigl[G(0)/G(K_0)-1\Bigr]\sim \frac{1}{\Mr^2}   
\Bigl[1-\Bigl(\frac{2\pi}{z}\Bigr)^2\Bigl(\gamma_2-1\Bigr)
\Bigr]\,.
\label{janosfact}
\ee
In our simulations the values of $2\pi/z$ are $\sim 1$ i.e.~not so 
small; nevertheless at such values the correction factor
on the rhs of (\ref{janosfact}) only deviates from 1 by the order
$10^{-3}$. This deviation is much smaller than the 
statistical accuracy of our simulations, 
and hence we ignore these additional effects in our analyses
of the lattice data.
\newsection{The Ising model}

The particular field theory we are considering in this section is that 
obtained from the Ising model in zero external field%
\footnote{One can obtain an infinite number of field theories from the
Ising model in the presence of an external field $H$ by taking the limit
$H\to0,\,T\to T_{\rm c}$ with $h=H/|T-T_{\rm c}|^{15/8}$ fixed.}
for $0<T-T_{\rm c}\to0$. The spin-spin correlation functions 
in the scaling limit are known exactly from the work of Wu et al
\cite{Wu}, and from this knowledge Sato, Miwa and Jimbo \cite{Sato}
found that the S-matrix operator was given by
\be
{\bf S}=(-1)^{{\bf N}({\bf N}-1)/2}\,,
\label{isingsmat}
\ee  
where ${\bf N}$ is the particle number operator. 
An energy independent phase
is not observable in a scattering experiment; the non-trivial
S-matrix (\ref{isingsmat}) reflects the fact that the off-shell 
spin-spin correlation functions are not that of a free field.
The continuum limit of the Ising model is also described by a 
free Majorana field, but this is non-local with respect to the
spin field; for a more detailed discussion we refer the reader 
to the lectures of McCoy \cite{McCoy}.

\newsubsection{Form factor determination}

The generalized form factors are given by \cite{Isingff}
\ba
&&\phantom{}^{\rm out}\langle
\th_1,...,\th_m|\sigma(0)|\th_{m+1},...,\th_N
\rangle^{\rm in}=
\nonumber\\
&&(2i)^{(N-1)/2}\prod_{1\le i<j\le m}T(|\th_i-\th_j|)
\prod_{1\le r\le m<s\le N}\frac{\cal P}{T(\th_r-\th_s)}
\prod_{m< k<l\le N}T(|\th_k-\th_l|)\,,
\label{isi1}
\ea
with $N$ an odd (positive) integer. We evaluate the dominant 
contribution to the coupling using Eq.~(\ref{grlead}). The non-integrated 
part (\ref{gammaA}) vanishes. For the integral (\ref{VCreal}) we need 
$f_1(\th)$, which is readily obtained from (\ref{isi1}),
\be
f_1(\th)=-2i\,T(2\th)/T^2(\th)\,.
\ee
Thus the dominant contribution to $\gamma_4$ is  
\be
\gamma_{4;121} = \frac{1}{2\pi} \int_0^{\infty} 
\rmd u \left[ \frac{T^2(2u)}{T^4(u)\ch^2 u} -
\frac{16}{u^2} \right]\;=-\frac{5}{2}-\frac{47}{6\pi}\;.
\label{isi2}
\ee
Numerically this gives $\gamma_{4;121} = -4.993427441(1)$ or $\gr \approx 
14.98$ in the leading approximation. 

The simplicity of the form factors (\ref{isi1}) also makes the Ising
model a good testing ground for the computation of the sub-leading 
contributions, to which we turn now. The evaluation of the spectral moments 
(\ref{norm1}), (\ref{gamdel}) is straightforward. For $m=3,5,7$ the 
results are given in Table \ref{Ispectralmoments}. 

\begin{table}[h]
\centering
\begin{tabular}[t]{r|r|r}
\hline
$m$ &$\gamma_{2;m}\sspace\quad\;\;\; $& 
$\delta_{2;m}\ \ \ \ \ \ \ \ $  \\[0.5ex]
\hline \hline    

3 & $ 8.1446256566(1) \times 10^{-4}$ & $1.094(1) \times 10^{-5}$\\[0.5ex]

5 & $7.96(1) \times 10^{-7}$ & $2.22(1) \times 10^{-10}$\\[0.5ex]

7 & $7.8(1) \times 10^{-10}$ & $4.6(1) \times 10^{-15}$\\[0.5ex]

\hline\end{tabular}
\caption{\footnotesize $m$-particle contributions to $\gamma_2, \delta_2$
in 
the Ising model}
\label{Ispectralmoments}
\end{table}

Table \ref{Ispectralmoments} suggests that the series (\ref{gamdel}) 
converge extremely rapidly and we would estimate 
\be
\gamma_2=1 + 8.15259(1) \times 10^{-4}\,,\sspace
\delta_2=1 + 1.094(1) \times 10^{-5}\;,
\label{gamdel2}
\ee
where the estimated errors come both from the numerical 
integration and from estimating the contributions of the higher particle 
terms. To get some check on this we may consider the ratio 
$\delta_2/\gamma_2$ for which from the leading terms (\ref{gamdel2}) 
we get $\delta_2/\gamma_2 =  0.999\,196\,336(11)$. This is in excellent 
agreement with the result $\delta_2/\gamma_2 = 0.999\,196\,33$ 
of Campostrini et al.~\cite{CPRV}, which they obtained by numerical 
evaluation of the exact formula for the 2-point function%
\footnote{this famous Fredholm determinant
(solving  the Painlev\'e III equation) is basically the summed up FF
series; see e.g.~\cite{BaBe}.}
of Wu et al. \cite{Wu}.

The evaluation of $\gamma_4$ is more involved. In order to gain insight into
the rate of decay of the higher particle contributions as well as their 
sign pattern we pushed the computation up to $k+l+m\leq 8$. 
The $k+l+m=8$ contributions in particular turned out 
to be a formidable computation despite the deceptive simplicity of the 
form factors. The computation is based on the formulae (\ref{dom1}), 
(\ref{sdom1}) and similar ones for $(m,2,m)$, with $m$ odd, and for 
$(1,l,3)$, with $l$ even. To give the reader a chance to follow the 
computations we have collected some intermediate results in Appendices B, C. 
The final results for the contributions of the $k$-$l$-$m$ intermediate 
states with $k+l+m\le 8$ to $\gamma_4$ are summarized in
Table~\ref{gamma4}. 

\begin{table}[h]
\centering
\begin{tabular}[t]{c|l}
\hline
$k,l,m$ & $\sspace \quad \gamma_{4;klm}\;\;\;\;\;\;\;$  \\[0.5ex]
\hline \hline

$1,2,1$ & $\quad -4.993427441(1)\phantom{5}$\\[0.5ex]

\hline

$1,2,3$ & $\quad \phantom{-}0.046310(1)\phantom{5}$\\[0.5ex]

$1,4,1$ & $\quad -0.002653(1)\phantom{5}$\\[0.5ex]

\hline

$3,2,3$ & $\quad \phantom{-}0.0002884(3)\phantom{5}$\\[0.5ex]

$1,4,3$ & $\quad -0.0000420(5)\phantom{5}$\\[0.5ex]

$1,2,5$ & $\quad \phantom{-}0.00002562(2)\phantom{5}$\\[0.5ex]

$1,6,1$ & $\quad -0.0000040(1)\phantom{5}$\\[0.5ex]

\hline
\end{tabular}
\caption{\footnotesize $k$-$l$-$m$-particles contributions to 
$\gamma_4$ in the Ising model}
\label{gamma4}
\end{table}
The rapid decay of the terms is manifest. Increasing $k+l+m$ 
by $2$ gives a contribution roughly 
two orders of magnitude smaller than the previous one. 
The sign pattern appears to follow the rule: ${\rm Sign}(\gamma_{4;klm})
={\rm Sign}(k+m-l-1)$. Further terms with larger differences $|k-l|, |l-m|$ 
are suppressed as compared to those with smaller ones. In view of Table
\ref{gamma4} we would thus (conservatively) 
estimate the $k+l+m \geq 10$ particle 
contributions to be $\leq 10 \%$ of the sum of the $k+l+m = 8$
contributions. 
This gives
\be
\gamma_4 = -4.90321(3)\;.
\ee
Inserting into (\ref{grx}) with $\gamma_2,\delta_2$ taken
from (\ref{gamdel2}) then yields our final result
\be
\gr = 14.6975(1)\,.
\label{grIsing}
\ee 
This amounts to a determination of $\gr$ to within $<0.001\%$.
For comparison
we collected the results of some previous determinations 
in Table~\ref{grIsingprev} below.
\bigskip
\smallskip

\begin{table}[h]
\centering
\begin{tabular}[t]{c|c}
\hline
Method          & Value for $\gr$ \\[0.3ex] 
\hline\hline 

High temperature& $14.6943(17)$ \cite{peliss}, $14.67(5)$ \cite{Baker}\\[0.5ex]

Borel summation & $15.5(8)$ \cite{Zinn} \\[0.5ex]

Monte Carlo     & $14.3(1.0)$ \cite{KimP}\\[0.5ex]

\hline
\end{tabular}
\caption{\footnotesize Previous determinations of $\gr$ in the Ising model}
\label{grIsingprev}
\vspace{1cm}
\end{table}

Finally we would like to mention that an analogous 4-point coupling 
$h_{\sc r}$ can be defined at criticality $T=T_{\rm c}$ by sending the 
magnetic field $H$ to zero. 
Of course in this case the definition of Binder's cumulant has to be
modified appropriately to take into account the fact 
that the field has non-vanishing vacuum expectation value.
Remarkably $h_{\sc r}$ can be computed 
exactly by taking advantage of the fact that the small $H$ 
behavior of the partition function is known exactly \cite{Delfino}. 
The final result is  $h_{\sc r}=-{609 \pi \over 4}=-478.307$.

\newsubsection{Recent Monte Carlo simulation of the Ising model}

Our Monte Carlo investigation of $\gr$ was performed on several
IBM RISC 6000 workstations at the Werner-Heisenberg-Institut.

\begin{figure}[htb]
\centerline{\epsfxsize=6.0cm\epsfbox{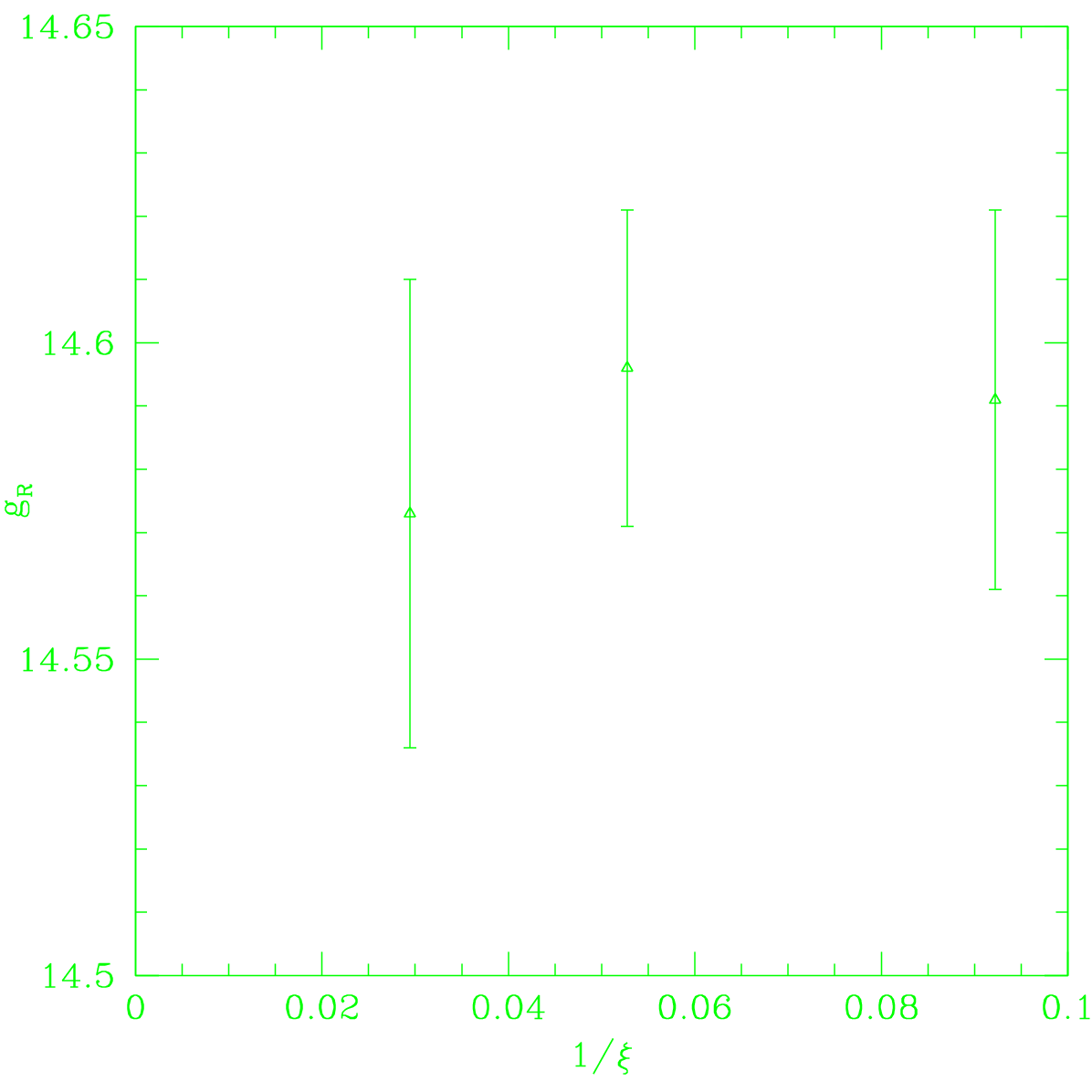}}
\caption{The data for $\gr$ at $z\approx 7.4$ for different lattice
spacings}
\label{grlatt}
\end{figure}

In this subsection $\xi^{\rm eff}$ is denoted simply by $\xi$.
We studied the dependence on the lattice spacing by running
at $\beta=.418$ ($\xi=10.839936$), $\beta=.4276$ ($\xi=18.924790$) 
and $\beta=.433345$ ($\xi=33.873923$) on lattices of size $L=80$,
$L=140$ and $L=250$, respectively. 
These values were chosen in such 
a way that they have almost exactly the same value of 
$z=L/\xi\approx 7.4$. Figure \ref{grlatt} shows that there is no 
significant dependence on the lattice spacing (i.e. $\xi$). Therefore 
we decided to use all the data together to study the finite size 
effects.

\begin{figure}[htb]
\centerline{\epsfxsize=8.0cm\epsfbox{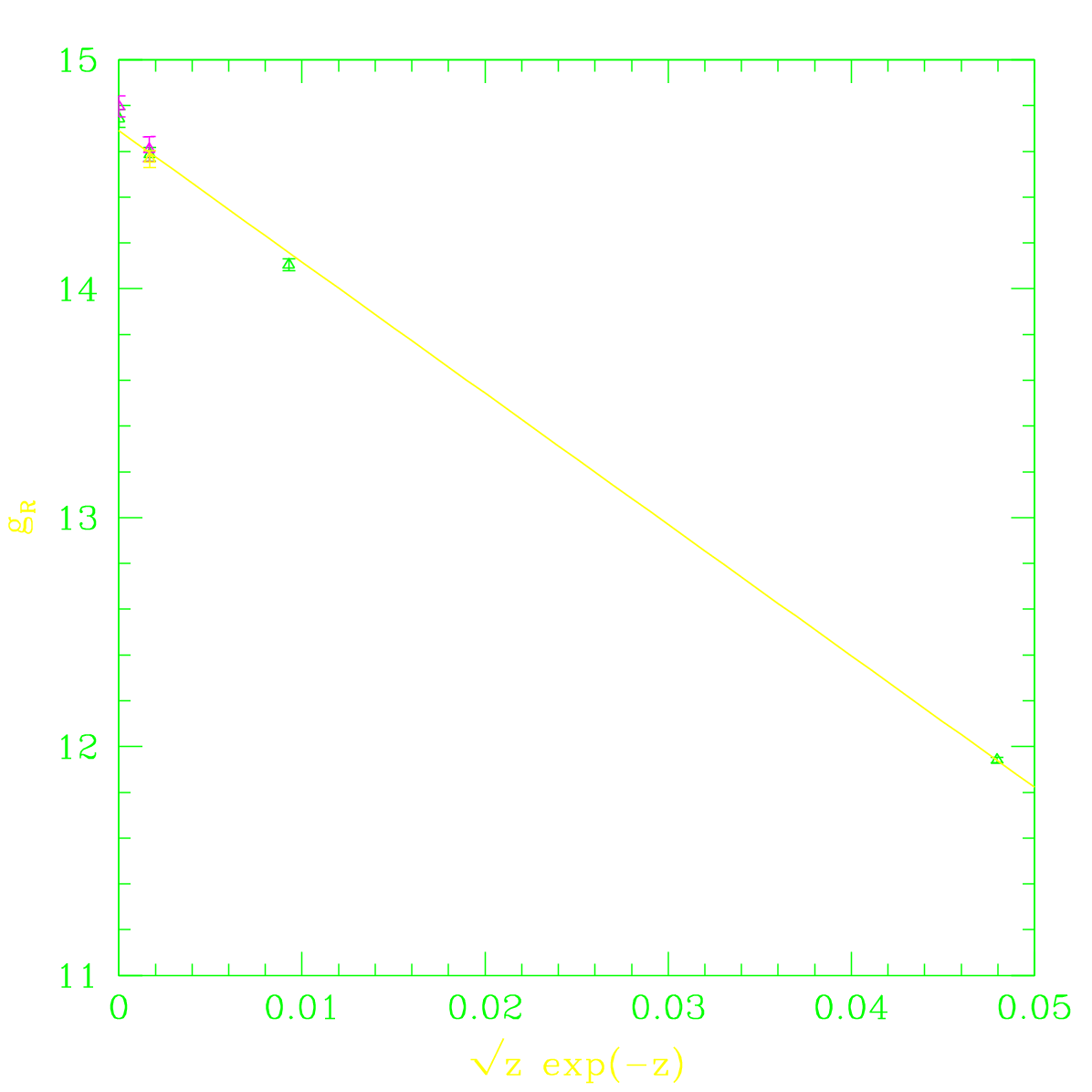}}
\caption{The data for $\gr$ vs $\sqrt z\exp(-z)$ together with the fit}
\label{grfss}
\end{figure}

We studied the finite size dependence by measuring in addition 
$\gr$ on lattices of size $L=40,\ 60,\ 80,\ 140$ at $\beta=.418$, 
($\xi=10.839936$). Finite size scaling works very well, i.e. the results
only depend on $z=L/\xi$. The dependence on $z$ is still quite well
described by Eq.~(\ref{grz}).
This can be seen in figure \ref{grfss}. A least square fit produces
\be
c=3.91(3),\sspace \hat{g}_{\rm R}(\infty)=14.69(2)\,.
\ee
The fit quality is not fantastic ($\chi^2=2.4$ per d.o.f.) but 
acceptable. So our final Monte Carlo estimate for $\gr$ is
\be
\gr=14.69(2)\,.
\label{isMCnew}
\ee

We report our numbers in Table \ref{MC1}.
In this table we also indicate the number of measurements. These were 
performed using the cluster algorithm as follows: one run consisted of 
100,000 clusters used for thermalization, followed by 20,000 sweeps
of the lattice used for measurements. Each run was repeated after 
changing the initial configuration. One such run was considered as one 
independent measurement. The error was computed out of this sample by 
using the jack-knife method.

\begin{table}[h]
\centering
\begin{tabular}[t]{l|c|c|c|l|c}
$\phantom{123}\beta$ & $L$ &\# of runs & $\phantom{123}\xi^{\rm eff}$ & $
\phantom{1234}\chi $ & $\gr$ \\
\hline \hline
.418    & $ 40 $ & $200$  & 10.839936 & 
\phantom{1}163.54(13)   &11.941(11)\\
\hline
.418    & $ 60 $ & $200$  & 10.839936 & 
\phantom{1}172.81(11)   &14.104(26)\\
\hline
.418    & $ 80 $ & $321$  & 10.839936 & 
\phantom{1}173.94(6)    &14.587(30)\\
\hline
.418    & $140 $ & $300$  & 10.839936 & 
\phantom{1}174.08(4)    &14.743(39)\\
\hline
.4276   & $140 $ & $100$  & 18.924790 & 
\phantom{1}455.34(35)   &14.610(54)\\
\hline
.4276   & $250 $ & $225$  & 18.924790 & 
\phantom{1}455.90(12)   &14.796(46)\\
\hline
.433345 & $250 $ & $202$  & 33.873923 &1254.48(72)   &14.567(37)\\ 
\hline
\end{tabular}
\caption{\footnotesize Ising data for $\chi=G_2(0)$ and $\gr(\beta,L)$}
\label{MC1}
\end{table}

Our estimated value for $\gr$ in Eq.~(\ref{isMCnew}) 
is in very good agreement with the
values from the analysis of the high temperature expansion
given in Table \ref{grIsingprev}; 
it is also consistent with the
value Eq.~(\ref{grIsing}) obtained from the 
form factor construction. 

\newpage
\newsection{The XY-model}

In this section we compute the leading contribution to the
four-point coupling in the two-dimensional O$(2)$ nonlinear
$\sigma$-model better known as the XY-model. 
Starting from the lattice formulation, after a chain of mappings
consisting of several steps the model is transformed to a system
equivalent to the two-dimensional Coulomb gas.
The continuum limit of the Coulomb gas model 
(corresponding to the Kosterlitz-Thouless critical point 
\cite{KoThou}) is thought to have a dual description
in terms of a Sine-Gordon model at the (extremal)
Sine-Gordon coupling $\beta^2=8\pi$. For a review of the
XY-model, see \cite{ZinnJ}.

In the following we will start by discussing the XY-model S-matrix.
The next step is to solve the Smirnov equations for the three-particle
form factors, which enter the formula for the leading term.
A general method for finding the Sine-Gordon form factors
is given in \cite{BaFrKaZa}. This extends the results of
Smirnov \cite{Smir}, where the form factors for an even number
of particles were found. The spin three-particle form factor
we are interested in is probably similar to the three-particle form 
factor of the fermion operator (corresponding to the equivalent massive
Thirring-model description), explicitly given in
\cite{BaFrKaZa}. Here  however  we need the three-particle form factor 
only for special rapidities and we found it simpler to obtain this 
special version by going back to the functional equations. It is then 
used to numerically evaluate the leading contribution to $\gr$.

\newsubsection{The XY-model S-matrix}

We will regard the XY-model as the $n=2$ member of the family
of O$(n)$ $\sigma$-models. Recall that the formulae (\ref{S123}), 
(\ref{tildeK}) have a smooth $n\to2$ limit;
this has been noted and commented on previously by Woo \cite{Woo}.
In this limit
\be
\sigma_1(\theta)=\frac{\theta}{(i\pi-\theta)}\,s_2(\theta)\,,\qquad\qquad
\sigma_2(\theta)=0\,,\qquad\qquad
\sigma_3(\theta)=s_2(\theta)\,
\label{S123O2}
\ee
and
\be
s_2(\theta)=-\exp\Big\{2i\int_0^\infty\frac{d\omega}{\omega}
\sin(\theta\omega)\,\tilde K_2(\omega)\Big\}\,,\qquad\qquad
\tilde K_2(\omega)=\frac{e^{-\frac{\pi\omega}{2}}}
{2\ch\frac{\pi\omega}{2}}\,.
\label{S0O2}
\ee

In this paper we will assume that the spectrum of the
XY-model in the (massive) continuum limit consists of
an O$(2)$ doublet of massive particles whose S-matrix
is given by (\ref{S123O2}) with (\ref{S0O2}). Of course,
taking the formal $n\to2$ limit of the bootstrap results
valid for $n\geq3$ would not be convincing in itself, but
(\ref{S123O2},\ref{S0O2}) actually coincide with the
$\beta^2\to8\pi$ limit of the Sine-Gordon S-matrix,
the prediction of the Kosterlitz-Thouless theory!
The S-matrix (\ref{S123O2}) and the corresponding scattering states
as a consequence have a ${\cal U}_{q=-1}$(su(2)) Hopf algebra
symmetry, which as a Lie algebra is isomorphic to su(2). 
The latter is an explicit symmetry in the alternative chiral 
Gross-Neveu formulation of the model \cite{ZinnJ}.

\newsubsection{The three particle form factor}

Next we calculate the three-particle form factor at the
special rapidities necessary to compute the leading contribution
(\ref{VCreal}). For this purpose we note that the equations
for the functions $k,l$ given in subsection 3.2 can relatively
easily be solved in this particular case $n=2$. We first note
that Eq.~(\ref{SmyKL3}) simplifies 
\ba
K(i\pi-\th) \is K(i\pi+\th)\,,\nonum
K(i\pi+\th) \is \frac{1}{2i\pi\th}\Bigl\{ (i\pi-\th)(i\pi-2\theta)L(i\pi-\th)
-(i\pi+\th)(i\pi+2\theta)L(i\pi+\th)\Bigr\}\,.
\label{SmyKL4}
\ea
Inserting (\ref{solL}) yields 
\be
K(-i\pi- \theta)=
\frac{3i\pi-2\theta}{3i\pi+2\theta}\cdot
\frac{i\pi-2\theta}{i\pi+2\theta}\,K(-i\pi+\theta)\,.
\label{Sigma2}
\ee
Luckily a term proportional to $K(i\pi- \theta)$ drops out here 
and one is left with the simple form (\ref{Sigma2}). This can easily 
be converted into the form (\ref{min1}) and solved as
\be
K(i\pi -\th) = (2\theta-5\pi i)(2\theta-7\pi i)\,
e^{D\big(\frac{\theta}{2}\big)}\,
\phi\Big(i\ch\frac{\theta}{2}\Big)\,.
\label{solSigma}
\ee
Here
\be
D(\theta)=\Delta_{\frac{1}{4}}(\theta)+
\Delta_{\frac{3}{4}}(\theta)
\ee
in the notation of Appendix D and $\phi(z)$ is a polynomial
function to be determined later.

Since $Y(\theta)$ already has the right singularity structure
the functions $K(\theta)$ and $L(\theta)$ are analytic in the
physical strip. The residue axioms determine their value at  
$\theta=0$ and $\theta=\frac{i\pi}{2}$ as
\ba
K(0)\is 0\,,\bspace \sspace \,L(0) =1\,, \nonum
K\Big(\frac{i\pi}{2}\Big)\is 
\frac{u_0}{u^2\Big(\frac{i\pi}{2}\Big)}\,,
\qquad\qquad
L\Big(\frac{i\pi}{2}\Big)=-s_2\Big(\frac{i\pi}{2}\Big)\,
K\Big(\frac{i\pi}{2}\Big)\;.
\label{resKL2}
\ea
So far we have established that the solution can be expressed
in terms of
\be
Y(\theta)=-2i\frac{\ch^3·\frac{\theta}{2}}
{\sh\frac{\theta}{2}\,\ch\theta}\,
e^{2\Delta(i\pi+\theta)+\Delta(2\theta)-\Delta(0)}
\label{solY}
\ee
and
\be
K(\theta)=(2\theta+3\pi i)(2\theta+5\pi i)\,
e^{D\big(\frac{i\pi-\theta}{2}\big)}\,
\phi\Big(\sh\frac{\theta}{2}\Big)\,.
\label{solK}
\ee
The polynomial $\phi(z)$ can be determined using the residue constraints
(\ref{resKL2}), which we can rewrite as
\ba
K(0)\is 0\,,\bspace \sspace \quad \;\;\,K^\prime(0)=\frac{2}{i\pi}\,,
\nonum
K\Big(\frac{i\pi}{2}\Big) \is 
e^{\Delta(0)-2\Delta\big(\frac{i\pi}{2}\big)}\,,\sspace
K\Big(-\frac{i\pi}{2}\Big)= -e^{\Delta(0)
-\Delta\big(\frac{i\pi}{2}\big)-\Delta\big(\frac{-i\pi}{2}\big)}\;.
\label{resK2}
\ea
Using (\ref{solY}) and (\ref{asyDelta}) one sees that for
real $\theta\to\infty$
\be
\vert Y(\theta)\vert\sim e^{-\theta}\,\theta^{\frac{3}{4}}\,.
\ee
This can be used to infer that the polynomial $\phi(z)$ can be at most 
second order, otherwise the integral contribution to the leading term 
would diverge. Taking into account that $K(0)=0$ and the requirement of
real analyticity one must have
\be
\phi(z)=i\phi_1\,z+\phi_2\,z^2\,,
\label{fz}
\ee
for real constants $\phi_1$ and $\phi_2$. Now it is easy to see that 
(\ref{resKL2}) determines $\phi_1$ as
\be
\phi_1=\frac{4}{15\pi^3}\,e^{-D\big(\frac{i\pi}{2}\big)}\,.
\label{solphi1}
\ee
In order to determine $\phi_2$ we employ the following identities 
\cite{Prud}
\ba
p(\alpha)&:=&\exp\Bigg\{\int_0^\infty\,\rmd\omega
\frac{\ch(\alpha\pi\omega)-1}{\sh(\pi\omega)}\,
e^{-\pi\omega}\Bigg\}=\frac{\alpha\pi}{2}\,
\frac{1}{\sin\Big(\frac{\alpha\pi}{2}\big)}\,,
\label{palpha}\\
q(\alpha)&:=&\exp\Bigg\{\int_0^\infty\,\rmd\omega
\frac{\ch(\alpha\pi\omega)-1}{\sh(\pi\omega)}\,
e^{-2\pi\omega}\Bigg\}=
\frac{1-\alpha^2}{\cos\Big(\frac{\alpha\pi}{2}\big)}\,,
\label{qalpha}
\ea
to obtain 
\be
\exp\left\{\Delta(0)-2\Delta\Big(\frac{i\pi}{2}\Big)
+D\Big(\frac{i\pi}{2}\Big)-D\Big(\frac{i\pi}{4}\Big) \right\}
=p(1)\,\frac{q^2(1)}{q\big(\frac{3}{2}\big)}=
\frac{16\sqrt{2}}{5\pi}\,.
\ee
This can be used to show that (\ref{resK2}) is satisfied
for the choice $\phi_2=0$. Thus $\phi(\sh \frac{\th}{2}) =
i \phi_1 \sh\frac{\th}{2}$, and since $4\th L(\th) = 
(i\pi - 2\th) K(\th) - (i\pi + 2\th) K(-\th)$, both 
$k(\th) = Y(\th) K(\th)$ and $l(\th) = Y(\th) L(\th)$ are 
known explicitly for the XY model.

\newsubsection{Calculation of the leading contribution}

Having all the ingredients at our disposal we can compute
the leading term (\ref{grlead}) of the intrinsic coupling. 
Firstly from (\ref{gammaAOn}) we have for $n=2$
\be
\gamma_{4;121}^{(I)}=\frac{4}{\pi}(\ln4-1)\,.
\ee
Further substituting the explicit results for the functions
$k,l$ obtained above into Eq.~(\ref{gammaC1}) and evaluating the
resulting expression numerically we obtain
\be
\gamma_{4;121}^{(II)}=-5.14902(1)\,,
\ee
and hence
\be
\gamma_{4;121}=
\gamma_{4;121}^{(I)}+\gamma_{4;121}^{(II)}=
-4.65718\,.
\label{gamma4O2}
\ee
Thus the leading contribution to the XY-model
four-point coupling is
\be
\gr=-2\frac{\gamma_4}{\gamma_2\delta_2}\approx
-2\gamma_{4;121}=9.314\,.
\label{grO2}
\ee

Since $\gamma_2\delta_2>1$ and since the next leading contributions
to $\gamma_4$ are probably positive (as they are in the Ising and
O(3) models), we expect that the true value of $\gr$ will be 
less than that given in (\ref{grO2}) (probably by $2-4\%$).

\newsubsection{Comparison with lattice results}

For the XY model with standard action Kim \cite{Kim} gives the value
\be
\gr=8.89(20)
\ee
for $\beta=1/0.98$. 
We are in the process of producing higher precision Monte Carlo data
for this model; so far we can only give a preliminary result,
obtained on  a lattice of size $L=500$ at $\beta=1.0174$:
\be
\gr=9.14(12)\,,
\ee
We will return to this issue in a separate publication, where we
intend to analyze the finite size corrections as well as the lattice
artifacts.

We also wish to mention the results from the high temperature 
expansion: Butera and Comi \cite{BC} obtain
\be
\gr=9.15(10)\,,
\ee
whereas Pelissetto and Vicari \cite{peliss} give
\be
\gr=9.01(5)\,.
\ee
So there is an overall rough agreement between the lattice and the form 
factor results, but the precision is not comparable to that
obtained for the Ising model.

\newpage
\newsection{The O(3) nonlinear sigma-model}

The O(3) nonlinear sigma model is an important testing ground for 
quantum field theoretical scenarios in nonabelian gauge theories.
The form factor technique has been particularly fruitful in studying
its possible off-shell dynamics and can be confronted with 
what can be achieved by perturbation
theory or numerical simulations \cite{BN}.
The intrinsic coupling has been computed before by a number of different 
techniques; we compare the results with ours at the end of this section.
The present form factor determination takes as usual the 
Zamolodchikov two-particle 
S-matrix \cite{ZZ} as its starting point; 
it is given by Eqs.~(\ref{OnS},\,\ref{S123}) with $n=3$ and
\be
s_2(\th)=\frac{\th-\pi i}{\th+\pi i}\,.
\label{o1}
\ee
The corresponding kernel (\ref{tildeK}) is simply given by
\be
\tilde K_3(\omega)=e^{-\pi\omega}\,.
\ee

\newsubsection{Form factor determination of {\boldmath $\gr$}}

Following the by now routine procedure we first collect the 
ingredients for the evaluation of the dominant $(1,2,1)$
contribution to the intrinsic coupling. 
From (\ref{gammaAOn}) one readily finds for $n=3$
\be
\gamma_{4;121}^{(I)}=\frac{4}{\pi}\,.
\ee
The O(3) form factors have been computed in \cite{Smir,BN,BNH}.
In particular the reduced 3-particle form factor ${\cal G}$ 
in Eq.~(\ref{red}) is given by
\be
{\cal G}^a_{a_1a_2a_3}(\theta_1,\theta_2,\theta_3)
=\tau_3(\theta_1,\theta_2,\theta_3)\Bigl\{
 \delta_{a_1}^a\delta_{a_2a_3}(\theta_3-\theta_2)
+\delta_{a_2}^a\delta_{a_1a_3}(\theta_1-\theta_3-2\pi i)
+\delta_{a_3}^a\delta_{a_1a_2}(\theta_2-\theta_1)\Bigr\},
\ee
where
\ba
\tau_N(\theta_1,\ldots,\theta_N)\is \prod_{1\le i<j\le N}
\tau(\theta_i-\theta_j)\,,\nonum
\tau(\th) \is \frac{\pi(\th - i\pi)}{\th(2\pi i - \th)} 
\tanh\frac{\th}{2}\,.
\ea
Correspondingly the functions $k,l$ parametrizing $f_b$ via Eq.~(\ref{kl}) 
are for $n=3$ explicitly given by
\be  
k(\th)=\frac{2\th}{\pi i-\th}l(\th)\,,\sspace 
l(\th)=\frac{\pi^3 \,T^2(2\th)\,\th(2\th-\pi i)}
{4T^4(\th)(\pi^2+\th^2)^2}\,.
\ee
Plugging this into the general formula Eq.~(\ref{gammaC1}) yields
\be
\gamma_{4;121}^{(II)} = \frac{1}{8\pi}\int_0^{\infty}\rmd u
\bigg\{  
\frac{\pi^6 u^2(4 u^2 + \pi^2)(2 u^2 + \pi^2)}{4(u^2 + \pi^2)^5}\,
\frac{T^4(2 u)}{T^8(u)\ch^2 u} - \frac{64}{u^2} \bigg\}\,.
\label{o5}
\ee 
Numerically we then obtain $\gamma_{4;121} = -4.16835492(1)$,
so that as a 
first approximation $\gr \approx - \frac{5}{3} \gamma_{4;121} =
6.9472$. This is already in rough agreement with other 
determinations in the continuum theory: the $1/n$, the $\eps$- and
the $g$-expansions \cite{peliss, FMPPT}. 
The leading order $1/n$ computations have been
performed in \cite{CPRVU}. For the spectral integrals the result is 
\be
\gamma_2 = 1 + 0.00671941\frac{1}{n} + O\Big(\frac{1}{n^2}\Big)\;,\sspace
\delta_2 = 1 + 0.00026836\frac{1}{n} + O\Big(\frac{1}{n^2}\Big)\;.
\label{o6}
\ee
and for the coupling \cite{CPRVgR}
\be
\gr = \frac{8\pi}{n}\left[ 1 - 0.602033\frac{1}{n} + 
O\Big(\frac{1}{n^2}\Big)\right]\;.
\label{o7}
\ee
which gives the approximation $\gr\approx 6.70$ for the case $n=3$.  
The results from the other methods are given in Table~\ref{o3tab}.
Considering the rather short series in each case it is amazing how 
well the estimates by the various methods agree.

For a more precise determination we now return to the form factor 
approach and examine the sub-leading contributions. 
Using the exact form factors \cite{BN} the results for the 3- and  
5-particle contributions to $\gamma_2$ and $\delta_2$ are readily 
evaluated and are listed in Table~\ref{o3spectralmoments}. 

\begin{table}[ht]
\centering
\begin{tabular}[t]{r|r|r}
\hline
$m$ &$\gamma_{2;m}\ \ \ \ \ \ \ \ $& 
$\delta_{2;m}\ \ \ \ \ \ \ \ $  \\[0.5ex]
\hline \hline    

3 & $1.67995(1) \times 10^{-3}$ & $3.46494(1) \times 10^{-5}$\\[0.5ex]

5 & $6.622(1)\times 10^{-6}$ & $7.114(1) \times 10^{-9}$\\[0.5ex]

\hline
\end{tabular}
\caption{\footnotesize $m$-particle contribution to $\gamma_2, \delta_2$ 
in the O(3) model}
\label{o3spectralmoments}
\end{table}

The size of the higher particle contributions to 
$\gamma_2$ and $\delta_2$ can roughly be estimated by an off hand 
extrapolation of Table~\ref{o3spectralmoments}; essentially they are
negligible to the desired accuracy. The latter could also be 
justified by referring to a more refined extrapolation scheme, based 
on the scaling hypothesis of ref.~\cite{BN}. In upshot we obtain
\be
\gamma_2=1.001\,687(1)\,,\sspace \delta_2=1.000\,034\,657(1)\,.
\label{O3gamdel2}
\ee
The computation of the sub-leading terms to $\gamma_4$ is 
much more involved. The starting point is again the formulae (\ref{dom2})
in subsection 3.1. Due to the complexity of the form factors however 
the computation is feasible only computer aided.
The  essential steps are given in appendices B,C. The computation
has been performed independently by subsets of the authors using
slightly different techniques. The final results for the 
contributions of the $k$-$l$-$m$ intermediate states 
with $k+l+m\le 6$ to $\gamma_4$  are listed in Table~\ref{O3gamma4}. 

\begin{table}[ht]
\vskip 0.5cm
\centering
\begin{tabular}[t]{r|l}
\hline
$k,l,m$ &$\phantom{1234}\gamma_{4;klm}$  \\[0.5ex]
\hline \hline

$1,2,1$ & $-4.16835492(1)$\\[0.5ex]

$1,2,3$ & $\phantom{-}0.051748(1)$\\[0.5ex]

$1,4,1$ & $-0.004065(1)$\\[0.5ex]

\hline
\end{tabular}
\caption{\footnotesize $k$-$l$-$m$-particles contribution to 
$\gamma_4$ in the O(3) model}
\label{O3gamma4}
\end{table}

The leading $1$-$2$-$1$ contribution is a factor $\sim 42$ greater in 
magnitude than the sum of $k$-$l$-$m$ contributions with $k+l+m=6$. 
It is difficult to bound the rest of the contributions, especially
since the signs appear to be alternating. The computation of
the states with $l+m+n=8$ would be quite an undertaking.
But assuming that the pattern in Table~\ref{O3gamma4} continues,
as it seems to be the case in the Ising model (see Table~\ref{gamma4}),
then we consider the assumption that the sum of the remaining 
contributions $k+l+m\ge 8$ is $\le 10\%$ of the sum of the $k+l+m=6$ 
contributions to be reasonable and we then obtain
\begin{equation}
\gamma_4=-4.069(10)\,,
\end{equation}
and hence our final result
\begin{equation}
\label{grffb}
\gr=6.770(17)\,.
\end{equation} 
This amounts to a determination of $\gr$ to within $0.3\%$. 
For comparison we give some results of other already 
published determinations in Table \ref{o3tab}.  
The first two are continuum methods while the last 
one is based on the lattice regularization.
We describe the two lattice techniques in 
somewhat more detail in the next subsection,
including in particular
our own recent Monte Carlo results.

\begin{table}[ht]
\centering
\begin{tabular}[t]{c|c}
\hline
Method             & Value for $\gr$ \\[0.5ex]
\hline \hline    

$g$-expansion  & 6.66(6) \cite{FMPPT} \\[0.5ex]

$\eps$-expansion  & 6.55(8) \cite{peliss} \\[0.5ex]

High temperature   &  $6.56(4)$ \cite{CPRVgR}, $6.6(1)$ \cite{CPRVhT}\\[0.5ex]


\hline\end{tabular}
\caption{\footnotesize Other determinations of $\gr$ in the O(3) model}
\label{o3tab}
\end{table}

\newsubsection{Lattice computations of {\boldmath $\gr$}}

{\em High temperature expansion:} 

The analyses of the high temperature expansion for 
the spectral moments 
give $\gamma_2=1.0013(2)$ \cite{CPRVU} 
and $\delta_2=1.000029(5)$ \cite{rossi}. 
The agreement with the FFB values Eq.~(\ref{O3gamdel2}) is acceptable;
note that these are smaller than that anticipated from
the leading order of the $1/n$ approximation, Eqs. (\ref{o6}).

The various Pad\'e approximations
show the coupling falling rapidly as $\beta$ increases in the
region of small $\beta$, then a region of rather flat behavior
after which these approximations show diverse behavior;
some analyses indicate that in fact there is a shallow minimum and
that the continuum limit is actually approached from below
(see e.g. refs.~\cite{CPRV,peliss}).
In ref.~\cite{CPRVhT} Campostrini et al.
quote for the case $n=3$ the result $\gr=6.6(1)$,
and in a more recent publication Pelissetto and Vicari
cite $6.56(4)$~\cite{peliss}.
Butera and Comi on the other hand are rather cautious, and
did not quote a value for the case $n=3$ in ref.~\cite{BC};
if pressed they would at present cite $\gr=6.6(2)$ \cite{butera}.

{\em Numerical simulations:}

Monte Carlo computations of $\gr$ have a long history,
see e.g. refs.~\cite{FMPPT,Kim}. In order to attempt to
match the apparent precision attained in the FFB approach, 
we recently performed new high-precision measurements.
These were performed on several IBM RISC 6000 workstations at the 
Werner-Heisenberg-Institut. In addition we made use of the SGI 2000
machine of the University of Arizona, especially for the very
time consuming simulations on large lattices.

Based on the fixed point action \cite{FP} 
we have measured $\gr$ at three different 
values of $\beta$: 0.70, 0.85 and 1.00, corresponding to 
correlation length $\xi\approx 3.2$, $6.0$ and $12.2$,
at the values of $z=L/\xi$ in the range $5.4 \ldots 8.2$.
The data and their analysis can be found in \cite{usgR},  
the final result is $\gr^{\rm FP} = 6.77(2)$. 

Monte Carlo measurements 
with the standard action were performed  
using a method similar to the cluster estimator of~\cite{lw}.
We have reported the analysis of such simulations already 
in our earlier paper \cite{usgR}.
But in the meantime we produced more data
and we take the opportunity to report them here.

The present status of the results of our simulations are given in
Table~\ref{MC3}.
In this table we also indicate the number of measurements. These were 
performed using the cluster algorithm as follows: one run consisted of 
100,000 clusters used for thermalization, followed by 20,000 sweeps
of the lattice used for measurements. Each run was repeated after 
changing the initial configuration. One such run was considered as one 
independent measurement. The error was computed out of this sample by 
using the jack-knife method.

\begin{table}
\centering
\begin{tabular}[t]{l|r|c|l|l|c|c}
\phantom{1}$\beta$ & $L$ &\# of runs & \phantom{123}$\xi$ & 
$\phantom{1234}\chi$ &
$\gr(\xi,L)$&$\gr(\xi,\infty)$\\
\hline \hline
1.5 & 80 &344& \phantom{1}11.030(7) & \phantom{12}175.95(11)  &6.553(16)&
6.616(16)\\
\hline
1.6 &140 &370& \phantom{1}18.950(14)& \phantom{12}447.13(34)  &6.612(15)&
6.668(15)\\
\hline
1.7 &250 &367& \phantom{1}34.500(15)& \phantom{1}1267.20(57)  &6.665(14)&
6.730(14)\\
\hline
1.8 &500 &382& \phantom{1}64.790(26)& \phantom{1}3838.76(1.50)&6.691(15)&
6.733(15)\\
\hline
1.9 &910 &127&122.330(74)&11883.0(6.4)&6.737(21)& 6.792(21)\\
\hline
1.95&1230& \phantom{1}68&167.71(17) &20901.4(19.0) &6.792(40)& 6.853(40)\\
\hline
\end{tabular}
\caption{\footnotesize O(3) data for $\xi,\,\chi=3G_2(0)$ and $\gr$}
\label{MC3}
\end{table}

Our measurements were taken at 6 different correlation lengths ranging
from about 11 to about 168 on lattices satisfying $L/\xi\approx 7$.
To study the finite volume effects, we took in addition data at
$\xi\approx 11$  for lattices of sizes $L$ with 
$L/\xi\approx 5.5,\ \ 9$ and 13. As discussed in \cite{usgR}, the
finite size effects are well described by the formula (\ref{grz}),
even at finite (large) correlation lengths.
In the O$(3)$ model the $n=\infty$ value $c=\sqrt{8\pi}$ fits very well.

\begin{figure}[htb]
\centerline{\epsfxsize=8.0cm\epsfbox{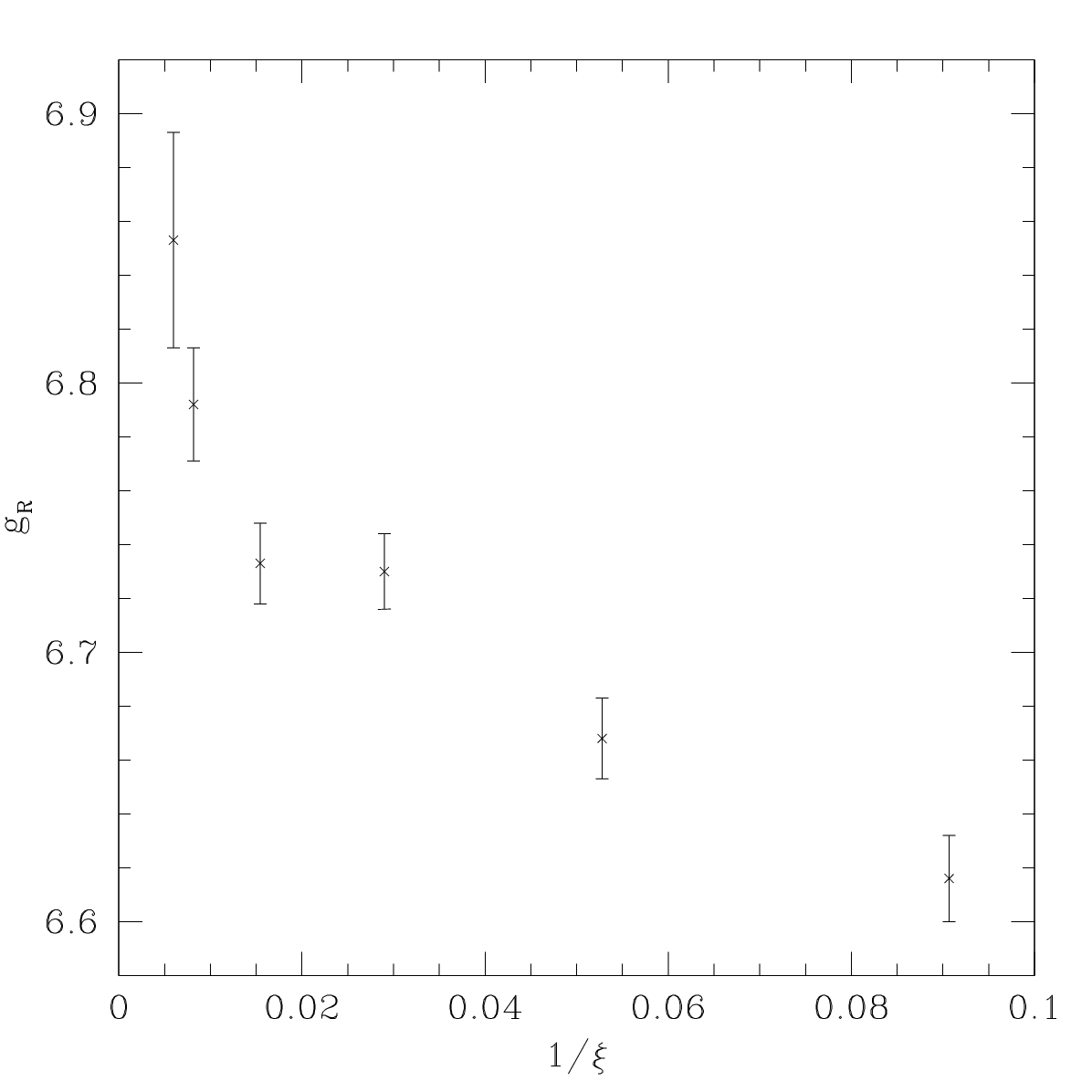}}
\caption{The extrapolated values of $\gr(\xi,\infty)$}
\label{grlatt3}
\end{figure}

But unlike the Ising model, the lattice artifacts are by no means
negligible. To study them, we first use Eq.~(\ref{grz}) to extrapolate
our data to $z=\infty$. In this extrapolation
we use the effective correlation length $\xi_{\rm eff}$ and neglect the
fact that this is not exactly equal to the exponential correlation
length. In Fig.~\ref{grlatt3} we plot those extrapolated values of $\gr$
against $1/\xi$ which we identify with $1/\xi^{\rm eff}$. 


Unfortunately there is no rigorous result concerning the nature
of the approach to the continuum limit.
At the time of our last analysis \cite{usgR} the data
point at the largest value of $\xi\sim 168$ was not available.
In that paper we fitted the data in the entire range from $\xi\sim 11$
to $\xi\sim 122$ using a Symanzik type ansatz of the form
$\gr (\xi)=\gr(\infty)\left[1+b_1\xi^{-2}\log\xi+b_2\xi^{-2}\right]$, 
and thereby obtained the result $\gr=6.77(2)$. 
When we now repeat the same fit for the new data,
which in particular includes the new point at $\xi\sim 168$,
the result is only slightly changed to $\gr=6.78(2)$
but the quality of the fit becomes poorer. The fact that 
the two data points closest to the continuum limit  
lie above $6.78$ is in this scenario interpreted 
as a statistical fluctuation.

On the other hand the present rather large central value
at $\xi\sim 168$ could be interpreted as an indication  
that the continuum limit is approached much slower than 
conventionally assumed, perhaps as slow as $1/\ln\xi$ 
(which may be expected in the O(2) model \cite{Hasenbusch})! 
If we adopt this viewpoint it is clear that,
although qualitative fits can be made, 
without further analytic information, our data are not sufficient 
to make a reliable quantitative extrapolation to the continuum limit.
However, independent of the assumed form of the approach to the
continuum limit, if the large value at $\xi\sim168$ is confirmed 
by more extensive studies it would practically establish a
discrepancy between the form factor and the lattice constructions
of the O(3) sigma-model. 
This point, which needs complete control over all
systematic effects, albeit extremely difficult on such large
lattices, is certainly worthy of further investigations.

\newsection{Conclusions}

A new technique to compute the intrinsic 4-point coupling
in a large class of two-dimensional QFTs has been developed and 
tested. Starting from the form factor resolution of the 
4-point function the termwise zero momentum limit turned out 
to exist, providing a decomposition of the coupling into 
terms with a definite number $(k,l,m)$ of intermediate 
particles. Based on the exactly known form factors these
terms can be computed practically exactly and in the models
mainly considered (Ising and O(3)) were found to be rapidly
decaying with increasing particle numbers. There is every reason
to expect that this trend continues, which allowed us to
equip the results with an intrinsic error estimate. 
The final results are
\ba
\mbox{Ising model:} \sspace \gr = 14.6975(1)\;,\nonum
\mbox{O(3) model:} \sspace \gr = 6. 770(17)\;.
\ea
They amount to a determination of $\gr$ to within $<0.001\%$ and
$0.3\%$, respectively. 

In addition we obtained the universal, model-independent 
formula (\ref{i4}) for the dominant contribution to the coupling, 
which typically
seems to account for about $98\%$ of the full answer. We illustrated 
its use in testing our proposed bootstrap description of the XY-model. 
It would surely also be interesting to apply it e.g.~to supersymmetric 
theories, where alternative techniques are hardly available. 

The comparison with the lattice determinations of $\gr$ is quite
impressive in the case of the Ising model, where there is also very
good agreement between the high temperature and the Monte Carlo
determinations. For O(2) we are so far lacking both precise Monte
Carlo and form factor data, but at this preliminary stage there is
rough agreement.  We intend to return to this model in a separate
publication.  

The situation in O(3) is not completely clear: There is a less than 
perfect agreement between the high temperature result and the new 
high precision Monte Carlo data, and there is also room for doubt 
about the agreement between Monte Carlo and form factor. We cannot 
resolve this question at the moment, mainly because even with our 
enormous amount of Monte Carlo data it is at the moment not clear what 
the correct extrapolation to the continuum is.

\vspace{1cm}

{\tt Acknowledgements:} 
This investigation was supported in part by the Hungarian National 
Science Fund OTKA (under T030099), and also by 
the Schweizerische Nationalfonds. The work of M.N. was supported 
by NSF grant 97-22097.

\newpage
\setcounter{section}{0}

\newappendix{General formula for the dominant term}
\setcounter{equation}{0}

Here we describe the generalization of the formula (\ref{grlead})
for the dominant 1-2-1 particle contribution to $\gr$ to 
general integrable QFTs without bound states and operators other 
than the `fundamental' field. The latter is particularly natural 
in the form factor approach because `fundamental' and `composite' 
operators are treated on an equal footing. Thus let $\cO_l$ be 
possibly distinct, possibly non-scalar but parity odd operators
$\cO_l$ and write $o_l$ for the quantum numbers labeling them. 
Parallel to (\ref{gcdef}) we define the Green functions by
\be
\tilde{S}_c^{o_1\ldots o_L}(k_1,\ldots,k_L) =(2\pi)^2 
\delta^{(2)}(k_1+\ldots+k_L)\,G^{o_1\ldots o_L}(k_1,\ldots,k_L)\,,
\label{Gcdef}
\ee
where $\tilde{S}_c^{o_1\ldots 0_L}(k_1,\ldots,k_L)$ is the Fourier
transform of the connected part of the Euclidean correlation function 
$\bra \cO_1(x_1) \ldots \cO_L(x_L)\ket$. The obvious generalization
of the intrinsic coupling is
\be
\gr = - \cN M^2 \frac{G^{o_1o_2o_3o_4}(0,0,0,0)}%
{\sum_{j<k} G^{o_jo_k}(0,0)^2}\;.
\label{Grdef}
\ee
Here $M$ is again the mass gap and the constant $\cN$ is conveniently
adjusted to normalize the 1-particle contribution to the denominator
to unity. If $\cF^o_a$ are the constant 1-particle form factors of $\cO$,
the leading 1-particle contribution to $G^{o_1o_2}(0,0)$ is just
$Z^{o_1o_2} := M^{-2}\cF^{o_1}_a C^{ab} \cF^{o_2}_b$, where $C^{ab}$ is 
the charge conjugation matrix associated with the given S-matrix 
(c.f.~below). Thus we take $\cN = \sum_{j<k} (Z^{o_jo_k})^2$. With these
normalizations the dominant 1-2-1 particle contribution to the coupling 
(\ref{Grdef}) is 
\ba
&\nspace &\gr\Big|_{1-2-1} = -\frac{1}{2}\sum_{s \in {\cal S}_4}
D^{o_{s1}o_{s2};o_{s3}o_{s4}} + 
\int_0^{\infty} \frac{\rmd u}{4\pi} \sum_{s \in {\cal S}_4}
\bigg[-\frac{4}{u^2}Z^{o_{s1}o_{s2}}Z^{o_{s3}o_{s4}}
\nonum
&\nspace & \;\;\;
+ \frac{1}{16 \ch^2 u}\,\cF_{a}^{o_{s1}} \cF_{b}^{o_{s4}} C^{aa_3}C^{bb_3}
\cF_{a_3a_2a_1}^{o_{s2}}(i\pi,-u,u) C^{a_2b_2}C^{a_1b_1}
\cF^{o_{s3}}_{b_3b_2b_1}(i\pi,-u,u)^* \bigg]\;.
\label{i4}
\ea
Here the symmetrization is over all elements of the permutation  
group ${\cal S}_4$. $D$ is defined in terms of the given bootstrap
S-matrix $S_{ab}^{cd}(\th)$ by 
\be
D^{o_1o_2;o_3o_4} =  -i\frac{d}{d\th} S_{cd}^{ab}(\th)\bigg|_{\th =0}\,
\cF^{o_1}_a\cF^{o_2}_b \,C^{cc'} C^{dd'} \cF^{o_3}_{c'} \cF^{o_4}_{d'}
\label{i5}
\ee
and $\cF^o_{abc}(\th_1,\th_2,\th_3)$ is the 3-particle form factor
of $\cO$. Taking the results for the O($n$) models as a guideline one
would expect that (\ref{i4}) typically yields about $98\%$ of the 
full answer for the coupling. 

In the following we describe the derivation of (\ref{i4}).
In contrast to that of (\ref{grlead}) we keep track here of 
the distributional terms like (\ref{b9}) and show explicitly that 
they cancel out in the final answer. In particular this illustrates 
that the use of the simplifying limit procedure (\ref{kappas}) is 
justified. 
 
To fix conventions we first recall the defining relations
of a generic bootstrap S-matrix. A matrix-valued meromorphic function
$S_{ab}^{dc}(\th),\;\th\in\C$, is called a two particle 
$S$-matrix if it satisfies the following set of equations. 
First the Yang-Baxter equation
\be
S_{ab}^{nm}(\th_{12})S_{nc}^{kp}(\th_{13})S_{mp}^{ji}(\th_{23})
=S_{bc}^{nm}(\th_{23})S_{am}^{pi}(\th_{13})S_{pn}^{kj}(\th_{12})\;,
\label{s1}
\ee
where $\th_{12}=\th_1-\th_2$ etc. Second 
unitarity (\ref{s2}a,b) and crossing invariance (\ref{s2}c)
\begin{subeqnarray}
&&S_{ab}^{mn}(\th)\,S_{nm}^{cd}(-\th)=\delta_a^d\delta_b^c\\
&&S_{an}^{mc}(\th)\,S_{bm}^{nd}(2\pi i-\th)=\delta_a^d\delta_b^c\\
&&S_{ab}^{dc}(\th)=C_{aa'}C^{dd'}\,S_{bd'}^{ca'}(i\pi -\th)\;,
\label{s2}
\end{subeqnarray}
where (\ref{s2}c) together with one of the unitarity conditions
(\ref{s2}a), (\ref{s2}b) implies the other. Further real analyticity and 
Bose symmetry
\be
[S_{ab}^{dc}(\th)]^* =S_{ab}^{dc}(-\th^*)\;,\sspace
S_{ab}^{dc}(\th)=S_{ba}^{cd}(\th)\;.
\label{s3}
\ee
Finally the normalization condition
\be
S_{ab}^{dc}(0)=-\delta_a^c\delta_b^d\;.
\label{s4}
\ee
The indices $a,b,\ldots$ refer to a basis in a finite dimensional
vector space $V$. Indices can be raised and lowered by means of 
the (constant, symmetric, positive definite) `charge conjugation matrix'
$C_{ab}$ and its inverse $C^{ab}$, satisfying $C_{ad}C^{db}=
\delta_a^b$. The S-matrix is a meromorphic function of $\th$.
Bound state poles, if any, are situated on the imaginary axis in 
the so-called physical strip $0\leq \mbox{Im}\,\th< \pi$. From crossing 
invariance and the normalization (\ref{s4}) one infers that 
$S_{ab}^{dc}(i\pi) = -C_{ab}C^{dc}$ is always regular, in contrast to  
$S_{ab}^{dc}(-i\pi)$ which may be singular.

Next we prepare the counterparts of Eqs (\ref{g9}), (\ref{g10}).
\ba
\bra 0|\cO_1|\mb\ket \bra \mb|\cO_2|0\ket
& \; \longleftrightarrow \;& I_m^{o_1o_2}(\th)\;,
\nonum
\bra 0|\cO_1|\kb\ket \bra \kb|\cO_2|\lb\ket\,
\bra \lb|\cO_3|\mb \ket \bra \mb|\cO_4|0\ket
&\; \longleftrightarrow \;& I_{klm}^{o_1o_2o_3o_4}(\omega|\xi|\th)\;,
\label{G8}
\ea
where
\ba
& \nspace & 
I_m^{o_1o_2}(\th) = \cF^{o_1}_A(\th) C^{AB} \cF^{o_2}_{B^T}(\th^T)\;,
\nonum
& \nspace & I_{klm}^{o_1o_2o_3o_4}(\omega|\xi|\th)=
\cF^{o_1}_A(\omega)\,C^{AB}\,\cF^{o_2}_{B^TC}(\omega^T|\xi)\,C^{CD}\,
\cF^{o_3}_{D^TE}(\xi^T|\th)\,C^{EK}\,\cF^{o_4}_{K^T}(\th^T)\,.
\ea
From the S-matrix exchange relations it follows that
$I^{o_1o_2}_m(\th)$ is a completely symmetric function
in $\th= (\th_1,\ldots,\th_m)$. Similarly
$I_{klm}^{o_1o_2o_3o_4}(\omega|\xi|\th)$ is symmetric
in each of the sets of variables $\omega = (\omega_1,
\ldots,\omega_k)$, $\xi = (\xi_1,\ldots,\xi_l)$ and
$\th = (\th_1,\ldots,\th_m)$. As before we denote by
$V_m(k_1,k_2)$ and $V_{klm}(k_1,k_2,k_3,k_4)$ the quantities
(\ref{g7}) with the integrations over the rapidities performed,
where the measure is inherited from (\ref{statesum}).
For simplicity we drop the operator labels $o_j$ in the 
notation. When evaluated at $k_j = (M\sh\kappa_j, 0)$ we write 
$v_{klm}(\kappa_1,\kappa_2,\kappa_3,\kappa_4)$, etc.

In the next step one inserts the expressions for the generalized 
form factors in terms of the ordinary form factors; see \cite{MNmod}
for an account in the present conventions. For $m=2$ one obtains 
explicitly
\ba
&& v_{121}(\kappa_1,\kappa_2,\kappa_3,\kappa_4) =
\frac{\cF_a^{o_1}C^{ab}C^{cd}\cF^{o_4}_d}{8\ch^2\kappa_1\ch^2\kappa_4}
\Bigg\{ \frac{4\pi \cF^{o_2}_m C^{mn} \cF^{o_3}_n\,C_{bc}}%
{\ch\kappa_2(\ch\kappa_1 + \ch\kappa_2)}\,\delta(\kappa_1 + \kappa_4)  
\nonum
&& \quad + \frac{4\pi \cF^{o_2}_m \cF^{o_3}_n}%
{\ch\kappa_2(\ch\kappa_1 + \ch\kappa_2)} S_{c\;b}^{mn}(i\pi - \kappa_1 +
\kappa_2)
\,\delta(\kappa_2 + \kappa_4)
\nonum
&& \quad + \frac{1}{\ch\kappa_2(\ch\kappa_1 + \ch\kappa_2)}\cF_m^{o_2}C^{mn}
\cF^{o_3}_{cb n}(\kappa_4 + i\pi - i\eps,-\kappa_1,-\kappa_2)^*
\label{dom4}
\\
&& \quad + \frac{1}{\ch\kappa_3(\ch\kappa_3 + \ch\kappa_4)}\cF_m^{o_3}C^{mn}
\cF^{o_2}_{bc n}(-\kappa_1 + i\pi - i\eps,\kappa_4,\kappa_3)
\nonum
&& \quad + \int_0^{\infty}\frac{\rmd u}{4\pi} \frac{1}{4 \ch^2 \frac{u}{2}
+
(\sh\kappa_1 + \sh\kappa_2)^2}\, C^{mk}C^{nl}\,
\nonum
&& \quad \times \cF^{o_2}_{bmn}\Big(-\kappa_1 + i\pi -i\eps,
\Lambda_* - \frac{u}{2}, \Lambda_* + \frac{u}{2}\Big) 
\cF^{o_3}_{ckl}\Big(\kappa_4 + i\pi -i\eps,
\Lambda_* - \frac{u}{2}, \Lambda_* + \frac{u}{2}\Big)^*\Bigg\}\,.
\nonumber
\ea
The $\kappa_i \ra 0$ limit of this expression can be evaluated on general
grounds. The key observation is that a three particle form factor   
has the following universal `small rapidity' expansion
\ba
&& \cF^o_{a_1a_2a_3}(\th_1 + i\pi - i\eps,\th_2,\th_3) =
\left[\frac{1}{\th_{12} - i\eps} - \frac{1}{\th_{13} - i\eps}\right]
\Big[ 2i (C_{a_1a_2} \cF^o_{a_3}+ C_{a_1a_3} F^o_{a_2}) \nonum
&& +  2 C_{a_1 c}( \th_{13} D_{a_2 a_3}^{c\;d} -
\th_{12} D_{a_2a_3}^{d\;c}) \cF_d^o\Big]\,,\sspace
\mbox{where} \;\;D_{ab}^{cd} = -i \frac{d}{d\th}
S_{ab}^{cd}(\th)\bigg|_{\th = 0}\;.
\label{dom5}
\ea
This expression is uniquely determined by the following properties:
(i) The numerator is linear and boost invariant in the rapidities.
(ii) It obeys the (linearized) S-matrix exchange relations in
$\th_2$ and $\th_3$. (iii) It has simple poles at $\th_{21} + i\eps$ and
$\th_{31} + i\eps$ with  residues dictated by the form factor
`residue equation' (see e.g. \cite{MNmod} for an account in the
present conventions). Using (\ref{dom5}) in (\ref{dom4}) one can    
compute the small $\kappa_i$ behavior of $v_{121}$. We denote by
$v_{121}^{(I)}$ the contribution from the non-integrated part and by
$v_{121}^{(II)}$ that from the integrated part. One finds
\be
v_{121}^{(I)} \stackrel{\cdot}{=} \frac{\pi}{4}Z^{o_1o_3}Z^{o_2o_4}
[\delta(\kappa_1 + \kappa_3) - \delta(\kappa_1 + \kappa_4)]
-\frac{1}{2}D^{o_1o_2;o_3o_4}\;,
\label{dom6}
\ee
where `$\stackrel{\cdot}{=}$' again indicates that both sides give the same
result for the symmetrized $\kappa_i \ra 0$ limit and
$D^{o_1o_2;o_3o_4} = D^{o_2o_1;o_4o_3} = D^{o_4o_3;o_2o_1}$ is given by
(\ref{i5}). 

Analyzing the small $\kappa_i$ behavior of the integral in (\ref{dom4})
(by splitting it according to 
$\int_0^{\infty} \rmd u = \int_0^{\eps} \rmd u +
\int_{\eps}^{\infty}\rmd u$, $\eps\ra 0^+$) one finds a distributional
term and a regular one. The result is
\ba
&& v_{121}^{(II)} = \frac{\pi}{4}[Z^{o_1o_4}Z^{o_2o_3} +   
Z^{o_1o_3}Z^{o_2o_4}]\delta(\kappa_1+ \kappa_4) \nonum
&& + \int_0^{\infty}\frac{\rmd u}{4\pi}\bigg[ \frac{1}{16 \ch^2 u}
\cF_{a}^{o_1} \cF_{b}^{o_4} C^{aa_3}C^{bb_3}
\cF_{a_3a_2a_1}^{o_2}(i\pi,-u,u) C^{a_2b_2}C^{a_1b_1}
\cF^{o_3}_{b_3b_2b_1}(i\pi,-u,u)^* \nonum
&& \bspace\sspace - \frac{2}{u^2}
(Z^{o_1o_4}Z^{o_2o_3} + Z^{o_1o_3}Z^{o_2o_4})\bigg]\;,
\label{dom8}
\ea
where the integrand is regular for $u \ra 0$. 

For the generalization of the $\Omega$ term (\ref{b9}) one 
obtains in the 1-particle approximation and in the $\kappa_i\rightarrow 0$ 
limit
\be
\Omega^{o_1o_2o_3o_4}(k_1,k_2,k_3,k_4)\bigg|_{k_j = (M\sh \kappa_j,0)}
= - \frac{\pi}{2}\delta(\kappa_1+\kappa_2)\,
Z^{o_1o_2} Z^{o_3o_4}\;.
\label{b12}
\ee

Finally, combining (\ref{dom6}),(\ref{dom8}) and (\ref{b12}) according to 
(\ref{b13}) one sees that, -- as promised in section 2.2 --
all distributional terms drop out when computing the right hand side of
(\ref{b13}). The final result thus does not depend on any prescription
how to take the $\kappa_i \ra 0$ limit and is given by (\ref{i4}),
as asserted.

\newpage
\newappendix{Computation of the {\boldmath $1-4-1$} contribution}
\setcounter{equation}{0}

We start from the general formula Eq.~(\ref{dom1}) with $l=4$.
For $\beta_1\ge\beta_2\ge\beta_3\ge\beta_4$ we have for one of 
the factors occurring in (\ref{dom2})
\ba
&\nspace &
\langle a,\alpha |S^c(0)|b_1,\beta_1;
b_2,\beta_2;b_3,\beta_3;b_4,\beta_4\rangle^{\rm in} =
\nonumber\\
&\nspace &\qquad {\cal F}^c_{ab_1b_2b_3b_4}
(i\pi+\alpha-i\epsilon,\beta_1,\beta_2,\beta_3,\beta_4)
\nonumber\\
&\nspace &\qquad 
+4\pi\Bigl\{\delta_{ab_1}\delta(\alpha-\beta_1){\cal F}^c_{b_2b_3b_4}
(\beta_2,\beta_3,\beta_4)
\nonumber\\
&\nspace & \qquad +\delta(\alpha-\beta_2){\cal F}^c_{db_3b_4}
(\beta_1,\beta_3,\beta_4)S_{b_1b_2,da}(\beta_1-\beta_2)
\nonumber\\
&\nspace &\qquad +\delta(\alpha-\beta_3){\cal F}^c_{deb_4}
(\beta_1,\beta_2,\beta_4)S_{b_2b_3,ef}(\beta_2-\beta_3)
S_{b_1f,da}(\beta_1-\beta_3)
\nonumber\\
&\nspace &\qquad +\delta(\alpha-\beta_4){\cal F}^c_{def}
(\beta_1,\beta_2,\beta_3)S_{b_3b_4,fg}(\beta_3-\beta_4)
\nonumber\\
&\nspace &\qquad \times
S_{b_2g,eh}(\beta_2-\beta_4)S_{b_1h,da}(\beta_1-\beta_4)\Bigr\}\,.
\ea
For the 5-particle form factor we introduce the
reduced form factor through
\be
{\cal F}^a_{a_1a_2a_3a_4a_5}
(\theta_1,\theta_2,\theta_3,\theta_4,\theta_5)=
T_5(\theta_1,\dots,\theta_5)
{\cal G}^a_{a_1a_2a_3a_4a_5}(\theta_1,\theta_2,\theta_3,\theta_4,\theta_5)\,.
\ee
Multiplying out we obtain
\be
I_{141}={\cal K}^{(I)}+{\cal K}^{(II)}+{\cal K}^{(III)}\,,
\ee
where we momentarily omit the arguments
$(\kappa_4,-\kappa_1,\beta_1,\beta_2,\beta_3,\beta_4)$. The shorthands
are: 
\ba
{\cal K}^{(I)} \is \sum_{b_1,b_2,b_3,b_4}{\cal F}^1_{1b_1b_2b_3b_4}
(i\pi-\kappa_1-i\epsilon,\beta_1,\beta_2,\beta_3,\beta_4)
\nonumber
\\
&& \sspace \times {\cal F}^1_{1b_1b_2b_3b_4}
(i\pi+\kappa_4-i\epsilon,\beta_1,\beta_2,\beta_3,\beta_4)^*\,,
\\[5mm]
{\cal K}^{(II)}\is 4\pi\Bigl\{ \delta (\beta_4+\kappa_1)
\bar{\cal K}^{(II)}(\kappa_4,-\kappa_1,\beta_1,\beta_2,\beta_3)
\nonumber\\
&&+\delta (\beta_4-\kappa_4)
\bar{\cal K}^{(II)}(-\kappa_1,\kappa_4,\beta_1,\beta_2,\beta_3)^*\Bigr\}
\nonumber\\
&&+(\beta_4\leftrightarrow\beta_3)
+(\beta_4\to\beta_2,\beta_2\to\beta_3,\beta_3\to\beta_4)
\nonumber\\
&&+(\beta_4\to\beta_1,\beta_1\to\beta_2,\beta_2\to\beta_3,\beta_3\to\beta_4)\,,
\\[5mm]
{\cal K}^{(III)} \is (4\pi)^2
\bar{\cal K}^{(III)}(\kappa_4,-\kappa_1,\beta_3,\beta_4)
\Bigl[ \delta (\beta_1+\kappa_1)\delta (\beta_2-\kappa_4)
+(\beta_1\leftrightarrow\beta_2)\Bigr]
\nonumber\\
&&+(\beta_2\leftrightarrow\beta_3)
+(\beta_2\to\beta_4,\beta_4\to\beta_3,\beta_3\to\beta_2)
\nonumber\\
&&+(\beta_1\to\beta_2,\beta_2\to\beta_3,\beta_3\to\beta_1)
+(\beta_1\to\beta_2,\beta_2\to\beta_4,\beta_4\to\beta_3,\beta_3\to\beta_1)
\nonumber\\
&&
+(\beta_1\to\beta_3,\beta_3\to\beta_2,\beta_2\to\beta_4,\beta_4\to\beta_1)\,,
\ea 
where
\ba
\bar{\cal K}^{(II)}(\alpha,\gamma,\beta_1,\beta_2,\beta_3)
\!\! \is \nspace \sum_{b_1,b_2,b_3}\!\!
{\cal F}^1_{b_1b_2b_3}(\beta_1,\beta_2,\beta_3)
{\cal F}^1_{11b_1b_2b_3}
(i\pi+\alpha-i\epsilon,\gamma,\beta_1,\beta_2,\beta_3)^*
\\[5mm]
\bar{\cal K}^{(III)}(\kappa_4,-\kappa_1,\beta_3,\beta_4)
\!\!\is \nspace \sum_{b_1b_2,b_3,b_4}\!\!
{\cal F}^1_{b_1b_3b_4}(\kappa_4,\beta_3,\beta_4)
{\cal F}^1_{b_2b_3b_4}(-\kappa_1,\beta_3,\beta_4)^*
S_{1b_2,b_1 1}(\kappa_1 \!+ \!\kappa_4)\,.
\ea 
Because of the symmetry in the $\beta_i$ arguments of 
${\cal K}^{(I)},\bar{\cal K}^{(II)},\bar{\cal K}^{(III)}$
one has 
\ba
&\nspace &
v_{141}^{(I)}(\kappa_1,\kappa_2,\kappa_3,\kappa_4)={1\over 12288\pi^3}
\int_{-\infty}^{\infty}\rmd \beta_1 
\int_{-\infty}^{\infty}\rmd \beta_2
\int_{-\infty}^{\infty}\rmd \beta_3
\int_{-\infty}^{\infty}\rmd \beta_4
\nonumber\\
& \nspace &
\phantom{v_{141}^{(I)}(\kappa_1,\kappa_2,\kappa_3,\kappa_4)=}
\times{\delta(\beta_1,\beta_2,\beta_3,\beta_4,\kappa_1,\kappa_2)
\over \sum_{k=1}^4\ch\beta_k}{\cal K}^{(I)}
(\kappa_4,-\kappa_1,\beta_1,\beta_2,\beta_3,\beta_4)\,,
\\[5mm]
& \nspace &
v_{141}^{(II)}(\kappa_1,\kappa_2,\kappa_3,\kappa_4)={1\over 768\pi^2}
\int_{-\infty}^{\infty}\rmd \beta_1
\int_{-\infty}^{\infty}\rmd \beta_2
\int_{-\infty}^{\infty}\rmd \beta_3
{1\over 1+\sum_{k=1}^3\ch\beta_k}
\\
& \nspace &
\quad \Bigl\{ \delta(\beta_1,\beta_2,\beta_3,\kappa_2)
\bar{\cal K}^{(II)}(\kappa_4,-\kappa_1,\beta_1,\beta_2,\beta_3)
+\delta(\beta_1,\beta_2,\beta_3,-\kappa_3)
\bar{\cal K}^{(II)}(-\kappa_1,\kappa_4,\beta_1,\beta_2,\beta_3)^*
\Bigr\}\,,
\nonumber\\[5mm]
&\nspace &
v_{141}^{(III)}(\kappa_1,\kappa_2,\kappa_3,\kappa_4)={1\over 64\pi}
\int_{-\infty}^{\infty}\rmd \beta_1 
\int_{-\infty}^{\infty}\rmd \beta_2
{\delta(\beta_1,\beta_2,\kappa_2,\kappa_4)
\over 2+\sum_{k=1}^2\ch\beta_k}
\bar{\cal K}^{(III)}(\kappa_4,-\kappa_1,\beta_1,\beta_2)\,.
\ea
The contribution $(III)$ is very simple; we can
set the $\kappa_i$ to zero to obtain 
\be
v_{141}^{(III)}(\kappa_1,\kappa_2,\kappa_3,\kappa_4)={1\over
128\pi}\int_{-\infty}^{\infty}\rmd \beta 
{1\over \ch\beta (1+\ch\beta)}
T^2(2\beta)T^4(\beta)k^{(III)}(\beta)
\,,
\ee
where we decomposed 
\be
\bar{\cal K}^{(III)}(0,0,\beta,-\beta)
=T^2(2\beta)T^4(\beta)k^{(III)}(\beta)\,.
\ee
Writing similarly 
\ba
&\nspace & \bar{\cal K}^{(II)}(\alpha,\gamma,\beta_1,\beta_2,\beta_3) =
k^{(II)}(\alpha,\gamma,\beta_1,\beta_2,\beta_3)
\nonum
&\nspace &  \qquad \times {1\over T(\alpha-\gamma)}
\prod_{1\le i<j\le 3} T^2(\beta_i-\beta_j)
\prod_{k=1}^3 {T(\beta_k-\gamma)\over
T(\beta_k-\alpha-i\epsilon)}\,,
\ea
one has 
\be
v_{141}^{(II)}=\sum_{j=1}^2 v_{141}^{(II,j)}\,,
\ee
with 
\ba
&\nspace&
v_{141}^{(II,1)}(\kappa_1,\kappa_2,\kappa_3,\kappa_4)={1\over 768\pi^2}
{1\over T(\kappa_1+\kappa_4)}
\int_{-\infty}^{\infty}\rmd \beta_1
\int_{-\infty}^{\infty}\rmd \beta_2
\int_{-\infty}^{\infty}\rmd \beta_3
\nonum
&\nspace &\quad \times
{1\over 1+\sum_{k=1}^3\ch\beta_k}
\prod_{1\le i<j\le 3}
T^2(\beta_i-\beta_j)
\nonum 
&\nspace &\quad \times \Bigl\{
\delta(\beta_1,\beta_2,\beta_3,\kappa_2)
k^{(II)}(\kappa_4,-\kappa_1,\beta_1,\beta_2,\beta_3)
\prod_{k=1}^3 T(\beta_k+\kappa_1)
{{\cal P}\over T(\beta_k-\kappa_4)}
\nonum
&\nspace &\quad -\delta(\beta_1,\beta_2,\beta_3,-\kappa_3)
k^{(II)}(-\kappa_1,\kappa_4,\beta_1,\beta_2,\beta_3)^*
\prod_{k=1}^3 T(\beta_k-\kappa_4)
{{\cal P}\over T(\beta_k+\kappa_1)}\Bigr\}\,,
\\[3mm]
&\nspace &v_{141}^{(II,2)}(\kappa_1,\kappa_2,\kappa_3,\kappa_4)={i\over 128\pi}
\int_{-\infty}^{\infty}\rmd \beta_1
\int_{-\infty}^{\infty}\rmd \beta_2
{\delta(\beta_1,\beta_2,\kappa_2,\kappa_4)\over
2+\sum_{k=1}^2\ch\beta_k}
\nonumber\\
&\nspace &\quad \times
T^2(\beta_1-\beta_2)
\prod_{k=1}^2 T(\beta_k-\kappa_4)
T(\beta_k+\kappa_1)
\nonumber\\
&\nspace &\quad \times\Bigl\{
k^{(II)}(\kappa_4,-\kappa_1,\beta_1,\beta_2,\kappa_4)
-k^{(II)}(-\kappa_1,\kappa_4,\beta_1,\beta_2,-\kappa_1)^*\Bigr\}\,.
\ea
In the latter term we can set the $\kappa_i$ to zero to obtain
\be
v_{141}^{(II,2)}(0,0,0,0)={-1\over 128\pi}
\int_{-\infty}^{\infty}\rmd \beta
{1\over \ch\beta (1+\ch\beta)}
T^2(2\beta) T^4(\beta)
{\rm Im}\Bigl[
k^{(II)}(0,0,\beta,-\beta,0)\Bigr]\,.
\ee

Lastly we turn to the $(I)$ contribution. There are many ways to 
manipulate the integral into a form more amenable to numerical
evaluation. Here we proceed as follows: Writing 
\ba
{\cal K}^{(I)}&=&
k^{(I)}(\kappa_4,-\kappa_1,\beta_1,\beta_2,\beta_3,\beta_4)
\prod_{1\le i<j\le 4}
T^2(\beta_i-\beta_j)
\nonumber\\
&& \times 
\prod_{k=1}^4 {1\over T(\beta_k+\kappa_1+i\epsilon)
T(\beta_k-\kappa_4-i\epsilon)}\,,
\ea
we replace the $1/(x\pm i\epsilon)$ distributions by a sum of
products of principal parts and delta functions, thereby obtaining
\be
v_{141}^{(I)}=\sum_{j=1}^3 v_{141}^{(I,j)}\,.
\ee
The three terms are: 
\pagebreak[2]
\ba
&\nspace &
v_{141}^{(I,1)}(\kappa_1,\kappa_2,\kappa_3,\kappa_4)={1\over 12288\pi^3}
\nonumber\\
&\nspace & \quad 
\int_{-\infty}^{\infty}\rmd \beta_1 
\int_{-\infty}^{\infty}\rmd \beta_2
\int_{-\infty}^{\infty}\rmd \beta_3
\int_{-\infty}^{\infty}\rmd \beta_4
{\delta(\beta_1,\beta_2,\beta_3,\beta_4,\kappa_1,\kappa_2)
\over \sum_{k=1}^4\ch\beta_k}
\nonumber\\
&\nspace & \quad 
k^{(I)}(\kappa_4,-\kappa_1,\beta_1,\beta_2,\beta_3,\beta_4)
\prod_{1\le i<j\le 4}
T^2(\beta_i-\beta_j)
\prod_{k=1}^4 {{\cal P}\over T(\beta_k+\kappa_1)}
              {{\cal P}\over T(\beta_k-\kappa_4)}\,,
\label{v141I1}
\\[5mm]
&\nspace &
v_{141}^{(I,2)}(\kappa_1,\kappa_2,\kappa_3,\kappa_4)={i\over 1536\pi^2}
{1\over T(\kappa_1+\kappa_4)}
\nonumber\\
&\nspace &\quad 
\int_{-\infty}^{\infty}\rmd \beta_1 
\int_{-\infty}^{\infty}\rmd \beta_2
\int_{-\infty}^{\infty}\rmd \beta_3
{1\over 1+\sum_{k=1}^3\ch\beta_k}
\prod_{1\le i<j\le 3}
T^2(\beta_i-\beta_j)
\nonumber\\
&\nspace &\quad \times\Bigl\{\delta(\beta_1,\beta_2,\beta_3,\kappa_2)
k^{(I)}(\kappa_4,-\kappa_1,\beta_1,\beta_2,\beta_3,-\kappa_1)
\prod_{k=1}^3 T(\beta_k+\kappa_1)
{{\cal P}\over T(\beta_k-\kappa_4)}
\nonumber\\
&\nspace &\quad +\delta(\beta_1,\beta_2,\beta_3,-\kappa_3)
k^{(I)}(\kappa_4,-\kappa_1,\beta_1,\beta_2,\beta_3,\kappa_4)
\prod_{k=1}^3 T(\beta_k-\kappa_4)
{{\cal P}\over T(\beta_k+\kappa_1)}
\Bigr\}
\\[5mm]
&\nspace & 
v_{141}^{(I,3)}(\kappa_1,\kappa_2,\kappa_3,\kappa_4)={-1\over 256\pi}
\int_{-\infty}^{\infty}\rmd \beta_1 
\int_{-\infty}^{\infty}\rmd \beta_2
{\delta(\beta_1,\beta_2,\kappa_2,\kappa_4)
\over 2+\sum_{k=1}^2\ch\beta_k}
\nonumber\\
&\nspace &\quad k^{(I)}(\kappa_4,-\kappa_1,\beta_1,\beta_2,-\kappa_1,\kappa_4)
T^2(\beta_1-\beta_2)
\prod_{k=1}^2
T(\beta_k-\kappa_4)T(\beta_k+\kappa_1)\,.
\ea

In the latter expression we can set the $\kappa_i$ to zero
to obtain
\be
v_{141}^{(I,3)}(0,0,0,0)={-1\over 512\pi}
\int_{-\infty}^{\infty}\rmd \beta 
{1\over \ch\beta (1+\ch\beta)}
T^2(2\beta)T^4(\beta)
k^{(I)}(0,0,\beta,-\beta,0,0)\,.
\ee

For $W^{(1)} := v_{141}^{(I,1)}$ we now invoke the identity
\be
{\prod_{1\le i<j\le 4} \sh (y_i-y_j)\over
\prod_{k=1}^4 \sh (y_k+x)}=
{1\over\ch(x)}\sum_{k=1}^4 {(-1)^k\ch(y_k)\over\sh(y_k+x)}
\prod_{1\le i<j\le 4,\,\,i\ne k\ne j} \sh (y_i-y_j)\,,
\ee
to get
\be
W^{(1)}=\lim_{\alpha_c\to\infty}
\Bigl[W^{(1)}[A](\alpha_c)+W^{(1)}[B](\alpha_c)\Bigr]\,,
\ee
with the notation 
\be 
W^{(1)}[X](\alpha_c)={1\over 192\pi^3}
\int_{-\alpha_c}^{\alpha_c}\rmd \alpha_1 G_X(\alpha_1)\,,
\sspace X =A,B\,.
\ee
Here 
\ba
G_A(\alpha_1)&=&
{{\cal P}\over T(\alpha_1)}
\int_{-\infty}^{\infty}\rmd \alpha_2
{{\cal P}\over T(\alpha_2)}
\int_{-\infty}^{\infty}\rmd \alpha_3
{\cal F}_A(\alpha_1,\alpha_2,\alpha_3)
\nonumber\\
& = &
{{\cal P}\over T(\alpha_1)}
\int_{0}^{\infty}\rmd \alpha_2
{{\cal P}\over T(\alpha_2)}
\int_{0}^{\infty}\rmd \alpha_3
\Bigl\{
{\cal F}_A(\alpha_1,\alpha_2,\alpha_3)
\nonumber\\
&&
-{\cal F}_A(\alpha_1,-\alpha_2,\alpha_3)
-{\cal F}_A(-\alpha_1,\alpha_2,\alpha_3)
+{\cal F}_A(-\alpha_1,-\alpha_2,\alpha_3)\,.
\Bigr\}\,,
\\[5mm]
G_B(\alpha_1) &=& {G(\alpha_1)\over\sh^2 {\alpha_1\over 2}}
-{4G(0)\over \alpha_1^2}\,,
\ea
where
\be
G(\alpha_1)=\ch^2 {\alpha_1\over 2}
\int_{-\infty}^{\infty}\rmd \alpha_2
\int_{-\infty}^{\infty}\rmd \alpha_3
{\cal F}_B(\alpha_1,\alpha_2,\alpha_3)\,.
\ee
In these formulae
\be
{\cal F}_X(\alpha_1,\alpha_2,\alpha_3)={1\over 16}
\left[ { k^{(I)}(0,0,\alpha_1,\alpha_2,\alpha_3,\alpha_4)
f_X(\alpha_1,\alpha_2,\alpha_3,\alpha_4)
         \prod_{k=1}^4\ch^2{\alpha_k\over2}
\over\ch \alpha_4 \Bigl(\sum_{m=1}^4 \ch \alpha_m\Bigr) 
\prod_{i<j} \ch^2{\alpha_i-\alpha_j\over2}}
\right]_{\alpha_4=\gamma}
\ee
where $\gamma$ is given through
$\sh\gamma =-\sum_{k=1}^3\sh\alpha_k\,$
and
\ba
f_A(\alpha_1,\alpha_2,\alpha_3,\alpha_4)&=&
-3\sh{\alpha_1-\alpha_3\over 2}
  \sh{\alpha_1-\alpha_4\over 2}
\nonumber\\
&&\phantom{-}\times
  \sh{\alpha_2-\alpha_3\over 2}
  \sh{\alpha_2-\alpha_4\over 2}
\sh^2{\alpha_3-\alpha_4\over 2}\,,
\\
f_B(\alpha_1,\alpha_2,\alpha_3,\alpha_4)&=&
\sh^2{\alpha_2-\alpha_3\over 2}
\sh^2{\alpha_2-\alpha_4\over 2}
\sh^2{\alpha_3-\alpha_4\over 2}\,.
\ea

For the $[B]$ contribution it is numerically convenient to decompose
\be
W^{(1)}[B](\alpha_c)\sim 
hG(0)+W^{(1)}[B0]+ W^{(1)}[B1](\alpha_c)\,,
\ee
where
\ba
W^{(1)}[B0]&=&{1\over 96\pi^3}
\int_0^1\rmd \alpha_1 \left(
{G(\alpha_1)-G(0) \over\sh^2 {\alpha_1\over 2}}
\right)\,,
\nonumber\\
W^{(1)}[B1](\alpha_c)&=&{1\over 96\pi^3}
\int_1^{\alpha_{cut}}\rmd \alpha_1 
{G(\alpha_1)\over\sh^2 {\alpha_1\over 2}}\,,
\ea
and
\be
h=-{1\over 96\pi^3} 
\Bigl\{ 4-\int_0^1\rmd \alpha_1
\Bigl[
{1 \over\sh^2 {\alpha_1\over 2}}
-{4 \over\alpha_1^2}
\Bigr]
\Bigr\}=-0.0014539754\,.
\ee

Finally we recombine
\be
v_{141}(\kappa_1,\kappa_2,\kappa_3,\kappa_4)
=\sum_{j=1}^3W^{(j)}(\kappa_1,\kappa_2,\kappa_3,\kappa_4)\,,
\ee
with
\ba
W^{(1)}&=&v_{141}^{(I,1)}\,,
\nonumber\\
W^{(2)}&=&v_{141}^{(I,2)}+v_{141}^{(II,1)}\,,               
\\
W^{(3)}&=&v_{141}^{(I,3)}+v_{141}^{(II,2)}
+v_{141}^{(III)}\,.
\nonumber
\ea

{\bf Case {\boldmath $n=1$}:}

Here we simply have (recall the 2-particle S-matrix $=-1$) 
\be
k^{(I)}= 16\,,\qquad k^{(II)}=-8i\,,\qquad k^{(III)}=-4\,,
\label{iskI}
\ee
from which one sees
\ba
v_{141}^{(I,2)}(0,0,0,0)&=&-v_{141}^{(II,1)}(0,0,0,0)\,,
\nonum
v_{141}^{(I,3)}(0,0,0,0)&=&v_{141}^{(III)}(0,0,0,0)
=-\frac12 v_{141}^{(II,2)}(0,0,0,0)\,.
\ea 
Thus 
\be
W^{(2)}=0=W^{(3)}\,,
\ee
so that for the Ising case we simply get $v_{141}=W^{(1)}$, 
with $W^{(1)}$ given by Eq.~(\ref{v141I1}) and $k^{(I)}$ by 
Eq.~(\ref{iskI}). This is, as expected, the same expression as 
that obtained directly with the form factor written as a product 
over principal parts as in Eq.~(\ref{isi1}).

{\bf Case {\boldmath $n=3$}:}

Firstly for $W^{(1)}$ we obtain
\be
W^{(1)}
=-0.0005420(1)\,.
\ee
Next for $W^{(2)}$ one has 
\ba
&\nspace &W^{(2)}(\kappa_1,\kappa_2,\kappa_3,\kappa_4)={1\over 768\pi^2}
{1\over T(\kappa_1+\kappa_4)}
\int_{-\infty}^{\infty}\rmd \beta_1
\int_{-\infty}^{\infty}\rmd \beta_2
\int_{-\infty}^{\infty}\rmd \beta_3
J(\beta_1,\beta_2,\beta_3)
\nonumber\\
&\nspace &\quad \times\Bigl\{
\delta(\beta_1,\beta_2,\beta_3,\kappa_2)
w^{(2)}(\kappa_1,\kappa_4,\beta_1,\beta_2,\beta_3)
\prod_{k=1}^3 T(\beta_k+\kappa_1)
{{\cal P}\over T(\beta_k-\kappa_4)}
\nonumber\\
&\nspace &\quad +\delta(\beta_1,\beta_2,\beta_3,\kappa_3)
\tilde{w}^{(2)}(\kappa_1,\kappa_4,\beta_1,\beta_2,\beta_3)
\prod_{k=1}^3 T(\beta_k+\kappa_4)
{{\cal P}\over T(\beta_k-\kappa_1)}
\Bigr\}\,,
\ea
where
\be
J(\beta_1,\beta_2,\beta_3)=
{1\over 1+\sum_{k=1}^3\ch\beta_k}
\prod_{1\le i<j\le 3}
T^2(\beta_i-\beta_j)\,,
\ee
and
\ba
&\nspace &w^{(2)}(\kappa_1,\kappa_4,\beta_1,\beta_2,\beta_3)=
\nonum
&\nspace &\quad  k^{(II)}(\kappa_4,-\kappa_1,\beta_1,\beta_2,\beta_3)
+{i\over 2}
k^{(I)}(\kappa_4,-\kappa_1,\beta_1,\beta_2,\beta_3,-\kappa_1)\,,
\\ 
&\nspace &\tilde{w}^{(2)}(\kappa_1,\kappa_4,\beta_1,\beta_2,\beta_3)=
\nonum 
&\nspace &\quad -k^{(II)}(-\kappa_1,\kappa_4,-\beta_1,-\beta_2,-\beta_3)^*
+{i\over 2}
k^{(I)}(\kappa_4,-\kappa_1,-\beta_1,-\beta_2,-\beta_3,\kappa_4)\,.
\ea 
Explicit calculation reveals the fact
\be
\tilde{w}^{(2)}(\kappa_1,\kappa_4,\beta_1,\beta_2,\beta_3)=
w^{(2)}(\kappa_4,\kappa_1,\beta_1,\beta_2,\beta_3)\,.
\ee
Now we expand $w^{(2)}$ for small $\kappa_i$:
\ba
&\nspace &w^{(2)}(\kappa_1,\kappa_4,\beta_1,\beta_2,\beta_3)
=w_0(\beta_1,\beta_2,\beta_3)
\nonumber\\
&\nspace & \quad +(\kappa_1+\kappa_4)w_1(\beta_1,\beta_2,\beta_3)
+(\kappa_1-\kappa_4)w_2(\beta_1,\beta_2,\beta_3)
+O(\kappa_i^2)\,.
\ea
In fact we do not require $w_2$.  
Note that the functions $w_i$ are real,
so that in particular 
\be
{\rm Im}\,w^{(2)}(0,0,\beta_1,\beta_2,\beta_3)=0\,,
\label{im0}
\ee
which is needed to avoid a singularity in $W^{(2)}$ for
$\kappa_i\to0$. Hence 
\be
W^{(2)}=W^{(2)}[A]+W^{(2)}[B]\,,
\ee
with 
\be
W^{(2)}[A]={1\over 192\pi^2}
\int_{-\infty}^{\infty}\rmd \beta_1
\int_{-\infty}^{\infty}\rmd \beta_2
{1\over\ch\beta_0}J(\beta_1,\beta_2,\beta_0)
w_1(\beta_1,\beta_2,\beta_0)\,,
\ee
where $\beta_0$ is determined through
$\sh \beta_0=-\sh \beta_1 -\sh \beta_2\,$,
and
\ba
&&W^{(2)}[B](k_1,k_2,k_3,k_4)={1\over 768\pi^2}
{1\over T(\kappa_1+\kappa_4)}
\int_{-\infty}^{\infty}\rmd \beta_1
\int_{-\infty}^{\infty}\rmd \beta_2
\int_{-\infty}^{\infty}\rmd \beta_3
Z(\beta_1,\beta_2,\beta_3)
\nonumber\\
&&\qquad \times\Bigl\{
\delta(\beta_1,\beta_2,\beta_3,\kappa_2)
\prod_{k=1}^3 T(\beta_k+\kappa_1)
{{\cal P}\over T(\beta_k-\kappa_4)}
+(\kappa_2\leftrightarrow\kappa_3,
\kappa_1\leftrightarrow\kappa_4)
\Bigr\}\,,
\ea
with
\be
Z(\beta_1,\beta_2,\beta_3)=
J(\beta_1,\beta_2,\beta_3)
w_0(\beta_1,\beta_2,\beta_3)\,.
\ee
We then see that $W^{(2)}[B]$ is a sum of two parts
\be
W^{(2)}[B]=W^{(2)}[B1]+W^{(2)}[B2]\,,
\ee
with
\ba
W^{(2)}[B1]&=&{1\over 64\pi^2}
\int_{-\infty}^{\infty}\rmd \beta_1
\int_{-\infty}^{\infty}\rmd \beta_2
{1\over\ch\beta_0}
Z(\beta_1,\beta_2,\beta_0)
{{\cal P}\over\sh \beta_1}\,,
\nonum
W^{(2)}[B2]&=&{1\over 384\pi^2}
\int_{-\infty}^{\infty}\rmd \beta_1
\int_{-\infty}^{\infty}\rmd \beta_2
{1\over\ch\beta_0}
{\partial\over\partial\beta_3}
\left({Z(\beta_1,\beta_2,\beta_3)\over\ch\beta_3}\right)_{\beta_3=\beta_0}\,.
\ea
Numerically this gives 
\ba
W^{(2)}[A]&=&4.41085(1)\times10^{-4}\,,
\nonumber\\
W^{(2)}[B1]&=&4.9600(1)\times10^{-5}\,,
\\
W^{(2)}[B2]&=&1.1503(1)\times10^{-5}\,,
\nonumber
\ea
and hence
\be
W^{(2)}=0.00050219(1)\,.
\ee

Finally we turn to the computation of $W^{(3)}$.
Due to Eq.~(\ref{im0}) it follows that
\be
v_{141}^{(II,2)}(0,0,0,0)=-2v_{141}^{(I,3)}(0,0,0,0)\,.
\ee
Now explicit computation yields 
\ba
k^{(I)}(0,0,\beta,-\beta,0,0)&=&
\pi^{-6}|\tau_3(0,\beta,-\beta)|^2
(40\beta^2+32\pi^2)\,,
\nonum
k^{(III)}(\beta)&=&-12\pi^{-6}|\tau_3(0,\beta,-\beta)|^2(\beta^2+\pi^2)\,.
\ea
So for $W^{(3)}$ we arrive at
\ba
W^{(3)}&=&-{\pi^5\over 64}\int_0^{\infty}{\sh^6{\beta\over2}
T^6(\beta)\over\ch^5\beta}
{(\beta^2+\pi^2)(4\beta^2+\pi^2)(\beta^2+2\pi^2)
\over\beta^6 (\beta^2+4\pi^2)^2}
\nonum
&=&-0.0004682756\,.
\ea

\newpage
\newappendix{Computation of the {\boldmath $1-2-3$} contribution}
\setcounter{equation}{0}

We use the results (\ref{v12m}-\ref{v12mx}) with $m=3$,
and begin with contribution $(IV)$:
\ba
&\nspace &v_{123}^{(IV)}(\kappa_1,\kappa_2,\kappa_3,\kappa_4)
\sim{1\over 1536\pi^2}
\int\rmd^3 \alpha
{\delta(\alpha_1,\alpha_2,\alpha_3,-\kappa_4)
\over \sum_{i=1}^3\ch\alpha_i}
\nonumber\\
&\nspace &\bspace   
\times {\cal G}^{(IV)}(-\kappa_2+i\pi_-,-\kappa_1+i\pi_-,
\alpha_1,\alpha_2,\alpha_3)\,.
\ea
Here
\be
{\cal G}^{(IV)}(A)= \sum_{a_1a_2a_3}
{\cal F}^1_{11a_1a_2a_3}(A)
{\cal F}^1_{a_1a_2a_3}(A')^*
= T_5(A)T_3(A')^*g^{(IV)}(A)\,,
\ee
where $A$ stands for $\theta_1,\theta_2,\theta_3,
\theta_4,\theta_5$ and $A'$ for $\theta_3,\theta_4,\theta_5$.
Note that $g^{(IV)}(A)$ is totally symmetric in the subset $A'$.

We decompose the $1/(x\pm i\epsilon)$ factors to obtain
terms involving products of principle parts and delta-functions;
only terms having less than three delta-functions contribute in the Lim
procedure i.e. 
\be
v_{123}^{(IV)}=\sum_{s=1}^3v_{123}^{(IV,s)}\,.
\ee
The terms are 
\ba
&& v_{123}^{(IV,1)}(\kappa_1,\kappa_2,\kappa_3,\kappa_4)
\sim {1\over 1536\pi^2}T(\kappa_1-\kappa_2)
\int\rmd^3 \alpha
{\delta(\alpha_1,\alpha_2,\alpha_3,-\kappa_4)
\over \sum_{i=1}^3\ch\alpha_i}
\nonumber\\
&& \qquad   
\times g^{(IV)}(i\pi,i\pi,\alpha_1,\alpha_2,\alpha_3)
\prod_{i<j}T^2(\alpha_i-\alpha_j)
\prod_k{{\cal P}\over T(\alpha_k+\kappa_1)}
{{\cal P}\over T(\alpha_k+\kappa_2)}\,,\\[5mm]
&& v_{123}^{(IV,2)}(\kappa_1,\kappa_2,\kappa_3,\kappa_4)
\sim -{i\over 128\pi}
\int\rmd^2 \alpha
{\delta(-\kappa_2,\alpha_1,\alpha_2,-\kappa_4)
\over 1+\sum_{i=1}^2\ch\alpha_i}
\nonumber\\
&&\qquad   
\times g^{(IV)}(i\pi,i\pi,0,\alpha_1,\alpha_2)
T^2(\alpha_1-\alpha_2)
\prod_k T(\alpha_k+\kappa_2)
{{\cal P}\over T(\alpha_k+\kappa_1)}\,,
\ea
while for the term involving the product of two delta-functions
one finds 
\be
v_{123}^{(IV,3)}(\kappa_1,\kappa_2,\kappa_3,\kappa_4)=O(\kappa^3)\,.
\ee
The contribution $(IV,1)$ is antisymmetric in 
$\kappa_1\leftrightarrow\kappa_2$ and
so it doesn't contribute in the sum over permutations. In the 
contribution $(IV,2)$ we can take the $\kappa\to 0$ limit and 
obtain
\be
v_{123}^{(IV)}(0,0,0,0)
\sim-{i\over 128\pi}
\int_{-\infty}^{\infty} \rmd \alpha
{T^2(2\alpha)\over \ch\alpha [1+2\ch\alpha]}
g^{(IV)}(i\pi,i\pi,0,\alpha,-\alpha)\,.
\ee

Now turn to the $(III)$ contribution:
\be
v_{123}^{(III)}(\kappa_1,\kappa_2,\kappa_3,\kappa_4)
\sim{1\over 64\pi}
\int\rmd^2 \beta
{\delta(\beta_1,\beta_2,\kappa_1,\kappa_2)
\over \sum_i\ch\beta_i
[1+\sum_j\ch\beta_j]}
{\cal G}^{(III)}(-\kappa_1+i\pi_-,\beta_1,\beta_2,\kappa_3) \,,
\ee
where
\ba
{\cal G}^{(III)}(\th_1,\th_2,\th_3,\th_4)&=&
\sum_{b_1b_2}
{\cal F}^1_{1b_1b_2}(\th_1,\th_2,\th_3)
{\cal F}^1_{b_1b_21}(\th_2,\th_3,\th_4)^*
\nonumber\\
&=&T_3(\th_1,\th_2,\th_3)T_3(\th_2,\th_3,\th_4)^*
g^{(III)}(\th_1,\th_2,\th_3,\th_4)\,.
\ea
Here one can set the $\kappa_i$ to zero to obtain
\be
v_{123}^{(III)}(0,0,0,0)
\sim{1\over 128\pi}
\int_{-\infty}^{\infty}\rmd \beta
{T^2(2\beta)\over \ch^2\beta[1+2\ch\beta]}
g^{(III)}(i\pi,\beta,-\beta,0)\,.
\ee

For the $(II)$ contribution:
\ba
&\nspace &v_{123}^{(II)}(\kappa_1,\kappa_2,\kappa_3,\kappa_4)
\sim{1\over 256\pi^2}
\int\rmd^2 \beta
{\delta(\beta_1,\beta_2,\kappa_1,\kappa_2)
\over \ch\beta_1+\ch\beta_2}
\nonumber\\
&\nspace &\qquad 
\int\rmd^2 \alpha
{\delta(\beta_1,\alpha_1,\alpha_2,-\kappa_4)
\over  \ch\beta_1+\sum_{i=1}^2\ch\alpha_i}
{\cal G}^{(II)}(-\kappa_1+i\pi_-,\beta_1,\beta_2,\alpha_1,\alpha_2)
\,,
\ea
where
\ba
{\cal G}^{(II)}(A)&=&\sum_b\sum_{a_1,a_2}
{\cal F}^1_{1bb}(A')
{\cal F}^1_{ba_1a_2}(B)
{\cal F}^1_{ba_1a_2}(B')^*
\nonumber\\
&=&T_3(A')T_3(B)T_3(B')^*g^{(II)}(A)
\,,
\ea
where $A$ stands for $\theta_1,\theta_2,\theta_3,\theta_4,\theta_5$;
$A'$ stands for $\theta_1,\theta_2,\theta_3$,
$B$ stands for $\theta_3+i\pi_-,\theta_4,\theta_5$,
and $B'$ stands for $\theta_2,\theta_4,\theta_5$.
Making the familiar decomposition of the singular distributions 
one obtains: 
\be
v_{123}^{(II)}=\sum_{s=1}^3 v_{123}^{(II,s)}\,,
\ee
with
\ba
&\nspace & v_{123}^{(II,1)}(\kappa_1,\kappa_2,\kappa_3,\kappa_4)
\sim {1\over 256\pi^2}
\int\rmd^2 \beta
{\delta(\beta_1,\beta_2,\kappa_1,\kappa_2)
\over \ch\beta_1+\ch\beta_2}
\\
&\nspace & \quad 
\int\rmd^2 \alpha
{\delta(\beta_1,\alpha_1,\alpha_2,-\kappa_4)
\over  \ch\beta_1+\sum_{i=1}^2\ch\alpha_i}
g^{(II)}(-\kappa_1+i\pi,\beta_1,\beta_2,\alpha_1,\alpha_2)
T^2(\alpha_1-\alpha_2)
\nonumber\\
&\nspace &\quad 
T(\beta_1-\beta_2)
\prod_i{{\cal P}\over T(\kappa_1+\beta_i)}
\prod_j T(\beta_1-\alpha_j)
{{\cal P}\over T(\beta_2-\alpha_j)}\,,\nonumber \\[5mm]
&\nspace & v_{123}^{(II,2)}(\kappa_1,\kappa_2,\kappa_3,\kappa_4)
\sim {i\over 64\pi}
\int\rmd^2 \beta
{\delta(\beta_1,\beta_2,\kappa_1,\kappa_2)
\over [\ch\beta_1+\ch\beta_2][1+\ch\beta_1+\ch\beta_2]}
\\
&\nspace &\quad 
g^{(II)}(-\kappa_1+i\pi,\beta_1,\beta_2,\beta_2,-\kappa_3)
T^2(\beta_1-\beta_2)
\prod_j T(\beta_j+\kappa_3)
{{\cal P}\over T(\beta_j+\kappa_1)}\,,\nonumber \\[5mm]
&\nspace & v_{123}^{(II,3)}(\kappa_1,\kappa_2,\kappa_3,\kappa_4)
\sim {i\over 256\pi}
\int\rmd^2 \alpha T^2(\alpha_1-\alpha_2)
\\
&\nspace &\quad 
\Bigg\{ {\delta(-\kappa_1,\alpha_1,\alpha_2,-\kappa_4)
\over  1+\sum_{i=1}^2\ch\alpha_i}
g^{(II)}(-\kappa_1+i\pi,-\kappa_1,-\kappa_2,\alpha_1,\alpha_2)
\prod_j T(\alpha_j+\kappa_1){{\cal P}\over T(\alpha_j+\kappa_2)}
\nonumber\\
&\nspace &\quad +{\delta(-\kappa_2,\alpha_1,\alpha_2,-\kappa_4)
\over  1+\sum_{i=1}^2\ch\alpha_i}
g^{(II)}(-\kappa_1+i\pi,-\kappa_2,-\kappa_1,\alpha_1,\alpha_2)
{{\cal P}\over T(\alpha_j+\kappa_1)}\Bigg\} \,,
\nonumber 
\ea
In the contributions $s=2,3$ we can set the $\kappa_i$ to zero
to obtain
\ba
v_{123}^{(II,2)}(0,0,0,0) &\sim &{i\over 128\pi}
\int_{-\infty}^{-\infty}\rmd \beta {T^2(2\beta)
\over \ch^2\beta [1+2\ch\beta]}
g^{(II)}(i\pi,\beta,-\beta,-\beta,0)\,,
\\[5mm]
v_{123}^{(II,3)}(0,0,0,0) &\sim & {i\over 128\pi}
\int_{-\infty}^{-\infty}\rmd \alpha 
{T^2(2\alpha) \over  \ch\alpha [1+2\ch\alpha]}
g^{(II)}(i\pi,0,0,\alpha,-\alpha) \,,
\ea

Finally for the $(I)$ contribution
\ba
&\nspace & v_{123}^{(I)}(\kappa_1,\kappa_2,\kappa_3,\kappa_4)
\sim{1\over 6144\pi^3}
\int\rmd^2 \beta
{\delta(\beta_1,\beta_2,\kappa_1,\kappa_2)
\over \ch\beta_1+\ch\beta_2}
\nonumber\\
&\nspace & \qquad \int\rmd^3 \alpha
{\delta(\alpha_1,\alpha_2,\alpha_3,-\kappa_4)
\over \sum_{i=1}^3\ch\alpha_i} {\cal G}^{(I)}
(-\kappa_1+i\pi_-,\beta_1,\beta_2,\alpha_1,\alpha_2,\alpha_3)\,,
\ea
with 
\ba
{\cal G}^{(I)}(A)&=&
\sum_b\sum_{a_1,a_2,a_3}
{\cal F}^1_{1bb}(A')
{\cal F}^1_{bba_1a_2a_3}(B)
{\cal F}^1_{a_1a_2a_3}(B')^*
\\
&=&T_3(A')T_5(B)T_3(B')^*g^{(I)}(A)\,.
\ea
Here $A$ stands for 
$\theta_1,\theta_2,\theta_3,\theta_4,\theta_5,\theta_6$;
$A'$ stands for $\theta_1,\theta_2,\theta_3$,
$B$ stands for $\theta_3+i\pi_-,\theta_2+i\pi,\theta_4,\theta_5,\theta_6$,
and $B'$ stands for $\theta_4,\theta_5,\theta_6$.

Then we rearrange to terms where after doing the $\beta_2$
integral the singularities in the $\beta_1$ integral all
have negative imaginary parts
\be
v_{123}^{(I)}=\sum_{s=1}^4 v_{123}^{(I,s)}\,.
\ee
The terms are: 
\ba
&\nspace &v_{123}^{(I,1)}(\kappa_1,\kappa_2,\kappa_3,\kappa_4)
\sim-{1\over 6144\pi^3}
\int\rmd^2 \beta
{\delta(\beta_1,\beta_2,\kappa_1,\kappa_2)
\over \ch\beta_1+\ch\beta_2}
\nonumber\\
&\nspace &\qquad \int\rmd^3 \alpha
{\delta(\alpha_1,\alpha_2,\alpha_3,-\kappa_4)
\over \sum_{i=1}^3\ch\alpha_i}
g^{(I)}
(-\kappa_1+i\pi,\beta_1,\beta_2,\alpha_1,\alpha_2,\alpha_3)
\nonumber\\
&\nspace &\qquad \times
{T^2(\beta_1-\beta_2)
\over T(\kappa_1+\beta_1+i\epsilon)
      T(\kappa_1+\beta_2-i\epsilon)}
{\prod_{i<j}T^2(\alpha_i-\alpha_j)
\over\prod_k T(\beta_1+i\epsilon-\alpha_k)
T(\beta_2-i\epsilon-\alpha_k)}\,,
\\[5mm]
&\nspace & v_{123}^{(I,2)}(\kappa_1,\kappa_2,\kappa_3,\kappa_4)
\sim {i\over 512\pi^2}
\int\rmd^2 \beta
{\delta(\beta_1,\beta_2,\kappa_1,\kappa_2)
\over \ch\beta_1+\ch\beta_2}
\nonumber\\
&\nspace &\qquad \int\rmd^2 \alpha
{\delta(\beta_1,\alpha_1,\alpha_2,-\kappa_4)
\over \ch\beta_1+\sum_{i=1}^2\ch\alpha_i}
g^{(I)}  
(-\kappa_1+i\pi,\beta_1,\beta_2,\beta_1,\alpha_1,\alpha_2)
\nonumber\\
&\nspace &\qquad \times
{T(\beta_1-\beta_2)T^2(\alpha_1-\alpha_2)
\prod_i T(\beta_1-\alpha_i)
\over T(\kappa_1+\beta_1+i\epsilon)
T(\kappa_1+\beta_2-i\epsilon)
\prod_j T(\beta_2-i\epsilon-\alpha_j)}
\,,
\\[5mm] 
&\nspace &v_{123}^{(I,3)}(\kappa_1,\kappa_2,\kappa_3,\kappa_4)
\sim{i\over 3072\pi^2}T(\kappa_1-\kappa_2)
\nonumber\\
&\nspace &\qquad \int\rmd^3 \alpha
{\delta(\alpha_1,\alpha_2,\alpha_3,-\kappa_4)
\over \sum_{i=1}^3\ch\alpha_i}
g^{(I)}
(-\kappa_1+i\pi,-\kappa_2,-\kappa_1,\alpha_1,\alpha_2,\alpha_3)
\nonumber\\
&\nspace &\qquad \times  
{\prod_{i<j}T^2(\alpha_i-\alpha_j)\over
\prod_k T(-\kappa_1-i\epsilon-\alpha_k)
T(-\kappa_2+i\epsilon-\alpha_k)} \,,
\\[5mm] 
&\nspace &v_{123}^{(I,4)}(\kappa_1,\kappa_2,\kappa_3,\kappa_4)	
\sim-{1\over 256\pi}T(\kappa_1-\kappa_2)
\nonumber\\
&\nspace &\qquad \int\rmd^2 \alpha
{\delta(-\kappa_2,\alpha_1,\alpha_2,-\kappa_4)
\over 1+\sum_{i=1}^2\ch\alpha_i}
g^{(I)}
(-\kappa_1+i\pi,-\kappa_2,-\kappa_1,-\kappa_2,\alpha_1,\alpha_2)
\nonumber\\
&\nspace &\qquad \times  
{T^2(\alpha_1-\alpha_2)\prod_i T(\kappa_2+\alpha_i)
\over
\prod_j T(\kappa_1+i\epsilon+\alpha_j)} \,.
\ea
As for the last contribution, it vanishes in the limit $\kappa_i\to 0$
\be 
v_{123}^{(I,4)}(0,0,0,0)=0\,.
\ee

For the contribution $(I,1)$ we now perform the $\beta_2$ integral
and shift the $\beta_1$ integral to larger imaginary part,
after which we can send all the $\kappa_i$ to zero to obtain
\ba
&\nspace &v_{123}^{(I,1)}(0,0,0,0)
\sim{1\over 512\pi^3}
\int_{-\infty+i\phi}^{+\infty+i\phi}\rmd \beta
{\ch^4 {\beta\over2}\over \ch^4\beta}
\nonumber\\
&\nspace & \qquad \int_0^{\infty}\rmd u_1\int_0^{\infty}\rmd u_2
T^2(u_1)T^2(u_2)T^2(u_1+u_2) M^{(3)}(u)^{-2}
\nonumber\\
&\nspace & \qquad \times 
g^{(I)}(i\pi,\beta,-\beta,\alpha_1,\alpha_2,\alpha_3)
\prod_k \Bigl({\ch\alpha_k+\ch\beta
         \over\ch\alpha_k-\ch\beta}\Bigr)\,,
\label{v123I1}
\ea
where the $\alpha_k$ are determined in terms of the $u$'s as in 
Eq.~(\ref{trans3}).

For the $(I,2)$ term we obtain
\be
v_{123}^{(I,2)}\sim v_{123}^{(I,5)}+v_{123}^{(I,6)}+O(\kappa)\,,
\ee
where
\ba
&\nspace &v_{123}^{(I,5)}(\kappa_1,\kappa_2,\kappa_3,\kappa_4)
\sim {i\over 512\pi^2}
\int\rmd^2 \beta
{\delta(\beta_1,\beta_2,\kappa_1,\kappa_2)
\over \ch\beta_1+\ch\beta_2}
\nonumber\\
&\nspace &\qquad \int\rmd^2 \alpha
{\delta(\beta_1,\alpha_1,\alpha_2,-\kappa_4)
\over \ch\beta_1+\sum_{i=1}^2\ch\alpha_i}
g^{(I)}  
(-\kappa_1+i\pi,\beta_1,\beta_2,\beta_1,\alpha_1,\alpha_2)
\nonumber\\
&\nspace &\qquad \times
T(\beta_1-\beta_2)T^2(\alpha_1-\alpha_2)
\prod_i{{\cal P}\over T(\kappa_1+\beta_i)}
\prod_j T(\beta_1-\alpha_j)
{{\cal P}\over T(\beta_2-\alpha_j)}\,,
\\[5mm]
&\nspace &v_{123}^{(I,6)}(\kappa_1,\kappa_2,\kappa_3,\kappa_4)
\sim -{1\over 128\pi}
\int\rmd^2 \beta
{\delta(\beta_1,\beta_2,\kappa_1,\kappa_2)
\over \sum_i\ch\beta_i[1+\sum_j\ch\beta_j]}
\nonumber\\
&\nspace &\qquad 
g^{(I)}  
(-\kappa_1+i\pi,\beta_1,\beta_2,\beta_1,-\beta_2,-\kappa_3)
T^2(\beta_1-\beta_2)
\prod_i T(\beta_i+\kappa_3)
{{\cal P}\over T(\beta_i+\kappa_1)} \,.
\ea
In the latter we can do the $\beta_2$ integral and set the $\kappa_i$
to zero to obtain
\be
v_{123}^{(I,6)}(0,0,0,0)\sim
-{1\over 256\pi}
\int_{-\infty}^{\infty}\rmd \beta
{T^2(2\beta) \over \ch^2\beta [1+2\ch\beta]}
g^{(I)}(i\pi,\beta,-\beta,\beta,-\beta,0) \,.
\ee
Finally 
\be 
v_{123}^{(I,3)}(0,0,0,0)
=-{1\over 256\pi}
\int_{-\infty}^{\infty}\rmd \alpha
{T^2(2\alpha)
\over \ch\alpha[1+2\ch\alpha]}
g^{(I)}(i\pi,0,0,\alpha,-\alpha,0) \,.
\ee

Summarizing the results we have
\be
v_{123}(0,0,0,0)=\sum_{j=1}^5V^{(j)}\,,
\ee
where the five terms are as follows: 
\be
V^{(1)}=v_{123}^{(I,1)}(0,0,0,0)\,,
\ee
given in Eq.~(\ref{v123I1}). Further
\ba
V^{(2)}&=&v_{123}^{(III)}(0,0,0,0)+v_{123}^{(IV)}(0,0,0,0)
\nonumber\\
&=&{1\over 128\pi}
\int_{-\infty}^{\infty}\rmd \beta
{T^2(2\beta)\over \ch^2\beta}g^{(2)}(\beta) \,,
\ea
where
\be
g^{(2)}(\beta)={g^{(III)}(i\pi,\beta,-\beta,0)-
ig^{(IV)}(i\pi,i\pi,0,\beta,-\beta)\ch\beta\over
1+2\ch\beta}\,.
\ee
Next
\ba  
V^{(3)}&=&v_{123}^{(I,5)}(0,0,0,0)+v_{123}^{(II,1)}(0,0,0,0)
\nonumber\\
&=&-{1\over 512\pi^2}  
\int_{-\infty}^{\infty}\rmd \beta
{\ch^2{\beta\over2}\over \ch^3\beta}
{{\cal P}\over T(\beta)}
\int\rmd^2 \alpha
{\delta(\beta,\alpha_1,\alpha_2)
\over  \ch\beta+\sum_{i=1}^2\ch\alpha_i}
g^{(3)}(\beta,\alpha_1,\alpha_2)
\nonumber\\
&& 
\times T^2(\alpha_1-\alpha_2)
\prod_j T(\alpha_j-\beta){{\cal P}\over
T(\alpha_j+\beta)}
\,,
\ea
where
\be
g^{(3)}(\beta,\alpha_1,\alpha_2)=
2g^{(II)}(i\pi,\beta,-\beta,\alpha_1,\alpha_2)
+ig^{(I)}(i\pi,\beta,-\beta,\beta,\alpha_1,\alpha_2)\,.
\ee
Further
\ba
V^{(4)}&=&v_{123}^{(I,3)}(0,0,0,0)+v_{123}^{(II,3)}(0,0,0,0)
\nonumber\\
&=&-{1\over 256\pi}
\int_{-\infty}^{\infty}\rmd \alpha
{T^2(2\alpha)
\over \ch\alpha[1+2\ch\alpha]}g^{(4)}(\alpha)\,,
\ea
where
\be
g^{(4)}(\alpha)=-ig^{(3)}(0,\alpha,-\alpha)\,.
\ee
Finally 
\ba  
V^{(5)}&=&v_{123}^{(I,6)}(0,0,0,0)+v_{123}^{(II,2)}(0,0,0,0)\,,
\nonumber\\
&=&-{1\over 256\pi}
\int_{-\infty}^{\infty}\rmd \beta
{T^2(2\beta)
\over \ch^2\beta [1+2\ch\beta]}g^{(5)}(\beta) \,,
\ea
where
\be
g^{(5)}(\beta)=-ig^{(3)}(\beta,-\beta,0)\,.
\ee

\pagebreak[2]
{\bf Case {\boldmath $n=1$}:}

Here we have simply
\ba
g^{(I)}(A) \is -16\,,\sspace  g^{(II)}(B)=8 i\,,\nonum
g^{(III)}(C)\is 4\,,\sspace \quad g^{(IV)}(D)=8 i\,,
\ea
and so
\be
g^{(2)}(\beta)= 4\,,\sspace g^{(r)}=0\,\,,r=3,4,5\,.
\ee
Thus
\ba
V^{(2)} \is {1\over 32\pi} \int_{-\infty}^{\infty}\rmd \beta
{\sh^2 \beta\over \ch^4\beta}={1\over 48\pi}\,,
\nonumber\\[5mm]
V^{(1)} \is -{1\over 128\pi^3}
\int_0^{\infty}\rmd^2 u
T^2(u_1)T^2(u_2)T^2(u_1+u_2)
{S(u_1,u_2)\over M^{(3)}(u)^2}\,,
\ea
with
\be
S(u_1,u_2)=
\int_{-\infty+i\phi}^{+\infty+i\phi}\rmd \beta
{[1+\ch\beta]^2\over \ch^4\beta}
\prod_{k=1}^3 \left({\ch\alpha_k+\ch\beta
         \over\ch\alpha_k-\ch\beta}\right) \,.
\ee
Numerically we find
\be
V^{(1)}=-0.000842721(1)\,.
\ee

{\bf Case {\boldmath $n=3$}:}

First we have
\be
V^{(1)}={\pi^9\over 32768}
\int_0^{\infty}\rmd^2 u
|\psi(u_1)\psi(u_2)\psi(u_1+u_2)|^2
M^{(3)}(u)^{-2}S(u_1,u_2)\,,
\ee
where
\be
|\psi(u)|^2={u^2+\pi^2\over u^2(u^2+4\pi^2)}T^4(u)\,,
\ee
and
\ba
&\nspace &S(u_1,u_2)=
\int_{-\infty+i\phi}^{+\infty+i\phi}\rmd \beta
{(1+\ch\beta)^4\over \ch^6\beta}
h^{(I)}(i\pi,\beta,-\beta,\alpha_1,\alpha_2,\alpha_3)
\nonumber\\
&\nspace &\qquad 
{(4\beta^2+\pi^2)\over (\beta^2+\pi^2)^3}
\prod_k {\alpha_k^2-\beta^2\over
(\alpha_k^2-\beta^2)^2+2\pi^2(\alpha_k^2+\beta^2)+\pi^4}
\left({\ch\alpha_k+\ch\beta
         \over\ch\alpha_k-\ch\beta}\right)^2 \,.
\ea
Numerically this gives
\be
V^{(1)}=-0.000844527(1)\,.
\ee

Doing the contractions yields 
\ba
g^{(III)}(i\pi,\beta,-\beta,0)&=&{\pi^6\over4}{(2\beta^2+\pi^2)\over
(\beta^2+\pi^2)^2}{T^2(2\beta)\over\beta^2}\,,
\nonum
g^{(IV)}(i\pi,i\pi,0,\beta,-\beta)&=&2ig^{(III)}(i\pi,\beta,-\beta,0)\,.
\ea
Thus
\be
g^{(2)}(\beta)=g^{(III)}(i\pi,\beta,-\beta,0)\,,
\ee
and 
\be
V^{(2)}={\pi^5\over512}
\int_{-\infty}^{\infty}\rmd \beta
{\sh^4 \beta\over \beta^2\ch^6\beta}
{(2\beta^2+\pi^2)(4\beta^2+\pi^2)\over(\beta^2+\pi^2)^2
(\beta^2+4\pi^2)}=0.0074380765\,.
\ee

Next by explicit computation one verifies 
\be
g^{(4)}(\alpha)=0\,,\sspace g^{(5)}(\beta)=-g^{(5)}(-\beta)\,,
\ee
and thus%
\footnote{Eq.~(\ref{V45}) is perhaps true for all $n$ but we have not
verified this conjecture}
\be
V^{(4)}=V^{(5)}=0\,.
\label{V45}
\ee

It remains to compute $V^{(3)}$. Shifting the $\alpha_1$ integral
we obtain the representation
\ba
&& V^{(3)}={1\over256\pi^2}\int_{-\infty}^{\infty}\rmd\beta
\int_{-\infty}^{\infty}\rmd\alpha {{\cal P}\over\sh{\beta\over2}}
{\ch^2{\beta\over2}\over\ch^3\beta}
{{\cal P}\over T(\alpha)}
\nonumber\\
&&\quad {g^{(3)}(\beta,\alpha_1,\alpha_2)\ch{\alpha_1\over2}
\ch{\alpha_2+\beta\over2}\over
\ch\alpha_2\ch{\alpha_1-\alpha_2\over2}[\ch\beta+\ch\alpha_1+\ch\alpha_2]}
T(\alpha_1-\alpha_2)T(\alpha_1-\beta)
T(\alpha_2-\beta)\ch{\alpha_1\over2}
\,,
\ea
where 
\be
\alpha_1=\alpha-\beta\,,
\ee
and $\alpha_2$ is determined by 
\be
\sh\alpha_2=-\sh\beta-\sh\alpha_1\,.
\ee
Numerically this gives
\be
V^{(3)}=-0.000125112(1)\,.
\ee

\vfill
\eject
\newpage
\newappendix{Building blocks of form factors}
\setcounter{equation}{0}

In form factor calculations one often encounters the problem of
finding an analytic function $f(\theta)$ satisfying
\ba
f(\theta)&=&\sigma(\theta)f(-\theta)\,,\nonum
f(i\pi-\theta)&=&f(i\pi+\theta)\,,
\label{min1}
\ea
for given $\sigma(\th)$. If $\sigma(\theta)$ has the Fourier 
representation
\be
\sigma(\theta)=e^{i\delta(\theta)}\,,
\qquad\qquad
\delta(\theta)=2\int_0^\infty\frac{\rmd\omega}{\omega}\,
\sin(\theta\omega)\,\tilde k(\omega)\,,
\label{Fourier1}
\ee
with some kernel function $\tilde k(\omega)$ then the
\lq minimal' solution of (\ref{min1}) is given by \cite{KaWe}
\be
f(\theta)=e^{\Delta(\theta)}\,,
\qquad\qquad
\Delta(\theta)=\int_0^\infty\frac{\rmd\omega}{\omega}\,
\frac{\ch\omega(\pi+i\theta)-1}{\sh\pi\omega}\,
\tilde k(\omega)\,.
\label{Fourier2}
\ee
The function $\Delta(\theta)$ has the following properties.
If $\tilde k(\omega)\sim
e^{-z\omega}$ ($z>0$) for $\omega\to\infty$ 
then $\Delta(\theta)$ is analytic
for $-z<{\rm Im}\,\theta<2\pi+z$ and for real 
$\theta\to\infty$
\be
{\rm Re}\,\Delta(\theta)\sim\Delta(i\pi+\theta)\sim
-\frac{\theta}{2}\,\tilde k(0)-\frac{\ln\theta}{\pi}\,
\tilde k^\prime(0)+{\rm const.}
\label{asyDelta}
\ee

We encountered in Section 6 the following special case:
for some (positive, real) parameter $\alpha$
\be
\sigma_\alpha(\theta)=e^{i\delta_\alpha(\theta)}=
\frac{(1+\alpha)i\pi+\theta}{(1+\alpha)i\pi-\theta}\,,
\label{deltaalpha}
\ee
corresponding to the kernel
\be
\tilde k_\alpha(\omega)=-e^{-\pi\omega(1+\alpha)}\,.
\label{tildekalpha}
\ee
We denote the corresponding solution by $\Delta_\alpha(\theta)$.

\end{document}